\documentclass[ aip, amsmath,amssymb, reprint]{revtex4-1}

\usepackage{color,xcolor}
\usepackage{graphicx}
\usepackage{dcolumn}
\usepackage{bm}
\usepackage{verbatim}

\usepackage[utf8]{inputenc}
\usepackage[T1]{fontenc}
\usepackage{mathptmx}
\usepackage{makecell}

\usepackage[colorlinks=true,bookmarks=false,citecolor=blue,urlcolor=blue]{hyperref}

\begin{document}
 

\title[Nonlinear topological photonics]{Nonlinear topological photonics}

\author{Daria Smirnova}

\affiliation{Nonlinear Physics Centre, Australian National University, Canberra ACT 2601, Australia}

\author{Daniel Leykam}

\affiliation{Center for Theoretical Physics of Complex Systems, Institute for Basic Science (IBS), Daejeon 34126, Republic of Korea}

\author{Yidong Chong}

\affiliation{Division of Physics and Applied Physics, School of Physical and Mathematical Sciences, Nanyang Technological University, Singapore 637371}

\author{Yuri Kivshar}
\email{yuri.kivshar@anu.edu.au}
\affiliation{Nonlinear Physics Centre, Australian National University, Canberra ACT 2601, Australia}

\date{\today}

\begin{abstract}
Rapidly growing demands for fast information processing have launched a race for creating compact and highly efficient
optical devices that can reliably transmit signals without losses. Recently discovered topological phases of light provide a novel ground for photonic devices robust against scattering losses and disorder. Combining
these topological photonic structures with nonlinear effects will unlock advanced functionalities such as nonreciprocity
and active tunability. Here we introduce the emerging field of nonlinear topological photonics and highlight recent
developments in bridging the physics of topological phases with nonlinear optics. This includes a design of novel photonic
platforms which combine topological phases of light with appreciable nonlinear response, self-interaction effects
leading to edge solitons in topological photonic lattices, nonlinear topological circuits, active photonic structures exhibiting
lasing from topologically-protected modes, and harmonic generation from edge states in topological arrays and
metasurfaces. We also chart future research directions discussing device applications such as mode stabilization
in lasers, parametric amplifiers protected against feedback, and ultrafast optical switches employing topological
waveguides.
\end{abstract}

\maketitle

\tableofcontents

\section{Introduction}

Topological insulators are a special, recently discovered class of solids which are insulating in a bulk but conducting at their surfaces due to the existence of scattering-resistant topological edge states. In recent years, it was revealed that the concepts of topological phases are not restricted by fermionic states and solid state systems, but they can also be realized in electromagnetic structures such as photonic crystals and metamaterials~\cite{Lu2016,Ozawa2019,Xie2018}. The rapidly growing interest in the study of topological effects in photonics is motivated by a grand vision of waveguiding and routing light within optical circuits in a manner that is robust against scattering by disorder, due to the inherent features of topological edge states. 

The initial study of photonic topological effects was largely inspired by direct analogies with similar effects discovered for solids. Topological effects in condensed matter systems arise from the presence of topologically nontrivial energy bands of the electron wavefunctions. Electromagnetic waves in periodic media also form band structures, which can likewise contain topologically nontrivial bands. However, there are a number of important distinctions between photonic systems and their condensed matter counterparts, such as the bosonic nature of photons and the presence of absorption and radiation losses that make photonic systems intrinsically non-Hermitian~\cite{Longhi_2017}. Thus, the concepts of topology have become a significant guiding scheme in the search for both novel designs of photonic devices as well as novel physical effects and their applications. Topological edge states have now been predicted and realized in a wide variety of photonic systems, which include gyromagnetic photonic crystals, arrays of coupled optical resonators, metamaterials, helical waveguide arrays, and microcavity polaritons~\cite{Lu2016,Ozawa2019,Xie2018}. 

\begin{figure*}
    \centering
    \includegraphics[width= 0.9\linewidth]{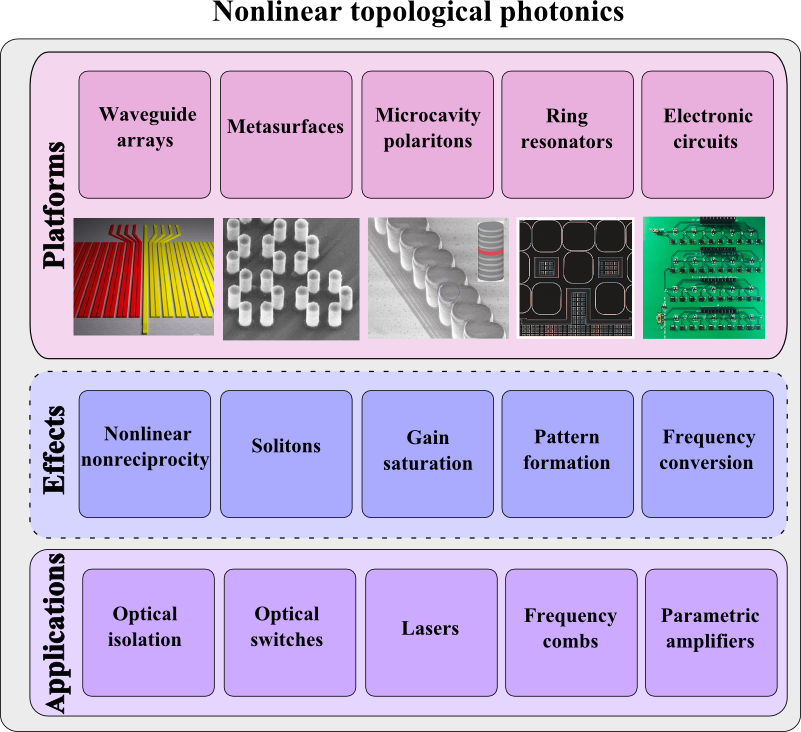} 
    \caption{Basic platforms, selected physical effects, and potential device applications of nonlinear topological photonics.}
    \label{fig:introduction}
\end{figure*}

The nonlinear regime is natural to consider at higher optical powers, and therefore the fundamental question arises: {\em What effects do nonlinearities have on topological phases and edge states, and vice versa?} In particular, the concept of band topology is inherently tied to linear systems---specifically, the existence of a bandgap structure---and the generalization to nonlinear systems is not straightforward. Nonlinear response in photonics and related fields such as Bose-Einstein condensates is expected to open a door towards advanced functionalities of topological photonic structures, including active tunability, genuine nonreciprocity, frequency conversion, and entangled photon generation~\cite{Zhou2017,Kartashov2017, Segev2018b,Chen2018,Mittal2018,Kruk2018,Leykam2018,Amo2018,Wang2019} (see Fig.~\ref{fig:introduction}). In addition, nonlinearities may provide a simple way to reconfigure and control topological waveguides~\cite{Barik2018NN,Shalaev2018NatNano}; in particular, they are required for ultrafast optical modulation~\cite{Husko2009,Eggleton2012}. Such studies are still in their initial stage, and will  uncover many surprises. 

Here, we review the recent advances in the emerging field of nonlinear topological photonics, focusing on the intersection between the studies of topological phases and nonlinear optics. We also describe the broader context of nonlinear effects in other engineered topological systems, including electronic and mechanical metamaterials. We omit discussions of Maxwell surface waves~\cite{Silveirinha2016,Bliokh2019} and the active research topic of using topological electronic materials for nonlinear optics applications~\cite{Hsieh2011,Zhao2012,Chen2014}. Our primary focus here is on artificial topological meta-structures that can be created using mature fabrication techniques and available platforms (silicon, lithography, etc), which are most feasible for near-term device applications.

This review paper is organized as follows. Section~\ref{sec:topo_intro} begins with a brief introduction into the field of topological photonics; for detailed reviews we suggest more comprehensive articles~\cite{Lu2016,Ozawa2019,Xie2018}. Section~\ref{sec:lattices_intro} describes representative topological photonic systems that can be employed for the study of nonlinear effects. In Sec.~\ref{sec:nonlinear_intro}, we discuss how to introduce nonlinear effects to topological photonic media. Section~\ref{sec:localized} reviews recent theoretical and experimental results on nonlinear localization in topological systems. Section~\ref{sec:electronics} discusses electronic circuit implementations of nonlinear topological systems. Nonlinear saturable gain leading to topological lasers is the subject of Sec.~\ref{sec:lasers}. We then discuss 
nonlinear nanophotonics in Sec.~\ref{sec:parametric}. Finally, Sec.~\ref{sec:outlook} concludes with a discussion of future prospects and open problems.

\section{Background}  
\label{sec:topo_intro}

Materials supporting topologically protected edge states were first discovered in condensed matter physics~\cite{bernevig2013}. Such states commonly occur in special types of solids under an applied magnetic field (the so-called {\em quantum Hall} phase~\cite{Thouless1982}) or due to the spin-orbit interaction (the {\em quantum spin-Hall} phase~\cite{Kane2005}). An example of two-dimensional (2D) topological insulators with spin-orbit interaction are hetero-structures with quantum wells~\cite{Bernevig2006,Konig2007}, whose spectrum of edge states was calculated back in the 1980s~\cite{Volkov1985,Gerchikov1990}.
Topological phases also appear in three-dimensional (3D) materials with strong spin-orbit coupling~\cite{Fu2007}. 

Topology is a field of mathematics concerned with the subtle global properties of objects. Topological properties are identified on the basis of continuous deformations: if some property is unaffected by such deformations, it is classified as topological and can be assigned {\it a topological invariant}. A {\it topological phase transition} is accompanied by a step-wise (quantized) change of this invariant.

For example, a closed 2D surface of a finite 3D object can be characterized by the {\it genus} $g$, which counts the number of holes in the object. Thus, a sphere has a genus of $g=0$, and a torus has a genus of $g=1$; these two objects can not transform continuously into each other. Topology can be formally linked to geometry, which describes local specifics, via the Gauss-Bonnet theorem stating that the genus can be calculated by integrating the Gaussian curvature over the entire surface. 

Under certain conditions, topological invariants can be assigned to the band structures of periodic crystalline materials. When two materials are topologically distinct, peculiar {\it boundary states} can arise at the physical interface between those media. This relationship between bulk topology and the existence of boundary states is called the {\it bulk-boundary correspondence}.

\begin{figure}
\centerline{\includegraphics*[width=0.9\linewidth]{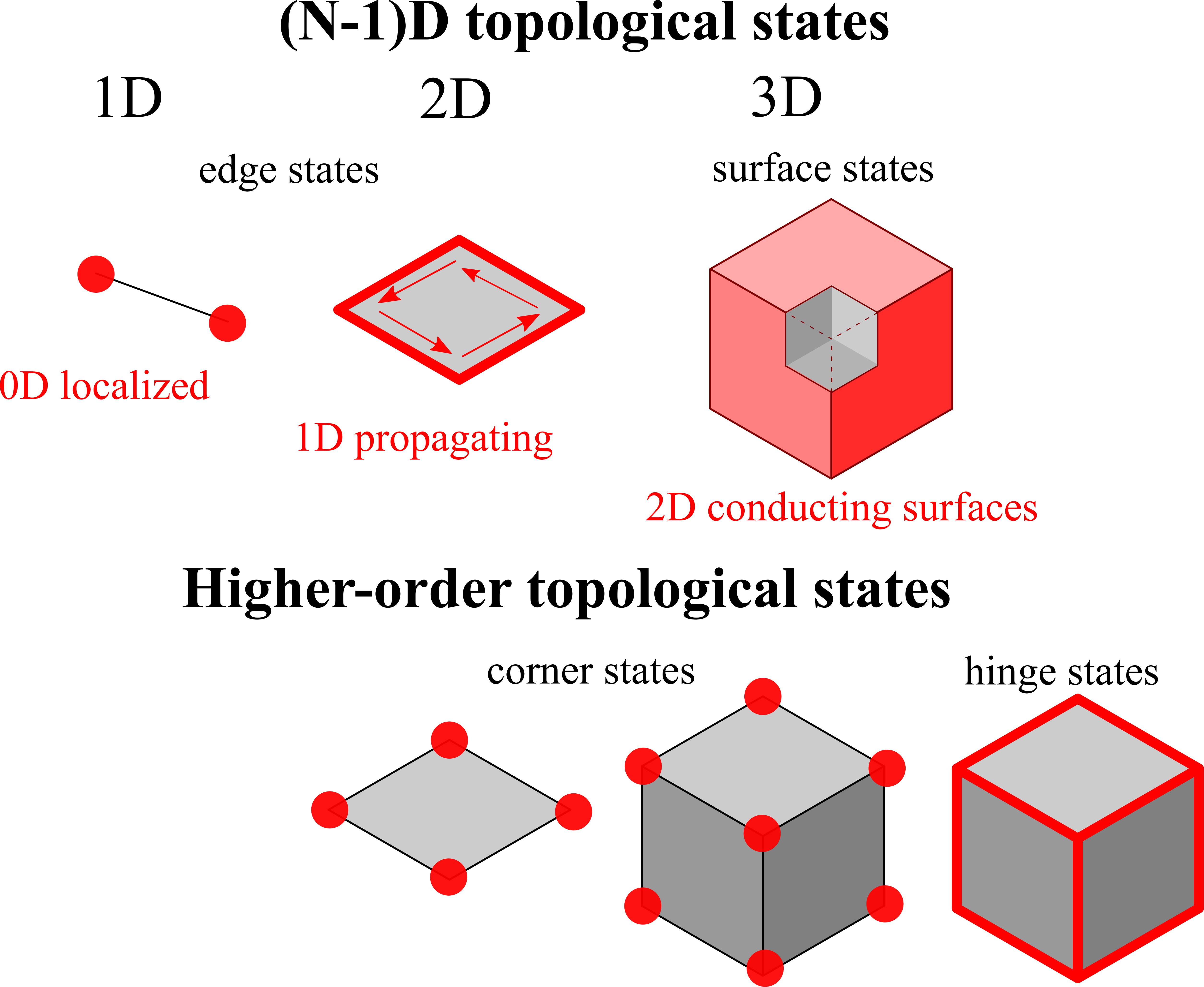}}
\caption{Topological classes of matter in different dimensions. Conventional topological insulators are insulating in a bulk but conducting via gapless states at their edges or surfaces. Higher-order topological insulators have topological states at corners or hinges. \label{cubes3}}%
\end{figure}

Topological phases can be classified according to their dimensionality~\cite{Manoharan2010}, as shown in Fig.~\ref{cubes3}. Topological boundary states can occur at the ends of one-dimensional (1D) systems, the edges of a two-dimensional (2D) systems, or the surfaces of a three-dimensional (3D) systems, as shown schematically in Fig.~\ref{cubes3} (upper row). Generally, an $N$ dimensional topological insulator has $N$ dimensional gaped bulk states and $(N-1)$ dimensional boundary states. For example, the boundaries of 1D systems are end points, and a 1D topological insulator has two end states whose energies are pinned to the middle of the band gap.

\begin{figure} [b]
\centerline{\includegraphics*[width=0.99\linewidth]{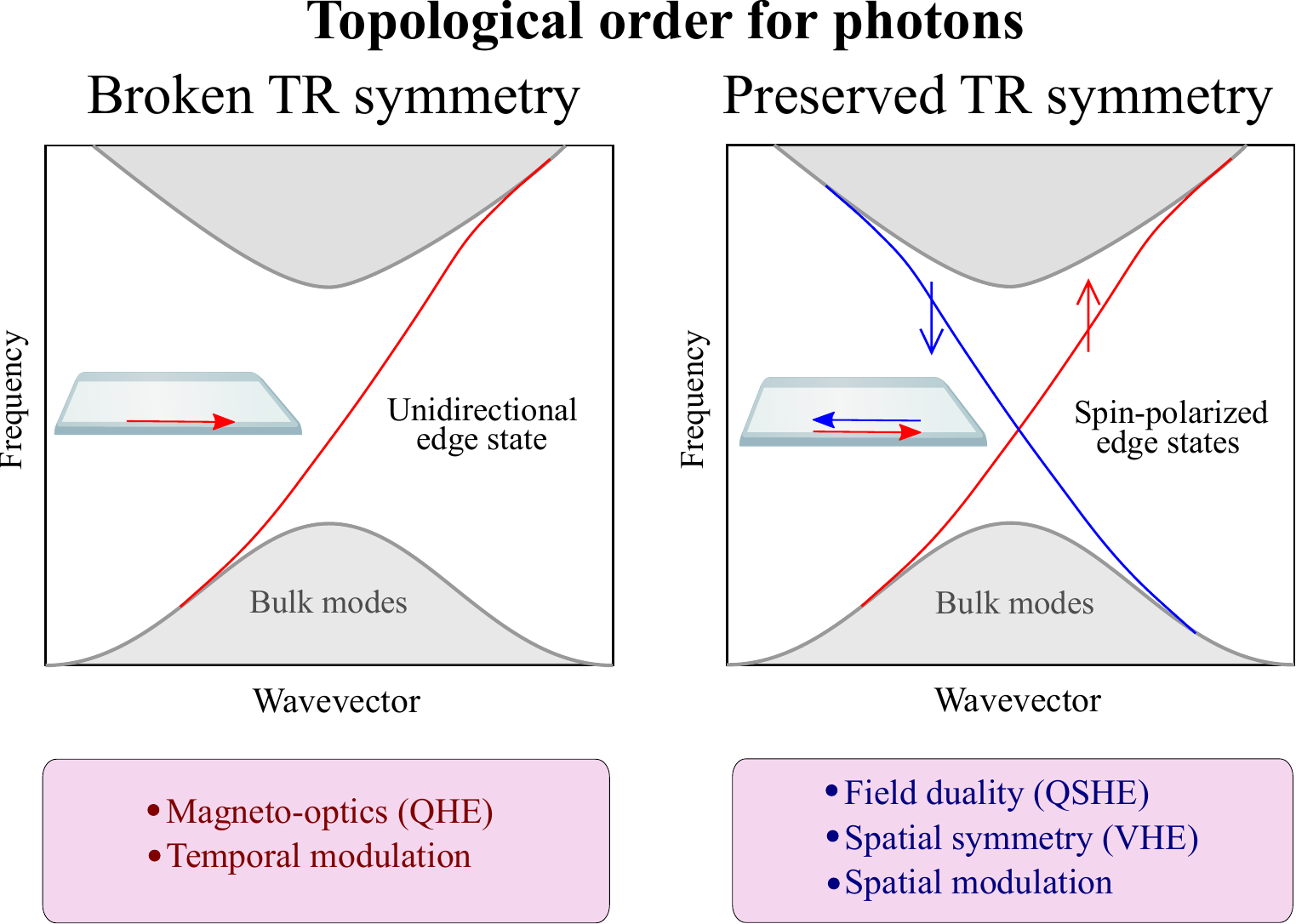}}
\caption{Schematic of the energy spectra for two distinct types of topologically nontrivial photonic systems. The structures are designed to emulate the quantum Hall effect (QHE) and quantum spin- and valley-Hall effects (QSHE and VHE). \label{fig:TopoPhases2D}}
\end{figure}

The 2D case is particularly notable, as there is an assortment of 2D topological phases with strikingly different properties and physical requirements. The simplest of these is the quantum Hall (QH) phase formed by a 2D electron gas in a static magnetic field. This topological phase requires time-reversal (TR) symmetry to be broken, and its topological boundary states are called chiral edge states and propagate unidirectionally along the boundary, immune to backscattering from disorder. Another 2D topological phase, the quantum spin Hall (QSH) phase, preserves TR symmetry but requires the existence of a spin degree of freedom with spin-orbit interactions. It supports pairs of counter-propagating edge modes with opposite spins, called helical edge states, which are protected from backscattering into each other. 
The analogue of QSH phase can be reproduced in topological crystalline insulators (TCI) respecting TR symmetry and a certain point-group symmetry of the lattice~\cite{Wu2015}. The interface between distinct inversion-breaking topological insulators in the valley-Hall (VH) phase also hosts valley-polarized edge states~\cite{Semenoff2008,Yao2009}.

\begin{figure*} [t!]
\centerline{\includegraphics*[width=0.8\linewidth]{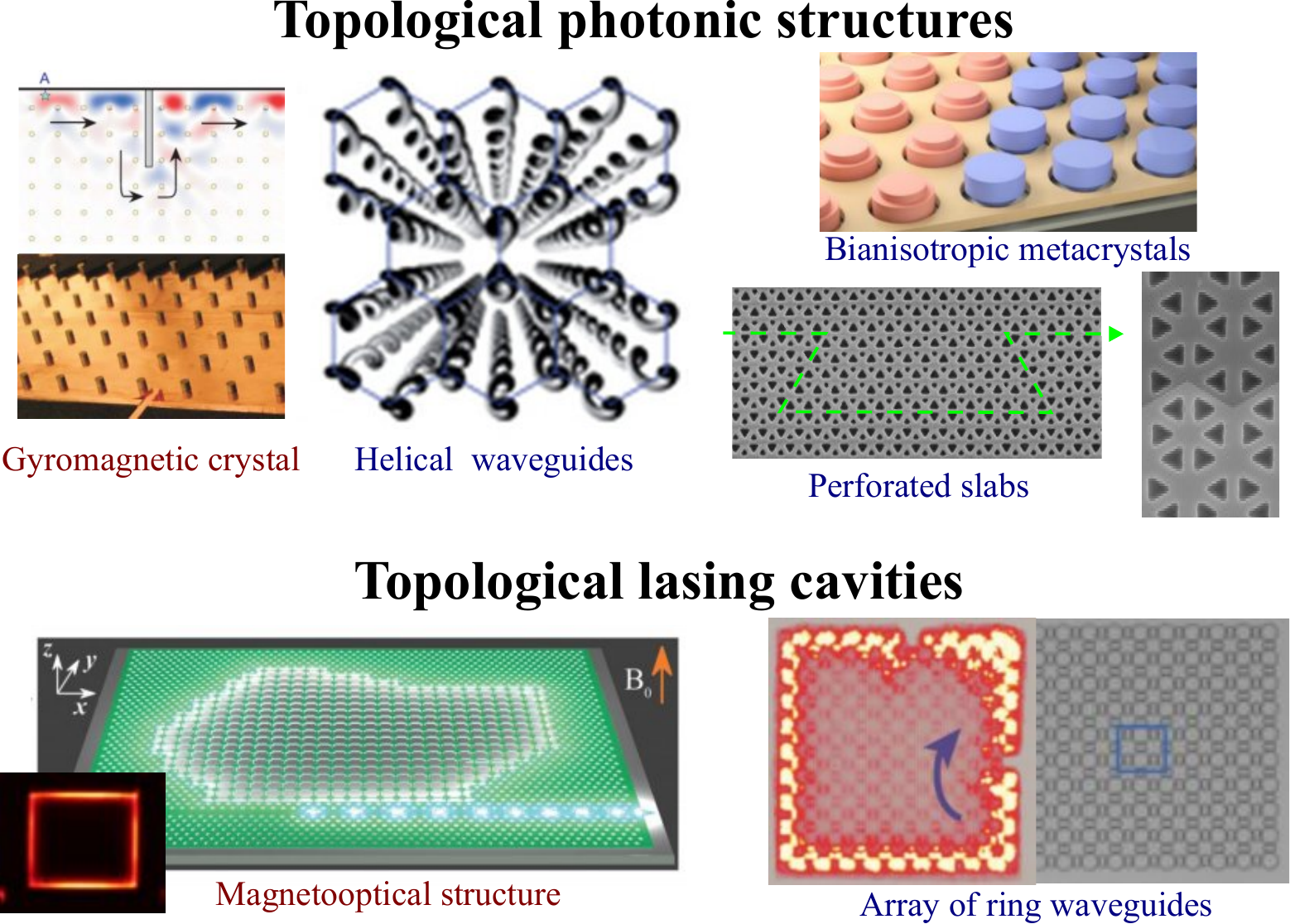}}
\caption{The TR-broken and TR-preserved approaches are illustrated by the images of photonic structures fabricated for specific applications in microwaves and optics (images adapted from Refs.~\cite{Wang2008,rechtsman2013photonic,Slobozhanyuk2019,Shalaev2018NatNano,Amo2018,Bahari2017,Segev2018b}). \label{TopoPhases2D_real}}
\end{figure*}

Recently, new classes of so-called {\em higher-order} topological insulators in dimensions $N > 1$ have been discovered~\cite{Benalcazar2017}. Higher-order topological insulators have $(N-1)$--dimensional boundaries that, unlike those of conventional topological insulators, do not exhibit gapless states but instead constitute topological insulators themselves. An $n$-th order insulator has gapless states on an $(N-n)$-dimensional subsystem. For instance, in three dimensions, a second-order topological insulator has gapless states on the 1D hinges between distinct surfaces, and a third-order topological insulator has gapless states on its 0D corners, as shown in Fig.~\ref{cubes3} (lower row). Similarly, a second-order topological insulator in 2D also has mid-gap corner states. 


There are numerous approaches to engineering photonic topological structures~\cite{Lu2014,Lu2016,Khanikaev2017,Sun2017,Xie2018,Rider2019}, which can be subdivided into TR-broken systems, which require an external magnetic bias, and TR-preserved routes, which do not. One may separately distinguish a group of Floquet TIs that involve temporal or spatial modulation. Some notable implementations of 2D topological photonics are illustrated in Fig.~\ref{fig:TopoPhases2D}, following Ref.~\onlinecite{Khanikaev2017}. In many cases, the structures are designed to emulate topological materials studied in condensed matter physics~\cite{Shen2013}. 

Photonic analogues of QH systems can be realized via gyroelectric or gyromagnetic photonic crystals, where the gyrotropy effect breaks TR symmetry. The first demonstration of backscattering-immune photonic topological edge states with the use of a gyrotropic microwave photonic crystal was performed by Wang \textit{et al.}~in 2009~\cite{Wang2009} following a theoretical proposal by Raghu and Haldane~\cite{Haldane2008,Raghu2008,Wang2008}. However, approaches with preserved TR are preferential in optics due to the difficulty of integrating magnetic materials with optical circuity, and the fact that magnetic responses are weak at optical frequencies. TR unbroken photonic topological systems have been designed using waveguide arrays~\cite{rechtsman2013photonic}, coupled resonators~\cite{hafezi2013imaging}, quasicrystals~\cite{Kraus2012,Verbin2013} and metacrystals~\cite{Slobozhanyuk2016}. 

Rechtsman \textit{et al.}~implemented a waveguide array that acts as a Floquet photonic topological insulator in the optical frequency domain~\cite{rechtsman2013photonic}. The waveguides are twisted so that inter-waveguide tunneling is accompanied by phase accumulation, similar to a gauge field; hence, the propagation of light in the array is similar to the time evolution of 2D electrons in a magnetic field. Notably, the waveguide array itself preserves TR symmetry; the sign of the effective magnetic field depends on the direction of propagation along the waveguide array axis.

Another breakthrough work~\cite{Hafezi2011} demonstrated edge states in the near-infrared (1.55 $\mu$m) regime in a lattice of coupled optical ring resonators. Here, each ring supports degenerate clockwise and counterclockwise modes, and a gauge field is implemented by auxiliary coupling rings with different optical path lengths. The overall structure obeys TR symmetry, with the sign of the effective magnetic field depending on whether the clockwise or counterclockwise mode is considered~\cite{Hafezi2011,hafezi2013imaging}.

Photonic QSH systems have been implemented based on lattice engineering~\cite{Wu2015}, field duality~\cite{Slobozhanyuk2016}, and other approaches. It turns out that because classical waves are not fermions, the standard electromagnetic TR symmetry is insufficient to generate a QSH phase; it is however possible to use other symmetries that play the role of TR, based on the constitutive relations~\cite{Khanikaev2013} or lattice symmetries~\cite{Wu2015}. For example, photonic QSH systems can be implemented in bianisotropic photonic crystals, with the magnetoelectric coupling serving in the role of spin-orbit interaction~\cite{Khanikaev2013,Ma2015,Slobozhanyuk2016SciRep}. Symmetry-protected topological states can also be realized using metacrystals~\cite{Slobozhanyuk2016,Slobozhanyuk2019} containing overlapping electric and magnetic dipolar resonances specially designed to satisfy electromagnetic duality.

Motivated by optical on-chip applications, there has been a concerted effort towards realizing topological photonics at the nanoscale. Presently, most experimental demonstrations have been based on VH and QSH (in particular, TCI) implementations in nonmagnetic photonic crystal slabs or nanoparticle arrays made of high-index dielectrics~\cite{Wu2015,Ma2016,Gorlach2018,Barik2018,Shalaev2018NatNano,He2018,Peng2018,Smirnova2019}. Strong optical resonances and low Ohmic losses make this all-dielectric platform~\cite{Kuznetsov2016} the most feasible for practical implementation of topological order for light at subwavelength scales. 

We emphasize a common feature of the structures listed above. They utilize ``{\em synthetic}'' fields induced by special structural features, which act like effective magnetic fields or spin-orbit interactions. Imperfections in real samples can therefore cause the topological properties to break down, so photonic topological edge states are only protected from scattering on defects of certain types, and are overall less robust than topological edge states in condensed matter systems.

Merging topological photonics with nonlinear optics provides many novel opportunities for scientific and technological exploration. Nonlinearity can enable on-demand tuning of topological properties via the intensity of light, and nonlinearity can break optical reciprocity to realize full topological protection.

\section{Topological lattice models}
\label{sec:lattices_intro}

\begin{figure} [t!]
\centerline{\includegraphics*[width=0.9\linewidth]{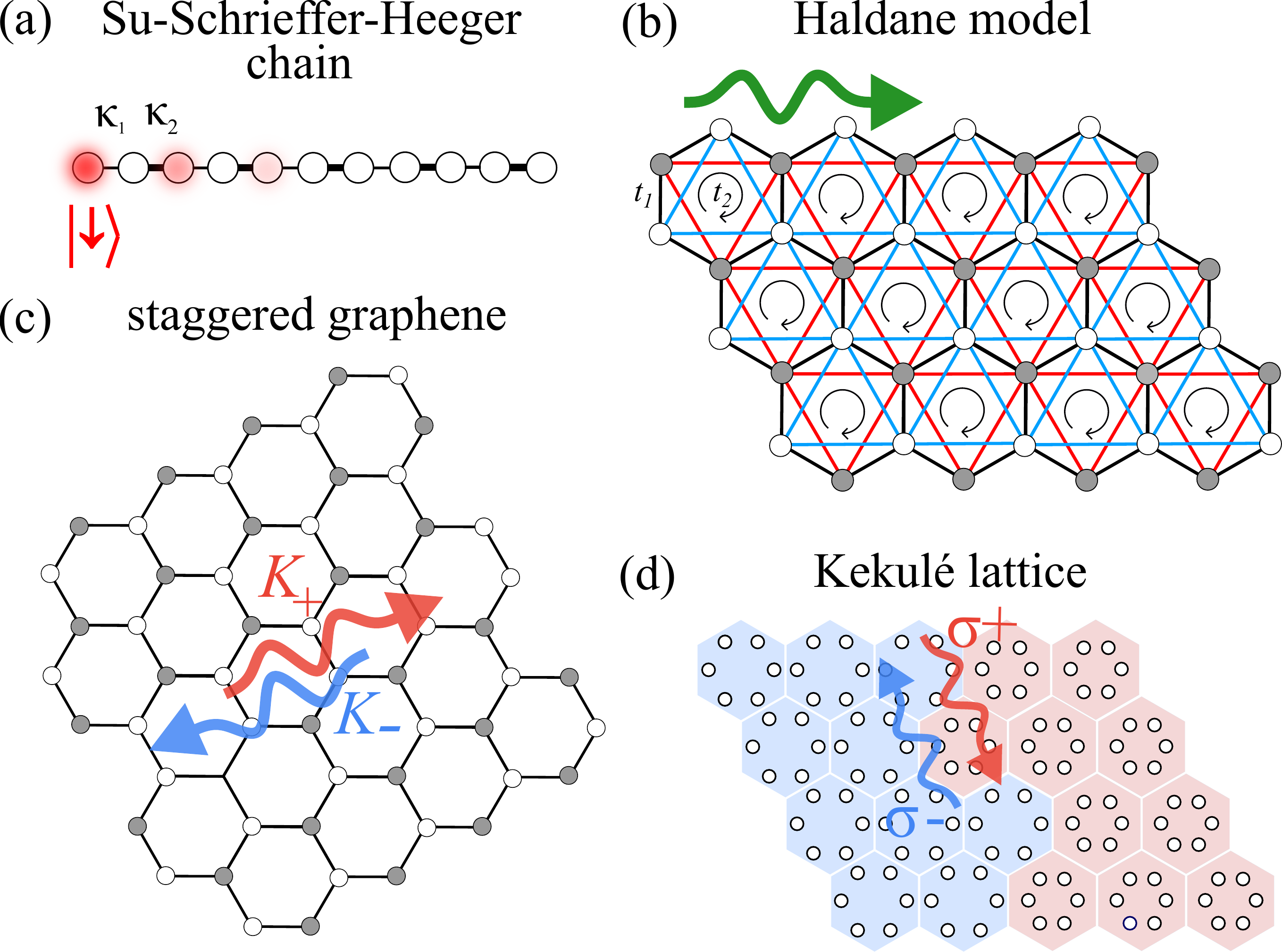}}
\caption{Schematics of topological lattices: (a) SSH array of dimers (1D chiral chain), (b) Haldane model on a honeycomb lattice (QHE), (c) staggered graphene (QVHE), and (d) hexagonal Kekule lattice of expanded/shrunken hexamers (QSHE). \label{fig:lattices}}%
\end{figure}

The key features of topological bands, including their interface states, can be understood by studying simple discrete lattice models.  This Section introduces the basic lattice designs known to exhibit topological transitions in the linear regime. They can be formulated in terms of abstract tight binding models for 1D and 2D arrays, illustrated in Fig.~\ref{fig:lattices}. 

Wave propagation dynamics in a discrete lattice can be described by an effective Hamiltonian ${\hat H}$ that includes couplings between different lattice sites. Performing a Fourier transformation, ${\hat H}$ is block diagonalized to a Bloch Hamiltonian $\hat{H} ( {\bf k} ) $, which is a function of the wave vector ${\bf k}$. Assuming $m$ modes per unit cell,  $\hat{H} ( {\bf k} ) $ is a $m \times m$ Hermitian matrix, which results in an $m$ band lattice model. Each eigenvalue ${\mathcal E}_n({\bf k})$ of  $\hat{H} ( {\bf k} ) $ gives the dispersion relation of a band in the lattice's band structure. The corresponding eigenvector ${\bf u}_{n}({\bf k})$ defines a Bloch wave ${\bf \psi}_n  = {\bf u}_{n}({\bf k}) e^{i {\bf k} {\bf r}}$. The topological properties of these Bloch waves can be related to physical observables. 

One fundamental property of the Bloch waves is their Berry phase~\cite{Berry1984}, which can be calculated as a line integral along a closed path in {\bf k} space as $\gamma_n =  \oint {\bf A}_n \cdot \text{d} {\bf k}$, where ${\bf A}_n ({\bf k})=  \langle \mathbf{u}_{n} | i \nabla_{\bf k}   | \mathbf{u}_{n}  \rangle  $ is the Berry connection. Roughly speaking, the Berry connection provides a measure of how the shape of the Bloch function changes along the given path. Using Stokes' theorem, the Berry phase can be alternatively computed as an integral over the area enclosed by the path: $ \gamma_n = \iint {\bf {F}}_n \text{d}^2 {\bf k} $, where $ {\bf {F}}_n =  \nabla_{\bf k} \times {\bf A}_n$ is the Berry curvature. 

The Berry connection and curvature are strongly reminiscent of the vector potential and magnetic field in the theory of electromagnetism. For example, the Berry connection is gauge-dependent: transforming the Bloch functions as ${\bf u}_{n}({\bf k}) \rightarrow {\bf u}_{n}({\bf k})e^{i\varphi({\bf k})}$ modifies the Berry connection as ${\bf A}_n \rightarrow {\bf A}_n - \nabla_{\bf k} \varphi ({\bf k})$. On the other hand, the Berry phase and Berry curvature are gauge-invariant quantities that can be related to physical observables. Integrating the Berry curvature over a 2D Brillouin zone yields the quantized Chern number $C_n$, which characterizes QH topological phases. The Chern number counts the winding (the number of complete turns) of the phase evolution of the eigenvector upon encircling the entire Brillouin zone. 

\begin{table*}[htp]
  \begin{center}
    \caption{ Continuum Hamiltonians of common 2D topological models employed in photonics. Here, the Pauli matrices $\hat{\sigma}$, $\hat{\tau}$, $\hat{s}$ act 
on the sublattice, valley and 
spin degrees of freedom, respectively. $m_{\text{T}}$, $m_{\text{I}}$ are the mass terms induced by TR and P symmetry reductions; $m_{\text{SO}}$ is responsible for the effective spin-orbit interaction. }
\begin{tabular}
   {|m{1.4in}|m{2.5in}| m{2.7in}| @{}m{0pt}@{}}
    \hline
    \centering{\textbf{Model}} & \centering{\textbf{Hamiltonian}} & \centering{\textbf{Topological invariant}} &\\[0ex] 
    \hline
   \center{ \textbf{Haldane}} & \center{$ \hat{H} =  v_{\text{D}} \left( \hat{\sigma}_x \hat{\tau}_z \delta k_x  + \hat{\sigma}_y \hat{\tau}_0 \delta k_y \right)+ \hat{\sigma}_z \left( \hat{\tau}_z m_{\text{T}} -  \hat{\tau}_0 m_{\text{I}}  \right)$}  & 
   \center{ Chern number $C  = \frac{1}{2} \left( \text{sgn} (m_{\text{I}} -m_{\text{T}} ) - \text{sgn} (m_{\text{I}} + m_{\text{T}}) \right)$}& \\[5ex]
    \hline
    \center{\textbf{Kane-Mele}} &  \center{$\hat{H} \!= \! v_{\text{D}}  \hat{s}_0 \left( \hat{\sigma}_x \hat{\tau}_z  \delta k_x  + \hat{\sigma}_y \hat{\tau}_0 \delta k_y \right)+ \hat{\sigma}_z \hat{\tau}_z  \hat{s}_z m_{\text{SO}}$ } & \center{ spin Chern number $C^{\text{spin}}= \text{sgn}\,(m_{\text{SO}})$}& \\[5ex]
    \hline
     \center{\textbf{Bernevig-Hughes-Zhang}} &  \center{$ \hat{H} =  v_{\text{D}} \left(\hat{\sigma}_x \hat{s}_x\delta k_x  + \hat{\sigma}_y \hat{s}_0 \delta k_y \right) + \hat{\sigma}_z \hat{s}_0 (m + \beta \delta k^2)$} &  \center{spin Chern number $C^{\text{spin}}=\frac{1}{2}\,(\text{sgn}\,m-\text{sgn}\,\beta)\:$}&\\[5ex]
    \hline
    \center{ \textbf{staggered graphene}}&  \center{$ \hat{H} =  v_{\text{D}} \left(\hat{\sigma}_x \hat{\tau}_z \delta k_x  + \hat{\sigma}_y \hat{\tau}_0 \delta k_y \right) - \hat{\sigma}_z \hat{\tau}_0 m_{\text{I}}$} &  \center{valley Chen number 
    $C^{\text{valley}} = \pm \frac{1}{2}\text{sgn} (m_\text{I})$ }&\\[5ex]
    \hline
  \end{tabular}
  \label{tabular::Hamiltonians}
  \end{center}
\end{table*}

In all one-band models, the Bloch functions are trivial (because they are independent of ${\bf k}$), so the simplest models of topological phases involve two bands. The most general two band Bloch Hermitian Hamiltonian can be written as $\hat{H} ( {\bf k} ) =  \mathbf{h}({\bf k}) \cdot {\hat{\sigma } }$ 
and has eigenvalues ${\mathcal E}_{\pm} ( k ) = \pm |\mathbf{h}(k) |$, where $\mathbf{\hat{\sigma}} = (\hat{\sigma}_x, \hat{\sigma}_y, \hat{\sigma}_z )$ is a vector of the three Pauli matrices. For many topological lattice models in the continuum limit, Dirac-like equations describing quasirelativistic dynamics can be recovered in the vicinity of the bandgap. In 2D, the continuum limit Hamiltonian has the form
\begin{equation}
{\hat H_D}(\delta \bm k)= v_{\text{D}} (\delta k_x \hat{\sigma}_x + \delta k_y \hat{\sigma}_y ) + m \hat{\sigma}_z \:, \label{eq:dirac}
\end{equation} 
where $v_{\text{D}}$ is a velocity parameter and $m$ is an effective mass. When $m=0$ Eq.~\eqref{eq:dirac} describes a conical intersection with linear dispersion relation ${\mathcal E}_{\pm} = \pm v |\bf k|$ resembling that of massless fermions~\cite{conical_review}. It can be generalized to a larger number of intersecting bands. For example, three intersecting levels are described by the effective continuum Hamiltonian 
\begin{equation} \label{eq:H_M}
{\hat H}_M(\delta \bm k)= v_{\text{D}} (\delta k_x \hat{S}_x + \delta k_y \hat{S}_y )\:,
\end{equation} 
expressed through spin-$1$ matrices; its three eigenvalues are ${\mathcal E}_{0}=0$, ${\mathcal E}_{\pm} = \pm v_{\text{D}} |\bf k|$, corresponding to a zero energy flat band and two linearly dispersing modes.

The simplest lattice that exhibits topological modes is Su-Schrieffer-Heeger model, which describes a 1D dimer chain with alternating weak and strong nearest-neighbor couplings $\kappa_{1,2}$. In second quantized notation, the Hamiltonian is
\begin{equation}
\hat{H}_{\text{SSH}} = - \sum_{j=1}^{N} \left( \kappa_1 \hat{a}_{j}^{\dagger} \hat{b}_j  + \kappa_2  \hat{a}_{j+1}^{\dagger} \hat{b}_j + h.c.  
  \right)\:,
\end{equation}
where $\hat{a}_j$, $\hat{b}_j$ ($\hat{a}^{\dagger}_j$, $\hat{b}^{\dagger}_j$) denote creation (annihilation) operators at $A$ or $B$ sublattices of the $j$th unit cell. The two bands ${\mathcal E}_{\pm}$ are separated by a gap when when $\kappa_1 \ne \kappa_2$. In a finite lattice whose terminations break the stronger coupling, there exist edge states in the middle of the band gap. These states are localized in one sublattice and decay exponentially away from the lattice edge, at a rate determined by the size of the gap. These states are protected in the sense that their frequency is pinned to zero and they cannot be destroyed by any perturbation that respects the chiral symmetry $\hat{\sigma}_z \hat{H}_{\text{SSH}} \hat{\sigma}_z = - \hat{H}_{\text{SSH}} $, as long as the two bands remain separated by a gap. The topological invariant associated with this protection is a 1D Berry phase called the Zak phase~\cite{Zak1989}, which takes the quantized values $\pi$ (in the nontrivial case, $\kappa_2> \kappa_1$) or $0$ (in the trivial case, $\kappa_2 < \kappa_1$).

The prototypical example of a 2D topological lattice model is the honeycomb lattice, which can be used to implement optical analogues of graphene (i.e., {\it photonic graphene}) \cite{Peleg2007,Haldane2008, Raghu2008,Plotnik2013,rechtsman2013photonic}. This lattice has a hexagonal Brillouin zone, whose inequivalent corners (called the $K_{\pm}$ points) host conical intersection degeneracies protected by TR and parity P (spatial inversion) symmetries. Breaking either symmetry lifts the degeneracies and opens a band gap. Breaking P creates a trivial gap, because the Berry curvatures at the $K_{\pm}$ points have opposite signs, yielding a vanishing Chern number. Breaking TR symmetry generates a nontrivial topological phase, as the Berry curvature having the same sign at the $K_{\pm}$ points.

A QH phase can be induced in a honeycomb lattice by breaking TR symmetry. In the Haldane model~\cite{Haldane1988}, this is accomplished by complex-valued next-nearest-neighbor (NNN) couplings. To ensure that the Brillouin zone is unaltered by the TR symmetry breaking, the couplings are staggered so that there is no net magnetic flux per unit cell: encircling one lattice plaquette clockwise (counterclockwise) gives a phase factor of $e^{i\phi}$ ($e^{-i\phi}$). The lattice Hamiltonian is
\begin{multline}
\hat{H}_{\text{Haldane}} = m_{\text{I}} \sum_{\langle i \rangle} \left(\hat{a}_{i}^{\dagger} a_i - \hat{b}_{i}^{\dagger} b_i  \right) -t_1 \sum_{\langle i,j \rangle} \left(  \hat{a}_{i}^{\dagger} \hat{b}_j   + \hat{b}_{j}^{\dagger} \hat{a}_i \right) \\
- t_2 \left[ e^{i\phi}  \sum_{\langle \langle i,j \rangle \rangle}  \hat{a}_{i}^{\dagger} \hat{a}_j +  e^{-i\phi} \sum_{\langle \langle i,j \rangle \rangle} \hat{b}_{i}^{\dagger} \hat{b}_j + h.c.  \right]\:, \label{eq:Haldane}
\end{multline}
where $\langle i,j \rangle$ and $\langle \langle i,j \rangle \rangle$
denote summations over the first and second nearest neighbors sites, respectively, $t_1$ and $t_2$ are the hopping amplitudes, and $m_{\text{I}}$ is a parameter that breaks inversion symmetry via a sublattice detuning. Near the $K_{\pm}$ points, the effective continuum Hamiltonian is
\begin{multline}
\hat{H}_{K_{\pm}}\!=\!\ 3 t_2 \cos{\phi} + v_D \left( \pm  \delta k_x \hat{\sigma}_x + \delta k_y \hat{\sigma}_y \right)  \\ + \left(   m_{\text{I}} \pm 3 \sqrt{3} t_2 \sin{\phi}\right) \hat{\sigma}_z \:,
\end{multline}
where $v_D = \sqrt{3} t_1 /2 $. Thus, the effective mass due to TR breaking has opposite signs at the two valleys, whereas the effective mass due to P breaking has the same sign at both valleys. The band structure of Eq.~\eqref{eq:Haldane} can be characterized by the Chern number, which is non-zero when the gap is dominated by the TR breaking terms; in this regime, chiral edge modes are guaranteed to exist along the boundary of the finite lattice. Systems similar to the Haldane model are known as {\it Chern insulators}. 

In 2004, Kane and Mele discovered a new topological phase, the QSH insulator~\cite{kane2014topological}. The Kane-Mele model is derived by incorporating spin and spin-orbit interactions into the honeycomb lattice model; in its simplest form, it is essentially two copies of the Haldane model, with opposite effective magnetic fields for each spin. Although the net Chern number is zero due to TR symmetry, one can formulate a spin Chern number $C^{\text{spin}}=\left(C_{\downarrow}-C_{\uparrow}\right)/2$ which is nonzero in the QSH phase.

A honeycomb lattice with preserved TR symmetry but broken P (space-inversion) symmetry is a VH insulator. The domain walls separating VH lattices that have opposite P breaking host chiral edge states~\cite{Yao2009}. For small P breaking, the Berry curvatures are strongly localized at the valleys, and the local integrals around $K_{\pm}$ valleys take non-zero quantized values of $\pm \pi$ for each band, which yields in a valley Chern number $C^{\text{valley}}=\pm 1/2$. Flipping the sign of the P breaking also flips the sign of the Berry curvature in each valley. Across a domain wall, there is a difference of $ \pm 1$ between valley Chern numbers, resulting in one family of topological edge states in each valley.

Another honeycomb lattice variant that is extremely useful for topological photonics is a topological crystalline insulator devised by Wu and Hu~\cite{Wu2015}. It involves clustering neighboring plaquettes of 6 lattice sites by alternately widening or narrowing the inter-site separations (see Fig.~\ref{fig:lattices}(d)). This clustering causes the $K_{\pm}$ points to be folded onto the center of the Brillouin zone (the $\Gamma$ point); the interaction of the overlaid Dirac cones causes a band gap to open. The corresponding effective Hamiltonian is of the Bernevig-Hughes-Zhang QSH Hamiltonian~\cite{Bernevig2006}. The model exhibits helical edge states at the boundaries between domains with shrunken (trivial) and expanded (nontrivial) clusters.

The continuum Hamiltonians and topological invariants discussed in this Section are listed in Table~\ref{tabular::Hamiltonians}. 

\section{Platforms for nonlinear photonics}
\label{sec:nonlinear_intro}

The linear lattice models discussed in the previous section are agnostic about length and frequency scales and the wave amplitudes involved. When studying nonlinear phenomena, however, this universality is lost. This Section provides an overview of nonlinear effects in photonic lattices.

Table~\ref{tab:platforms} summarizes different platforms for nonlinear topological photonics with characteristic frequencies ranging from the optical range ($10^{14}$ Hz) to microwaves ($10^9$ Hz) and electronics ($10^6$ Hz).
The main platforms that have been used to explore linear topological photonics are arrays of coupled waveguides, microring resonators and photonic crystals. Since nonlinear problems are generally much harder to solve, platforms where the full set of Maxwell's equations can be well approximated by simpler coupled-mode or tight-binding lattice models are preferred for studying nonlinear topological photonics.

Much of the older literature on nonlinear effects in lattices was written before topological effects came into focus, and therefore mainly dealt with the consequences of band gaps and discreteness, while overlooking the role of topology. For instance, many works have studied how nonlinearity affects non-topological surface states, which are typically generated by defects or localized potentials along the boundary of a lattice. For instance, a semi-infinite array of coupled quantum wells can form surface states if the energy of the first well is detuned from the energy of the other wells. Such threshold conditions are typical for (topologically trivial) Tamm surface states~\cite{Dyakov2012}. In the nonlinear regime, it has been shown that self-trapping can overcome surface repulsion, inducing localized modes near the edge of a discrete lattice above a certain power threshold~\cite{Kivshar2008}. On the other hand, topological lattice models support edge states even in the low-amplitude limit and do not require any threshold perturbation to exist.

\begin{table*}[htp]
  \begin{center}
   \caption{Examples of platforms for nonlinear topological photonics}
 
  \label{tab:platforms}
    \begin{tabular}{| *5{>{\centering\arraybackslash}m{1.1in}|} @{}m{0pt}@{}}
    \hline
     \makecell{\textbf{Platform}} & \textbf{Material} & \textbf{Spectral range} & \textbf{Power, Duration} &
    \textbf{Source}  &\\[0ex] 
    \hline
    \textbf{Optical Waveguides} & \makecell{AlGaAs \\  Fused silica \\
    Photorefractives \\ Lithium niobate \\ Chalcogenides} &  \makecell{1530 nm \\ 800 nm \\ 488 nm \\ 1550 nm \\ 1040 nm } & \makecell{500 W, 100 fs \\ 1 MW, 100 fs \\ mW, CW \\ 4 kW, 9 ps \\ 10kW, 300 fs} &  \makecell{[\cite{Eisenberg1998}] \\ {[\cite{Szameit2005}]} \\ {[\cite{Fleischer2003}]} \\ {[\cite{Iwanow2004}]} \\ {[\cite{Li2014}]}} &\\[0ex]
    \hline
    \textbf{Optical Resonators} & \makecell{ Si microrings \\
    InP PhC slab \\ GaAs/InAs PhC slab \\ Si Metasurfaces \\ Fibre loops \\ Exciton-polaritons} & \makecell{1550 nm \\ 1587 nm \\ 1040 nm \\ 1550 nm \\ 1555 nm \\ 780 nm} & \makecell{85 $\mu$W, CW \\ 200 $\mu$W, CW \\ 150 $\mu$W, 20 ns \\ 200 mW, 300 fs \\ 120 mW, 50 ns \\ 10 mW, CW} & \makecell{{[\cite{Fan2012}]} \\ {[\cite{Yu2015}]} \\ {[\cite{Ota2018}]} \\ {[\cite{Smirnova2018}]} \\  {[\cite{Wimmer2015}]} \\ {[\cite{Tanese2013}]}} \\
    \hline
    \textbf{Metamaterials} & \makecell{Split ring resonators \\ Circuit QED \\ RF circuits } & \makecell{ 1.5 GHz \\ 5 GHz \\ 100 MHz } & \makecell{1 W, CW \\ single photon, 100 ns \\ 300mW, CW }& \makecell{{[\cite{Dobrykh2018}]} \\ {[\cite{Ma2019}]} \\ {[\cite{hadad2018self}]}} \\
    \hline
  \end{tabular}

  \end{center}

\end{table*}

Nonlinear effects naturally emerge in waveguide lattices due to the intrinsic nonlinearity of the host medium. For example, the intensity-dependent refractive index of cubic nonlinear materials enters into tight binding models as a nonlinear on-site potential. One of the advantages of waveguide lattices is that even though the bulk material nonlinearity can be quite weak, the important parameter governing the dynamics is the ratio of the nonlinearity to the linear coupling coefficient. Therefore, provided one has access to a sufficiently long propagation distance and effects such as absorption remain negligible, one can reduce the coupling to increase the effective nonlinearity and observe effects such as optical switching and spatial solitons.

Nonlinear photonic waveguide lattices have a long history, dating back to the seminal prediction of optical discrete solitons by Christodoulides and Joseph in 1988~\cite{Christodoulides1988}. The first experiments by Eisenberg \textit{et al.}~in 1998~\cite{Eisenberg1998} used femtosecond laser pulses in a cubic nonlinear 1D AlGaAs waveguide array. The following decade saw several breakthroughs, including the observation of discrete solitons in photorefractive crystals using continuous wave beams~\cite{Fleischer2003}, laser-written waveguide arrays in fused silica glass~\cite{Szameit2005}, and quadratic nonlinear lithium niobate waveguides~\cite{Iwanow2004}. For details, see Refs.~\onlinecite{Lederer_review,Denz_review,Chen_review}.

The main challenge in generalizing these previous experiments to topological waveguide lattices is that there is a trade-off between ease of fabrication and ease of observing nonlinear effects. For example, AlGaAs, lithium niobate and photorefractive waveguide arrays have strong nonlinearity, but are presently limited to simple 1D topological lattices such as the SSH model. Alternatively, fused silica glass waveguides created using laser writing can be readily form 2D topological lattices, but the nonlinearity is much weaker, demanding shorter pulses with higher peak powers and increasing the complexity of experiments and modelling. For example, beam shaping is required to avoid material damage when exciting the waveguides and modelling should take into account effects such as material dispersion and two photon absorption~\cite{Lahini2007}. Moreover, many theoretical proposals are based on models of nonlinear coupling, which is negligible in this platform.

Optical cavities, supporting whispering-gallery, Fabry-P\'erot or Mie-type resonances, are able to efficiently trap light. Therefore, optical resonator lattices, such as microring arrays and particle metasurfaces, can enhance nonlinear effects and thus significantly lower optical power requirements, but at the expense of operating bandwidth. Additionally, the ability to tailor the pump beam or embed different materials onto the resonators gives access to a variety of nonlinear effects. For example, continuous wave operation leads to strong thermal nonlinearities due to absorption-induced heating of microresonators~\cite{Fan2012}, while two-photon absorption results in nonlinear resonance shifts due to free carrier dispersion~\cite{Yu2015}. Unfortunately, the mechanisms that provide the strongest self-action effects are also intrinsically lossy. Losses can be compensated via integration of gain media such as quantum wells~\cite{Ota2018,Wimmer2015}. 

Most experiments with nonlinear topological resonator lattices have focused on pump-probe setups, which are easier to analyze and support effects such as lasing (Sec.~\ref{sec:lasers}) and harmonic generation (Sec.~\ref{sec:parametric}). For self-action effects (e.g.~bistability and nonlinear non-reciprocity) it has been preferable to use only a few nonlinear elements to avoid complications such as multistability or instability~\cite{Fan2012,Yu2015}. There are two exceptions where self-action effects are observable in nonlinear propagation dynamics: pulse propagation in coupled fibre loops~\cite{Wimmer2015,Bisianov2018} and exciton-polariton condensates in microcavities~\cite{Tanese2013}.

In the microwave regime, nonlinearities are much harder to realize than in optics (unlike most other photonic phenomena). One approach is to insert nonlinear electronic lumped elements, like varactor diodes, into microwave metamaterials such split ring resonators~\cite{Dobrykh2018}. This can yield mean-field nonlinear effects under a pump power of $\approx 1$ W. Single photon nonlinearities are accessible by coupling microwave cavities to superconducting qubits~\cite{Anderson2016,Owens2018,Ma2019}, although this introduces the additional complication of cryogenic operating temperatures.

Other approaches to combine nontrivial topology with nonlinearity include {\it electronic circuits} (reviewed in Sec.~\ref{sec:electronics}), {\it photonic nanostructures} (see Sec.~\ref{sec:parametric}) and {\it mechanics}. 
The first mechanical implementation of the QSHE was experimentally demonstrated by Susstrunk and Huber~\cite{Susstrunk2015} in a lattice of mechanical pendula, with operating frequency in Hz range. Based on this model, nonlinear Duffing oscillators connected by linear springs can support unidirectional nonlinear traveling edge waves~\cite{Snee2018}. Another nonlinear topological phononic crystal -- 1D array which consists of masses connected with two alternating types of nonlinear springs -- was analysed in Ref.~\cite{Chaunsali2019}. It was numerically shown that by increasing the excitation amplitude the lattice makes a topological transition giving rise to different families of nonlinear solutions.

\section{Localized nonlinear states} 
\label{sec:localized}

\begin{figure} [t!]
\centerline{\includegraphics*[width=0.9\linewidth]{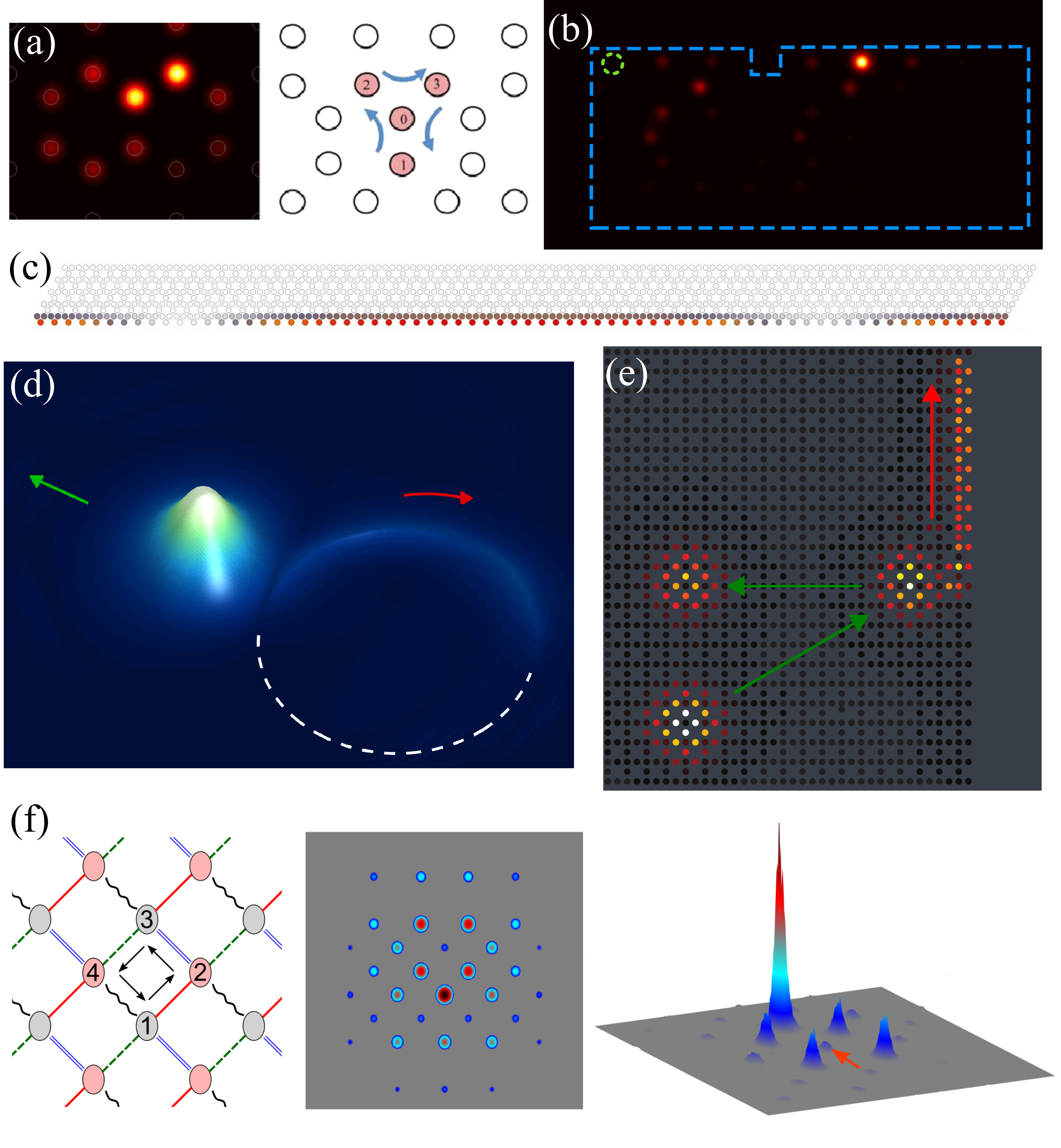}}
\caption{(a) Bulk soliton in a honeycomb lattice of helical waveguides alongside 
a sketch of the intraperiod rotation of the intensity~\cite{Plotnik2013}. The wavepacket rotates around the nonlinearity-induced defect as an edge state in a clockwise direction. (b) Single-site localized edge soliton propagating unidirectionally along the edge of the Floquet lattice of coupled helical waveguides embedded in a medium with local Kerr nonlinearity~\cite{Leykam2016}. Shown is a numerically calculated profile of the output intensity after propagation around a missing edge waveguide, when the waveguide circled in green is excited at the input. 
(c) Dark (left) and grey (right) edge solitons (seen as whitened dips in intensity) at the boundary of Kagome lattice strip of coupled microcavity pillars~\cite{Gulevich2017}. In the weakly nonlinear regime, the solitons  in the exciton-polariton condensate are constructed from wavepackets of topological edge modes with the envelope described by Schr\"odinger equation. (d,e) Excitation of chiral topological edge modes by scattered traveling gap solitons at (d) domain walls created by mass inversion in a 2D continuum Dirac model~\cite{Smirnova2019LPR}; (e) 
at the pointy edge of the Lieb lattice~\cite{Marzuola2019arxiv}.
(f) Observation of topological gap solitons in a square lattice of helical waveguides femtosecond-laser-written in glass~\cite{Mukherjee2019arxiv}: Schematic of the lattice (left); Calculated intensity profile of the soliton whose maximum exhibits cyclotron-like rotation jumping four sites (1-4) sequentually (middle); Measured output intensity distribution (right) at the propagation distance of 1.5 driving period for the input power 3.32 mW. 
\label{fig:solitons}}
\end{figure}

Nonlinear generalizations of linear topological models support peculiar mechanisms for field localization, leading to phenomena such as topological gap solitons and nonlinear edge states (bulk and edge solitons)~\cite{Lumer2016,Solnyshkov2017,Smirnova2019LPR}, embedded solitons~\cite{Leykam2016}, and semi-vortex solitons~\cite{RingSoliton2018}, as depicted in Fig.~\ref{fig:solitons}. Notably, these solitons have nontrivial vorticity and pseudospin structure. The formation of topological solitons in the bulk can be viewed as self-induced domain walls similar to the localization of edge states at the boundary between domains with distinct topological invariants, as discussed in Refs.~\cite{Lumer2016,RingSoliton2018}.

These phenomena pose an interesting challenge to our understanding of band topology. Strictly speaking, the concept of band topology is tied to linearity, which is necessary for the existence of a Bloch Hamiltonian and band structure, as discussed in Sec.~\ref{sec:lattices_intro}. Some authors have explored correcting the definition of the Berry phase in order to describe the band structures of weakly nonlinear Bloch modes with fixed homogeneous intensities~\cite{BerryPhaseNL2010,Bomantara2017}. However, localized nonlinear states---solitons---represent strong modifications to an underlying topological structure, or even the creation of topological order from a trivial system.

From a practical point of view, localized nonlinear states may be extremely useful for tunable topological photonics~\cite{Dobrykh2018,Smirnova2019LPR,Solnyshkov2018,Chen2018}. They may also be accompanied by novel effects such as the spontaneous breakdown of Lorentz reciprocity, wherein the light intensity itself determines whether the light can propagate via an edge state~\cite{Hadad2016,Chaunsali2019}.

In lattice models, nonlinearity can be introduced either into the on-site energy or the coupling between lattice sites. The resulting behaviour may be non-universal, sensitive to either the form of the nonlinearity or the particular lattice geometry. Non-perturbative studies of specific lattices have been mostly limited to numerical simulations, due to the scarceness of exact solutions to nonlinear problems~\cite{Plotnik2013,Leykam2016,Kartashov2017,Solnyshkov2017}. For weak nonlinearities, the formation of edge solitons was understood in the traditional framework of scalar effective nonlinear Schrodinger equations, where the nonlinearity compensates the linear edge state dispersion~\cite{Ablowitz2014,Kartashov2016,Gulevich2017}.

More universal insights can be obtained using the continuum nonlinear Dirac model, through the perspective of phase portraits and bifurcation analysis. This approach is able to describe bulk solitons and nonlinear edge states in a variety of 1D and 2D nonlinear photonic lattices, including SSH, honeycomb, and Kagome lattices. For instance, treating ${\hat H_D}$ in Eq.~\eqref{eq:dirac} as an operator, which contains spatial derivatives, and incorporating nonlinear corrections as a field-dependent operator ${\hat H_{N\!L}}$, yields a nonlinear equation for the evolution of the spinor wavefunction $\Psi = [\Psi_{1},\Psi_{2}]$:
\begin{equation}
i \partial_t \Psi = ({\hat H_D}(\delta \bm k) + {\hat H_{N\!L}}) \Psi \:. \label{eq:H_D_NL}
\end{equation}
This can be tackled analytically for various types of nonlinearity, including the one most commonly encountered in optics, a local cubic nonlinearity of the form ${\hat H_{N\!L}}= - g \bigl[{ |\Psi_{1}|^{2} }, 0; 0, {|\Psi_{2}|^{2} }\bigr]$. 
By contrast to relativistic field theory, nonlinear Dirac equations in photonics appear as effective equations and they are not restricted by the Lorentz invariance. 
As compared to the nonlinear Schr\"odinger equation, the existence and stability analysis of solitons in Dirac models is more subtle because of the absence of the rigorous Vakhitov-Kolokolov criterion~\cite{Maravero2016,Maraver2017b,Maraver2018}.

The simplest structure linking topology and nonlinearity is a nonlinear version of the 1D SSH model. Such a model can be implemented in arrays of resonant elements with nonlinear couplings~\cite{Hadad2016,Hadad2017}. This formally corresponds to off-diagonal nonlinearity in Eq.~\eqref{eq:H_D_NL}. It exhibits a self-induced topological transition, in which the nonlinearity drives the lattice into a different topological phase supporting edge states. However, in this model the edge states are not truly localized as they sit on a nonzero intensity background. Though this model has been implemented in electronic circuits, as discussed in Sec.~\ref{sec:electronics}, it is challenging to realize in optics, where local on-site Kerr nonlinearities are more feasible. 

Bulk solitons and edge states in a model with on-site Kerr nonlinearity have been studied theoretically \cite{Solnyshkov2017} and experimentally \cite{Dobrykh2018}. Later, in Ref.~\onlinecite{Smirnova2019LPR}, it was theoretically explicated that the bulk solitons and nonlinear edge states in this setting have a closely related origin; mutual transformations between edge and bulk states, forbidden in linear limit, can occur in the nonlinear regime. It has been theoretically predicted that traveling bulk solitons in both 1D and 2D topological settings are capable of exciting the edge states by reflecting off the topologically nontrivial edge~\cite{Smirnova2019LPR,Marzuola2019arxiv}. Recent experiments using coupled optical fibre loops have shown that indeed this class of nonlinearities can be used to couple between localized topological edge states and nonlinear bulk modes~\cite{Bisianov2018}.

\begin{figure} [t!]
\centerline{\includegraphics*[width=0.9\linewidth]{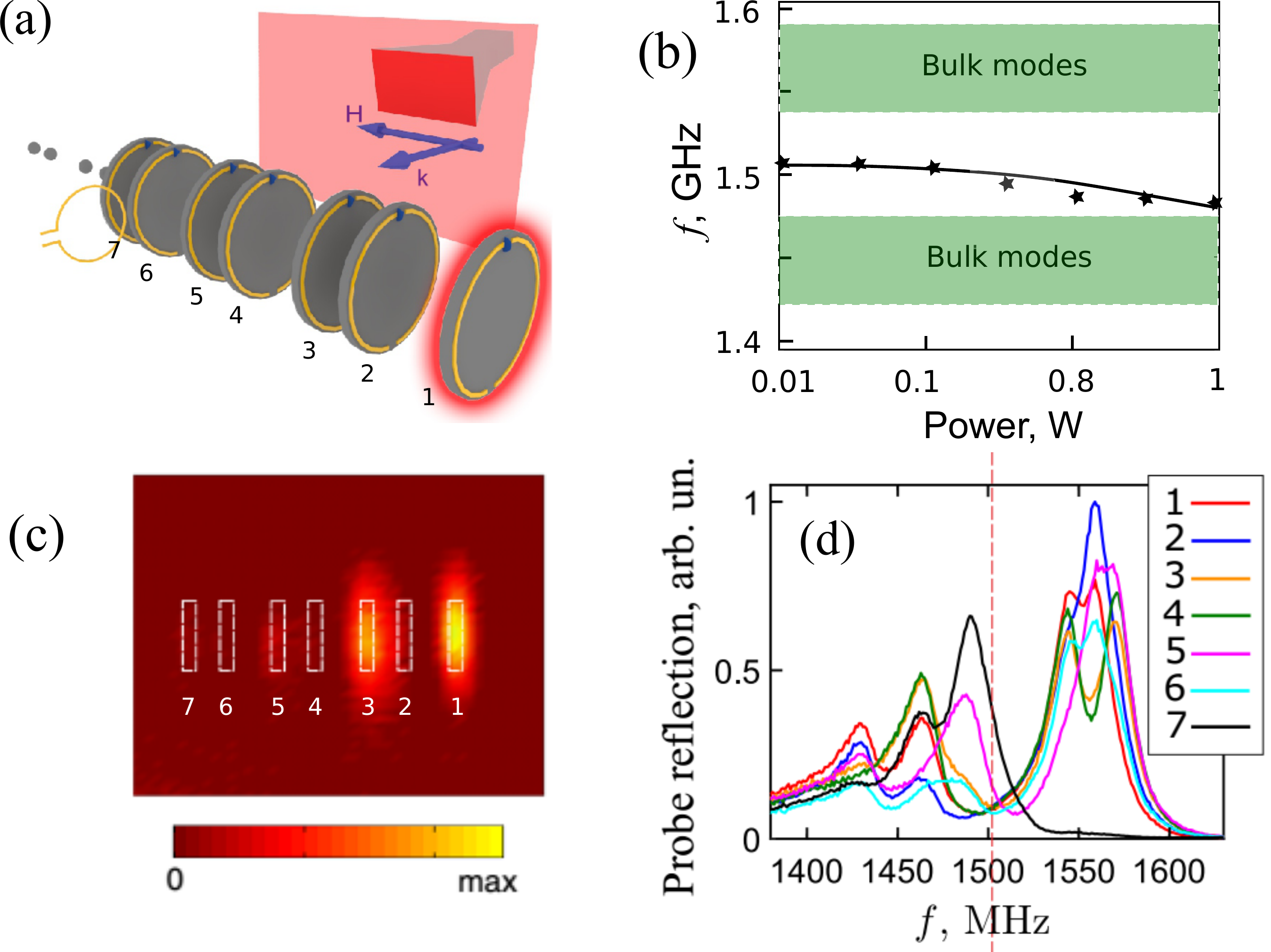}}
\caption{Nonlinear tuning of microwave topological edge states. (a) Experimental setup: SSH array of nonlinear microwave resonators with source (horn antenna) and receiver (loop antenna). (b) Measured nonlinear shift
of the edge state frequency.
(c) Measured magnetic field distribution. 
(d) 
Spatially resolved pump-probe reflection spectrum for power 0.8~W. Curves 1-7 correspond to the resonators from left to right.
\label{fig:tuning}}
\end{figure}

In a microwave experiment, Dobrykh \textit{et al.}~demonstrated nonlinearity-induced tuning of the electromagnetic topological edge states in topological arrays of coupled nonlinear resonators with alternating weak and strong couplings~\cite{Dobrykh2018}. 
An SSH array was made of $N=7$ broadside-coupled split-ring resonators with the magnetic dipole resonance at the frequency $f_{0}\approx 1.5~$GHz. The Kerr-type nonlinear tunability of the frequency was introduced by varactor diodes mounted inside the gap of each SRR. 
The experiment was conducted in the pump-probe setup. 
The monochromatic homogeneous pump came from a
rectangular horn antenna, while the probe signal was measured near each resonator by a small loop antenna. 
The model can be captured by the nonlinear lattice model
\begin{equation}
 \dfrac{d a_{n}}{d t}=-\gamma a_{n} - i |a_{n}|^{2}a_{n}+t_{n,-}a_{n-1}+
 t_{n,+}a_{n+1}+ P  \:,
\end{equation}
where $a_{n}$ is a normalized amplitude of the $n$-th oscillator ($n=1\ldots N$), $\gamma$ is a damping coefficient,
and $P$ is an amplitude of resonant homogeneous pump, $t_{n,-}$, $t_{n,+}$ are alternating weak and strong nearest-neighbor couplings.
With increasing the power, homogeneous pump becomes localized at the edge and induces a nonlinear blue shift for the edge state, as shown in Fig.~\ref{fig:tuning}. 

2D nonlinear lattice models have been studied theoretically, but still remain challenging to implement in optics experiments. They require nonlinear couplings~\cite{Gerasimenko2016,Hadad2016,Hadad2017,Hadad2018,Zhou2017,RingSoliton2018,Wang2019}, exciton-polaritons in strong external magnetic fields~\cite{Bleu2016,Kartashov2016,Kartashov2017,Solnyshkov2017,Gulevich2017,Li2018}, or longitudinally-modulated waveguide arrays (instabilities and radiation losses) ~\cite{Plotnik2013,Lumer2016,Leykam2016}.

Very recently, the observation of topological gap solitons has been reported in a square lattice of laser-written periodically-modulated waveguides, which emulates Floquet topological phase. The nonlinearity arises from the optical Kerr effect of the ambient glass. Under the paraxial approximation, the $z$ propagation of light through this photonic lattice is captured by the discrete equation which includes the linear tight-binding Hamiltonian with nearest-neighbour evanescent coupling and diagonal on-site nonlinearity:
\begin{equation}
\label{nlse}
i\frac{\partial a_n}{\partial z}=\sum_{\left\langle n' \right\rangle} H_{nn'} a_{n'} - |a_n|^2 a_n\; .
\end{equation}
In the nonlinear dispersion given by the dependence of quasienergy (propagation constant) on power, a family of gap solitons bifurcates from the linear modes and shows maximal localisation in the vicinity the mid-gap quasienergy. In accord with their chiral nature, solitons residing in the topological band gap exhibit continuous cyclotron-like rotation.  
The solitons were probed in propagation using single-site excitation in the input. see Fig.~\ref{fig:solitons}(f). The characteristic peak in the degree of localization vs power was observed that distinguishes topological gap solitons from trivial solitons in static lattices of straight waveguides, where localization continuously grows up and then saturates at very high nonlinearity~\cite{Mukherjee2019arxiv}.  

\section{Nonlinear circuits}
\label{sec:electronics}

Electronic circuits have recently emerged as a convenient and accessible platform for studying the combination of nonlinearity with
band topology \cite{Simon2015,Albert2015, hadad2018self,Rosenthal2018,Lee2018,Imhof2018,Li2018b, Wang2019,Liu2019,SerraGarcia2019,  Zangeneh-Nejad2019}. Key advantages include the ease with which such circuits can be designed and fabricated using circuit simulators, printed circuit boards (PCBs), and other commodity technologies; the
fact that they can be characterized using inexpensive laboratory equipment such as function generators and oscilloscopes; the
availability of strongly nonlinear circuit elements; and the exciting prospect of using circuit wiring to implement complex geometries (like M\"obius strips \cite{Simon2015}) that are practically impossible to realize on other platforms.  Such systems include circuits implemented
on breadboards or PCBs, typically operating in the 0.1--500 MHz frequency range \cite{Simon2015, hadad2018self, Imhof2018, Wang2019, Liu2019, SerraGarcia2019, Zangeneh-Nejad2019}, as well as electromagnetic structures (such as microstrip resonator arrays) with attached lumped circuit elements, which can operate at GHz frequencies \cite{Li2018b}.

Topological edge states were first demonstrated in electronic circuits by the Simon group in 2015 \cite{Simon2015}, using a linear non-dissipative circuit simulating the Hofstadter model (quantum Hall effect on a 2D square lattice) \cite{Hofstadter1976}.  Such LC circuits, which contain only linear inductors and capacitors, are symmetric under time-reversal (T), similar to condensed matter systems in the absence of magnetic effects or photonic systems without magneto-optic media.  In order to simulate a quantum Hall system, which requires breaking T, the lattice in Ref.~\onlinecite{Simon2015}
was designed to have multiple identical sublattices whose interconnections replicate the effects of the complex inter-site couplings associated with a magnetic vector potential \cite{Albert2015}; this ensured that the states of the target quantum Hall system, including the crucial topological edge states, are a multiply-degenerate subset of the states of the T-symmetric circuit.
A variety of T-symmetric topological phases have also been realized with LC circuits without using this sublattice trick, including linear 1D and 2D Su-Schrieffer-Heeger models \cite{Lee2018, Liu2019}, topological crystalline insulators \cite{Li2018b}, higher-order topological insulators \cite{Imhof2018, SerraGarcia2019}, and intrinsically non-Hermitian topological lattices \cite{Rosenthal2018}.

The time-domain dynamics of any linear non-dissipative LC circuit can be described using linear second-order equations of motion, expressed in terms of the voltages at the nodes of the circuit \cite{Vool2017}.
These equations are derived by systematically combining (i) the voltage-current relations for the individual circuit elements, and
(ii) Kirchhoff's laws, which state that charges do not accumulate within circuit nodes and that voltages are single-valued.  The circuit's normal modes of oscillation, in the absence of an external drive, then correspond to the eigenvectors of a Hermitian generalized eigenproblem (which can be transformed into a standard eigenproblem by
Cholesky factorization); the eigenvalues are real and correspond to the squares of the normal mode eigenfrequencies.  Alternatively, the special case of a circuit comprised of weakly-coupled high-Q LC resonators can be mapped to a Hermitian eigenproblem governed by a tight-binding Hamiltonian, whose eigenvalues correspond to the normal mode detunings from the center frequency.  In either case, one can use the lattice Hamiltonians and their eigenspaces to compute standard topological band invariants for the LC circuits.

In typical circuit experiments, the normal modes are studied using weakly-coupled probes such as pickup coils \cite{Simon2015, Li2018b}, or direct connections to an oscilloscope or network analyzer \cite{Rosenthal2018, Liu2019}.  Lee \textit{et al.}~have also developed a rigorous frequency-domain formalism for analyzing circuits with explicit current sources and sinks \cite{Lee2018}.  Their approach is similar to the Green's function method in electrodynamics and quantum mechanics, and involves calculating an admittance matrix whose zero modes manifest as a diverging impedances observable in parametric sweeps of the driving frequency.

The most commonly-used method for introducing nonlinearity into an LC circuit is to use varactors, also known as varicap diodes \cite{hadad2018self, Wang2019, SerraGarcia2019, Zangeneh-Nejad2019}. These two-port circuit elements are essentially diodes operated in
reverse bias; with increasing reverse bias voltage, the thickness of the diode's depletion region increases and its effective capacitance decreases.  For alternating-current (AC) operation, a nonlinear capacitor can be implemented by a pair of varactors arranged back-to-back, such that neither varactor can be forward-biased.  The resulting capacitance decreases with the magnitude of the voltage across the circuit element, independent of its sign.  Furthermore, a constant bias voltage can be applied to the individual varactors to
tune their effective capacitance; this method was recently used by Serra-Garcia \textit{et al.}~to perform high-quality observations of a topological transition in a quadrupole topological insulator circuit \cite{SerraGarcia2019}.

\subsection{Su-Schrieffer-Heeger circuits}

\begin{figure}
    \centering
    \includegraphics[width=\columnwidth]{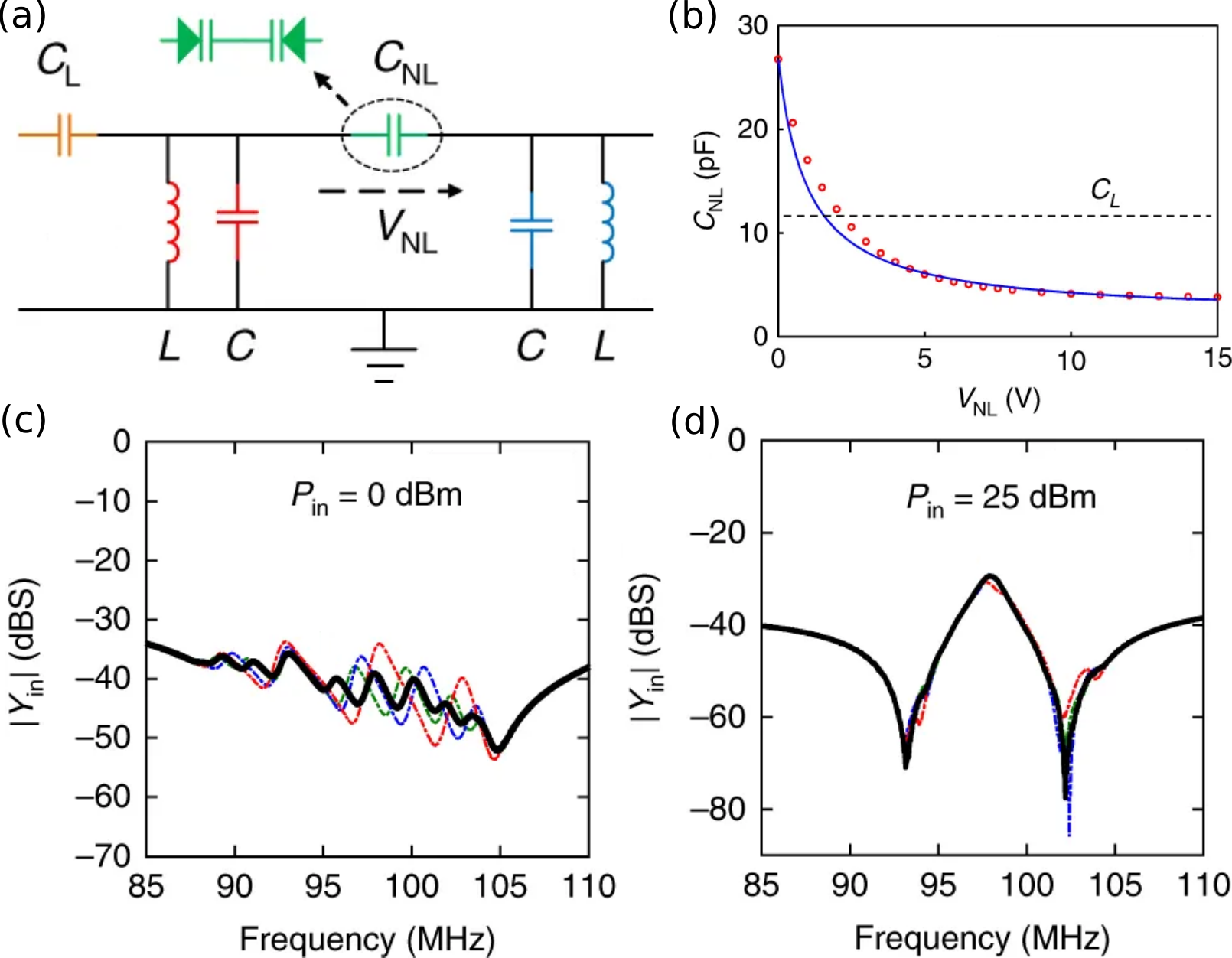}
    \caption{Nonlinear electronic circuit analogue of the Su-Schrieffer-Heeger model. (a) Schematic of unit cell consisting of sites (LC resonators) coupled via linear $C_{\mathrm{L}}$ and nonlinear $C_{\mathrm{NL}}$ capacitors. (b) Nonlinear capacitance $C_{\mathrm{NL}}$ implemented using back-to-back varactor diodes. (c,d) Measured input admittance spectra at the edge of a 6-cell chain. (c) In the linear limit the chain is trivial and multiple peaks (resonances) associated with bulk modes are observed. (d) In the nonlinear regime a self-induced edge state emerges, visible as a single dominant resonance in the middle of the spectrum. Solid black line corresponds to ideal chain while thin colored curves are measured spectra in presence of various defects, demonstrating robustness to disorder. Adapted from Ref.~\onlinecite{hadad2018self}.}
    \label{fig:electronics}
\end{figure}

Hadad \textit{et al.}~used such nonlinear capacitors to implement a circuit analogue of a nonlinear Su-Schrieffer-Heeger (SSH) model
\cite{hadad2018self}.  They fabricated a dimerized 1D lattice of LC resonators with two resonators (sites) per unit cell, shown in Fig.~\ref{fig:electronics}(a). The intra-cell and inter-cell couplings of the SSH model were implemented by two types of capacitors, one of which was nonlinear [Fig.~\ref{fig:electronics}(b)]. In the usual linear SSH model, the ratio of the two coupling strengths determines the topological phase.  In a nonlinear SSH model, it was theoretically predicted \cite{Hadad2016} that even if the lattice is topologically trivial in the linear (zero-intensity) limit, a nonlinear coupling ratio can drive a topological phase transition via the formation of a soliton-like self-induced boundary state.  Such a soliton was indeed observed in the circuit experiment, in the form of an input admittance peak appearing at a mid-gap frequency when the lattice was driven above a threshold power level, as shown in Fig.~\ref{fig:electronics}(c,d). The resonance frequency was shown to be insensitive to disorder introduced by deliberately shorting different resonators to ground, consistent with the topological protection in the underlying linear SSH model.

Subsequently, Wang \textit{et al.}~studied a similar nonlinear SSH-like circuit, focusing on the use of the topological boundary state to enhance harmonic generation \cite{Wang2019}.  They implemented a significantly longer 1D lattice, with 40 sites and broad operating frequency bands.  Such a circuit can be viewed as a type of left-handed nonlinear transmission line (NLTL) \cite{NLTL}, and supports an SSH-like bandgap with a mid-gap topological boundary state at the fundamental harmonic, as well as propagating-wave modes at higher harmonics.  When the circuit was excited at the boundary, cross-phase modulation between the two types of modes gave rise to strongly enhanced generation of third- and higher-harmonic signals, five times higher than in a standard (non-dimerized) NLTL and two orders of magnitude higher than in the lattice's topologically trivial configuration.

Zangeneh-Nejad and Fleury \cite{Zangeneh-Nejad2019} have extended nonlinear topological circuits to the class of high-order topological insulators \cite{Benalcazar2017, Imhof2018, SerraGarcia2019}.  The 2D lattice they studied hosts a nontrivial topological phase characterized by quantized Wannier centers (with quantized values of the bulk polarization) and robust mid-gap corner modes \cite{Ezawa2018}, with the topological transition governed by the ratio of intra-cell to inter-cell couplings (similar to the SSH case). The nonlinear circuit was again implemented by using back-to-back varactors for the inter-cell connections, and self-induced corner states were observed above a certain power threshold.

\subsection{Other directions} 

While back-to-back varactors have the advantage of realizing an extremely simple Kerr-like nonlinearity, they are not the only nonlinear circuit elements available.  The alternatives, however, introduce an additional complication: they are typically not only nonlinear but also ``active'' (i.e., energy-non-conserving or non-Hermitian).  As we will discuss in Sec.~\ref{sec:lasers}, the combination of nonlinearity, non-Hermiticity, and band topology to form topological lasers is an active and largely-unsettled area of research, and electronic circuits may serve as a key playground for future experimental investigations.  

Recently, Kotwal \textit{et al.}~have taken the first steps in this direction by performing a theoretical analysis of 1D and 2D topological circuits with nonlinear negative-resistance elements such as van der Pol circuits or tunnel diodes \cite{Kotwal2019}.  They uncovered an extremely rich set of behaviors, such as SSH-like boundary states that exhibit self-sustained limit cycle oscillations, which can induce synchronized bulk oscillations that mediate the interactions between different boundaries.  The topological features of the underlying lattice seem to make the self-sustained oscillations insensitive to lattice deformations.

Based on the methods experimentally demonstrated to date, there are numerous opportunities to use nonlinear circuits to study further topological phenomena.  For instance, not all of the predicted properties of topological solitons have been definitively observed in circuit experiments, such as the frequency detuning and non-exponential decay profiles of 1D solitons~\cite{Hadad2016}. It is presently unclear whether or to what extent the sublattice trick, which proved useful for simulating T-breaking in linear circuits, can co-exist with nonlinear circuit elements \cite{Simon2015, Albert2015}.  Achieving real or effective T-breaking in a 2D electronic circuit would enable intriguing applications such as robust traveling wave amplification \cite{Peano2016}.  It would also be interesting to explore how nonlinearities affect topological phenomena that rely intrinsically on non-Hermiticity, which have already been studied in linear circuits with resistive elements \cite{Rosenthal2018}.

\section{Topological lasers}
\label{sec:lasers}

\subsection{Motivation and general approaches} 

Topological photonics has exciting potential applications for the design of lasers, as it provides a systematic way to control the number and degree of localization of spectrally-isolated edge and defect modes in photonic structures. For example, mid-gap modes of 1D topological lattices are optimally localized within the band gap, which allows for the tight confinement of lasing modes~\cite{StJean2017,StJean2017,Parto2018,Zhao2018,Ota2018,Han2019}. In 2D systems, backscattering-immune edge modes hold promise for the design of ring cavities supporting large modal volumes and single mode operation regardless of the cavity shape~\cite{Bahari2017,Segev2018b,Klembt2018}. In both cases, the resulting modes are protected against certain classes of fabrication disorder, offering improved device reliability. 

At a fundamental level, topological lasers are interesting as a platform for exploring the interplay between nonlinearity and topology. Once a mode rises above the lasing threshold it becomes crucial to account for nonlinear gain saturation, which is what enables the system to relax towards a steady state. The high optical intensity within the laser cavity can also lead to other nonlinear effects such as Kerr self-focusing. Nonlinearities in conventional lasers are known to lead to a rich variety of phenomena including chaos and instabilities, so it is interesting to ask how these effects interact with the topological features of the photonic structure.

Since 2017, several experiments have demonstrated lasing of topological edge modes in both 1D and 2D lattices. The experiments can be divided into two classes: (1) photonic lattices of coupled resonators with structural periods somewhat larger than the operating wavelength, and (2) photonic crystals with structural periods comparable to the operating wavelength. These systems have been modelled as either class A or class B lasers~\cite{laser_book}.

In class A lasers such as quantum cascade lasers, the photon lifetime is much longer the gain medium's polarization and population inversion, which are adiabatically eliminated leaving a nonlinear wave equation involving only the optical field amplitude $\psi$. Under the tight binding approximation, this results in a discrete set of equations of the form
\begin{equation}
i \partial_t \psi_n = \hat{H}_L \psi_n + \frac{g_n (i + \alpha)}{1+|\psi_n|^2 / I_{\mathrm{sat}}} \psi_n, \label{eq:classA}
\end{equation}
where $n$ indexes the weakly coupled resonators forming the tight binding lattice and $\hat{H}_L$ is an effective Hamiltonian accounting for all the linear effects such as absorption $\gamma$, coupling between the resonators $J$, and disorder $W$. The second term is nonlinear and describes the saturation of the gain induced by the pump $g_n$, governed by a characteristic intensity scale $I_{\mathrm{sat}}$. The linewidth enhancement factor $\alpha$ accounts for carrier-induced shifts of the ambient refractive index, which can lead to self-focusing or defocusing behaviour. This model assumes frequency independent gain, a good approximation for tight binding lattices which typically have a narrow bandwidth. 

Semiconductor gain media such as quantum dots that are typically integrated with photonic nanostructures are class B lasers. In these lasers, the free carriers providing the gain have a much longer lifetime than the photons and their dynamics must be taken into account, resulting in coupled equations of the form~\cite{longhi2018b,longhi2018}
\begin{align}
    i\partial_t \psi_n &= \hat{H}_L \psi_n + N_n (i+\alpha) \psi_n, \\
    \tau \partial_t N_n &= R_n - N_n - (1+2N_n)|\psi_n|^2,
\end{align}
where $N_n(t)$ is the normalized excess carrier population, $R_n$ is the normalized excess pump rate, and $\tau$ is the ratio between the carrier and photon lifetimes ($\approx 100-1000$ for semiconductor lasers and $\ll 1$ for class A lasers). A significant challenge presented by class B models is that while topological protection can be readily implemented in the photonic part of the field, the carrier populations $N_n$ are not coupled directly to one another and do not share this protection. Slow carrier dynamics are a well-established source of instabilities in coupled semiconductor laser arrays; as the carrier lifetime is increased, the stable steady states of the class A limit become unstable and are replaced by limit cycles and eventually chaotic dynamics~\cite{laser_book}.

In both classes of lasers, the lasing modes can be obtained numerically using standard iterative methods to solve for stationary states of nonlinear wave equations (e.g.~Newton's method), and seeking real frequency solutions. The initial guess for these iterative schemes is typically chosen to be the profile of the mode of interest at its threshold, obtained by solving a linear eigenvalue problem, followed by standard linear stability analysis to determine if the lasing modes are stable. The solution can be further verified by taking a direct numerical solution of the governing equations starting from small random field amplitudes as the initial condition; this can result in convergence to the stationary solution (in the case of stable single mode lasing), persistent oscillations between competing lasing modes (multimode lasing), or more complex dynamics such as irregular pulsations and chaos~\cite{laser_book}. 

\subsection{Lasing of 1D edge modes}

The first examples of topological lasers were based on the 1D Su-Schrieffer-Heeger (SSH) model. Interest in this type of topological laser was sparked by a 2013 theoretical study of the SSH model with staggered linear gain and loss~\cite{Schomerus2013OL}, which modeled a 1D topological photonic crystal under inhomogeneous pumping. For sufficiently weak gain/loss, the bulk modes overlap with both the gain and loss regions due to the chiral symmetry of the SSH model. This results in bulk modes with vanishing net gain. At the ends of the array or at domain walls, there exist topological modes localized to a single sublattice; pumping this sublattice gives the modes nonzero net gain, allowing them to lase before the bulk modes. The topological modes inherit a certain robustness to disorder, since they reside in the middle of the band gap and are spectrally isolated from other modes. As the gain/loss is increased, the bulk band gap becomes smaller and eventually closes. The bulk modes then start to localize onto the pumped sublattice and compete with the topological modes, resulting in multiple modes rising above the lasing threshold~\cite{Weimann2016}.

\begin{figure}
    \centering
    \includegraphics[width=\columnwidth]{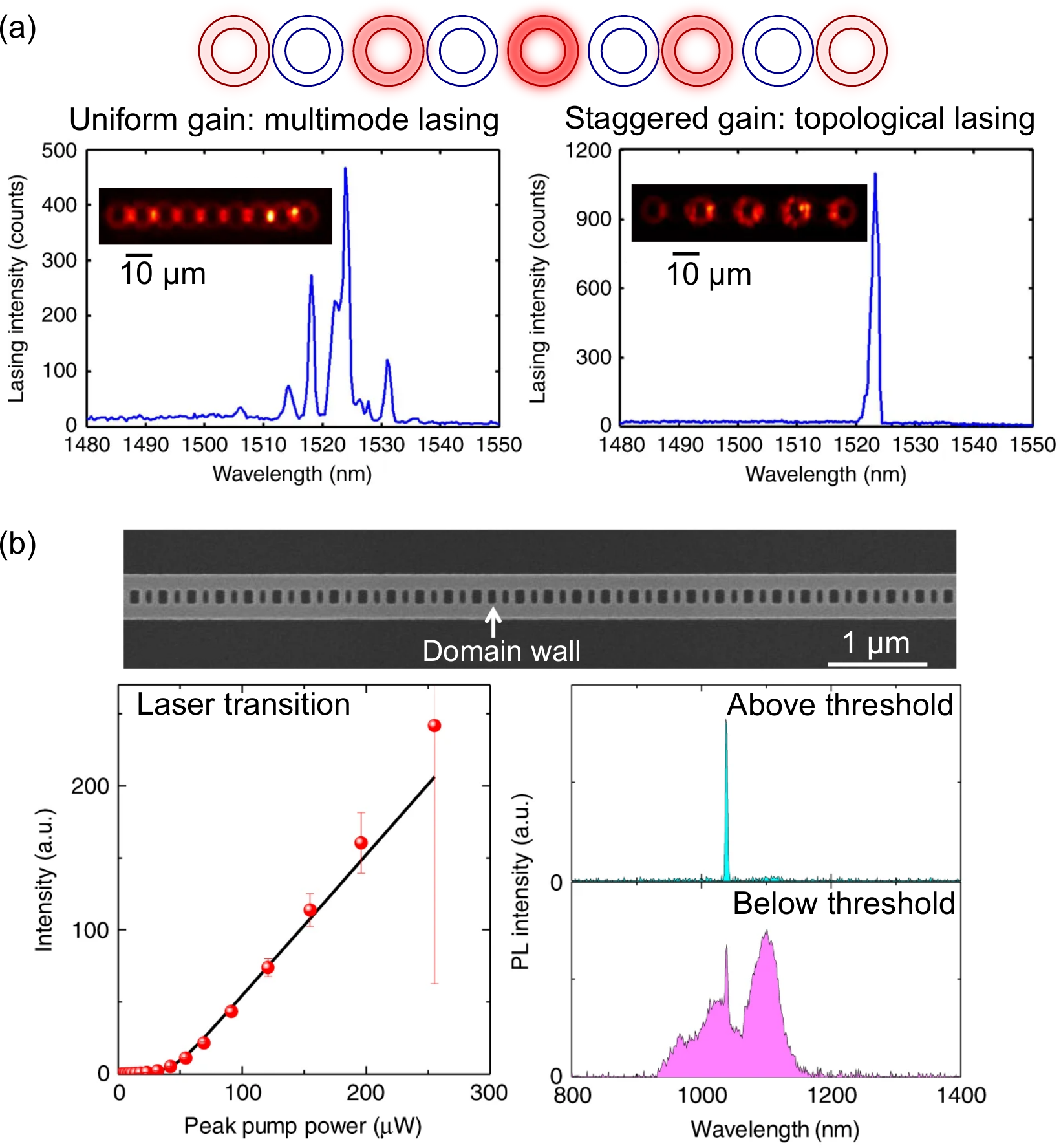}
    \caption{Topological lasers based on the 1D Su-Schrieffer-Heeger chain. (a) Photonic lattice of coupled silicon microring resonators exhibiting multimode or single mode lasing depending on the gain profile, adapted from Ref.~\onlinecite{Zhao2018}. (b) Lasing at a topological domain wall in a GaAs nanobeam  photonic crystal, adapted from Ref.~\onlinecite{Ota2018}. Left panel: Emission intensity vs pump power showing a transition to lasing at the threshold power of about 46~$\mu$W. Top right: Narrow spectral line of the topological mode emission above the threshold at the pump power 150~$\mu$W. Bottom right: Broad spectrum of the emission below the threshold at the pump power 5~$\mu$W.}
    \label{fig:laser1}
\end{figure}

These predictions were observed in a trio of photonic lattice experiments in 2017~\cite{StJean2017,Parto2018,Zhao2018}. St.-Jean \textit{et al.}~\cite{StJean2017} employed a zigzag polariton lattice of micropillars, while Parto \textit{et al.}~\cite{Parto2018} and Zhao \textit{et al.}~\cite{Zhao2018} both used ring resonator lattices with embedded InGaAsP/InP quantum wells as the gain medium. The latter is illustrated in Fig.~\ref{fig:laser1}(a). Uniform pumping results in spatially delocalized multimode emission due to competition between bulk modes, while pumping a single sublattice results in single mode lasing of the topological interface state as the bulk band gap remains open. Robustness of the edge modes to certain classes of perturbations was also demonstrated. 

Second generation designs based on nanoscale photonic crystals are now emerging. In 2018, Ota \textit{et al.}~reported lasing at $\lambda \approx 1040$nm in a protected defect mode at a topological domain wall of a GaAs nanobeam photonic crystal with embedded InAs quantum dots~\cite{Ota2018}, shown in Fig.~\ref{fig:laser1}(b). Their design supports strongly-confined defect modes with modal volumes as small as $0.23(\lambda/n)^3$, quality factors up to $Q \approx 59,700$, and spontaneous emission coupling factor $\beta \sim 0.03$. Similarly, Han \textit{et al.}~\cite{Han2019} used nanocavities based on L3 defects in a hexagonal InAsP/InP photonic crystal to achieve $Q \approx 35,000$ and $\beta \approx 0.15$ at 1550nm. These values are however comparable to conventional photonic crystal cavities; the main benefit of the topological design is the ability to systematically control the $Q$ factor and mode volume via the size of the bulk band gap while preserving single mode operation. So far, these experiments have been limited to optical pumping by ultrashort pulses at powers relatively close to the lasing threshold, with observations largely explained in terms of linear modes.

The SSH model also provides a simple testbed for exploring nonlinear dynamics of topological lasers and understanding whether there can be meaningful topological effects in the nonlinear regime. For example, if the linewidth enhancement factor is neglected ($\alpha = 0$), under inhomogeneous pumping the SSH model exhibits a dynamical charge conjugation symmetry, a nonlinear and non-Hermitian analogue of the chiral symmetry protecting the linear topological edge states~\cite{Malzard2018b}. The charge conjugation symmetry protects stationary zero modes localized to the pumped sublattice, with the number of these modes only changing at nonlinear bifurcations, which can be considered a nonlinear topological transition~\cite{Malzard2018}. Above a critical power, the zero modes become unstable and give birth to symmetry-protected time-periodic oscillatory modes at Hopf bifurcations. While $\alpha \ne 0$ breaks the charge conjugation symmetry, the spectral isolation of the nonlinear modes means they can persist for sufficiently weak symmetry-breaking perturbations. Similar behaviour is observed for other forms of nonlinearity~\cite{Cancellieri2019} and in 2D analogues of the SSH model such as the Lieb lattice~\cite{Malzard2018}.

The SSH model can also form the basis for a class of topology-inspired large volume single mode lasers by introducing \emph{non-Hermitian coupling}. For example, asymmetric non-Hermitian coupling $J_{n,n\pm 1} \propto \exp( \pm h)$ describes the preferential hopping of the optical field from site $n$ to $n+1$, equivalent to an imaginary effective gauge field $h$. In a finite lattice this does not affect the energy spectrum because the gauge field can be removed by the gauge transformation $\psi_n \rightarrow \psi_{n} \exp(h n)$. However, this transformation changes the eigenmodes' localization: all modes start to localize to one end of the lattice. When edge states exist (e.g., in the SSH model), the non-Hermitian localization competes with the localization $\xi$ of the topologically-protected edge states. At a critical imaginary gauge field strength $h = \xi$, this leaves all but one of the modes localized to one edge of the system, with the remaining (topologically-protected) zero mode delocalized over the whole lattice and therefore able to saturate the gain at all the pumped sites ~\cite{longhi_supermode}. 

The (Hermitian) SSH model is not the only way to design novel topological lasers. More recently, the idea of non-Hermitian topological phases has been developed~\cite{Gong2018}, which can be used to design disorder-robust delocalized modes in 1D systems using non-Hermitian coupling. As a second example, symmetric non-Hermitian coupling $J_{n,n\pm 1} \propto e^{i h}$ describes effective gain dependent on the modal wavenumber, i.e. the relative phase between the optical field at neighbouring lattice sites~\cite{longhi2018b,longhi2018}. This phase-dependent gain can promote single mode lasing in simple quasi-1D ring-shaped lattices. Another recent proposal by Longhi~\cite{Longhi2019} has predicted a non-Hermitian topological transition from single mode lasing to multi-mode lasing in a mode-locked laser.

\subsection{Lasing of 2D edge modes}

\begin{figure}
    \centering
    \includegraphics[width=\columnwidth]{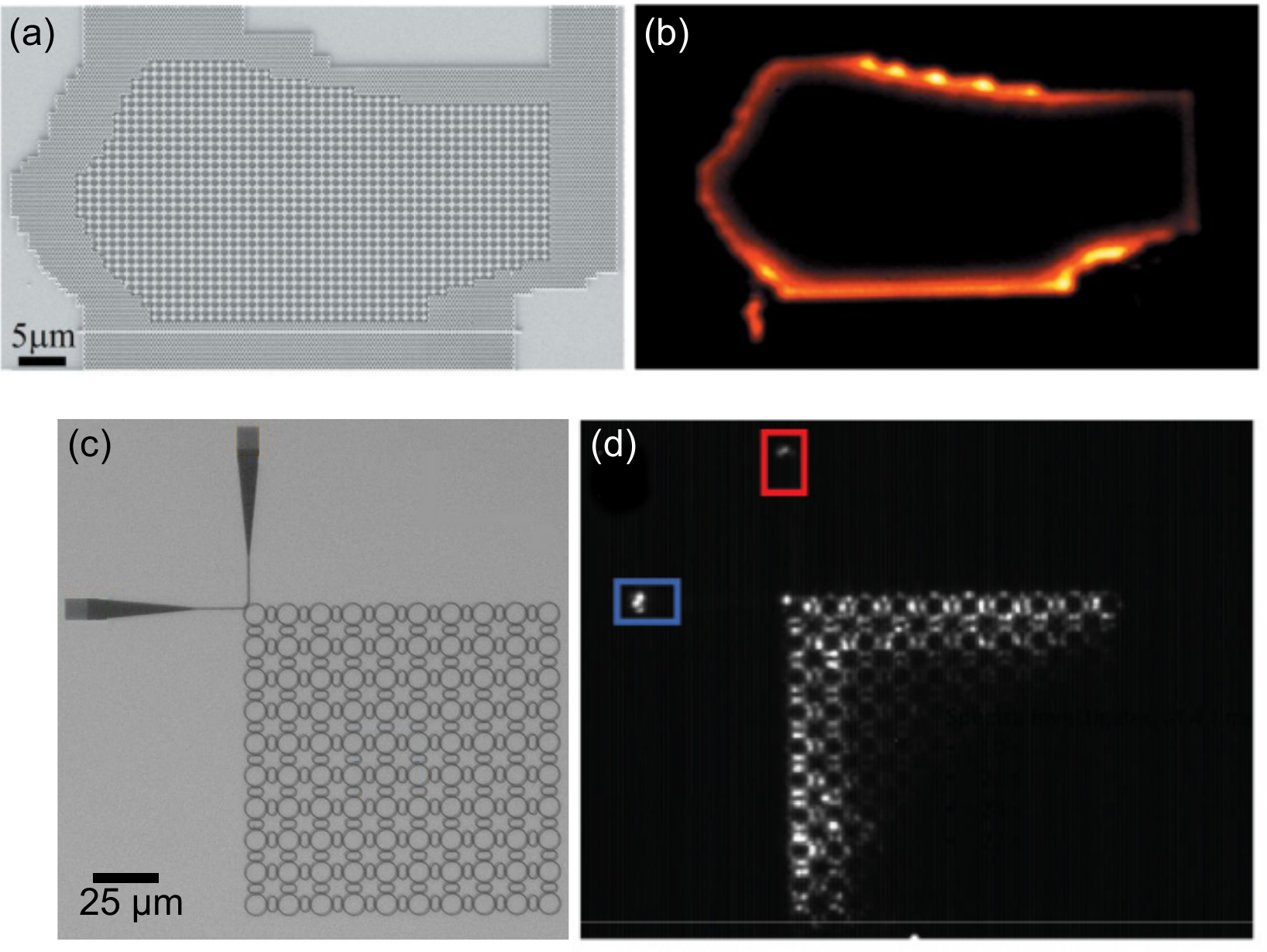}
    \caption{2D topological lasers on photonic quantum Hall edge states. Top row: (a,b) InGaAsP/YIG photonic crystal laser. (a) InGaAsP photonic crystal bonded on gyrotropic yttrium iron garnet (YIG). Under an external magnetic field, the inner domain forms a Chern insulator. (b) Intensity profile of the chiral lasing edge mode at the wavelength 1064~nm. Adapted from Ref.~\onlinecite{Bahari2017}. Bottom row: (c,d) InGaAsP ring resonators laser. (c) Photonic lattice of coupled InGaAsP microring resonators exhibiting spin momentum-locked edge states; (d) Chiral lasing via asymmetric pump: Despite preservation of time reversal symmetry, asymmetric pumping induces chiral lasing due to nonlinear gain saturation, resulting in an imbalance in intensity from the output ports coupled to the two spin states (highlighted in red and blue). Adapted from Ref.~\onlinecite{Segev2018b}.}
    \label{fig:laser2}
\end{figure}

The first experimental demonstration of lasing of 2D topological edge states by Bahari \textit{et al.}~in 2017~\cite{Bahari2017} used a photonic crystal embedded on a YIG substrate, shown in Fig.~\ref{fig:laser2}(a). Breaking time reversal symmetry via the magneto-optic effect creates a Chern insulator phase with a band gap hosting protected chiral edge modes. Despite the resulting topological band gap ($42$ pm) being very small due to weakness of magneto-optical effects at the operating wavelength 1530nm, as well as the entire device being pumped, the lasing profile shown in Fig.~\ref{fig:laser2}(b) is strongly localized to the edge and insensitive to its shape. This unexpected observation still awaits a theoretical explanation: based on the behaviour of the 1D SSH model, uniform pumping should have led to multi-mode lasing of bulk modes. In a follow-up study, the same group demonstrated the generation of high charge ($|l| \sim 100$) optical vortex beams using circular-shaped topological domain walls~\cite{Bahari_arxiv}.

Around the same time, Harari \textit{et al.}~\cite{Harari2018} studied theoretically class A laser models of 2D optical ring resonator lattices exhibiting topological edge states. They found that a pump localized to the edge sites was required to suppress bulk mode lasing and induce stable single mode lasing of the edge states. This single mode lasing also persisted in the presence of moderate disorder and weak symmetry-breaking perturbations that spoil the topological protection in the linear regime. In comparison, in similar non-topological models disorder tends to induce mode localization, resulting in multi-mode lasing involving modes localized at different positions along the edge. Ring resonator lattices incorporating gain and lasing of the topological edge states was then realized by Bandres \textit{et al.}~\cite{Segev2018b}; see Fig.~\ref{fig:laser2}(c,d). Interestingly, the combination of nonlinear gain saturation with spatial asymmetry (induced by either asymmetric pumping, or incorporating S bends into the ring resonators) resulted in observable optical non-reciprocity: preferential lasing of a single mode handedness or chirality, even though the underlying structure is non-magnetic and respects time-reversal symmetry, meaning that the linear edge modes occur in counter-propagating pairs with opposite spin. This is visible in Fig.~\ref{fig:laser2}(d) as an imbalance between the intensities at the two output ports.  

The final set of 2D experiments to date was based on a honeycomb lattice for exciton-polaritons combined with a strong magnetic field~\cite{Klembt2018}. The spin-orbit coupling of the polariton condensate combined with a magnetic field-induced Zeeman shift created a Chern insulator phase, although the resulting band gap was small, making it difficult to observe strong localization of the edge states. A subsequent theoretical study by Kartashov and Skryabin~\cite{KartashovLaser} of the governing class A gain model verified the existence of stable nonlinear lasing modes in this platform. They additionally found that above the polariton lasing threshold, self-action terms such as $\alpha$ lead to frequency shifts of the edge modes towards the bulk band edge. As the lasing mode approaches the band edge dynamical instabilities first develop, and then at higher powers the edge mode delocalizes due to resonance with bulk modes. Thus, the topological protection of the edge state lasing mode does not persist in the nonlinear regime.

In class B models, the slow carrier dynamics are another source of instability~\cite{longhi2018b}. While the photonic field remains localized to the edge and protected against disorder-induced backscattering, due to the carrier dynamics a limit cycle forms rather than a stationary state. In this limit cycle, a localized excitation circulates around the edge of the array due to competition between different edge modes with slightly different energies and similar effective gain. Because of the slow carrier response, the ``winner takes all'' effect of the saturable gain in the class A laser is not available to strictly enforce single mode operation. Due to this mode competition, the details of the dynamics and emission spectra become sensitive to the particular disorder realization.

Finally, we mention a very recent theoretical study of a class A laser model by Secli et al.~\cite{Secli2019} considering the effect of initial noise and fluctuations on the relaxation to stable lasing states. The precise energy of the lasing mode selected from one of the topological edge states spanning the gap is sensitive to the initial fluctuations. Moreover, the fluctuations along the edge are also protected, resulting in an ultraslow relaxation time. These effects may have a detrimental impact on the performance of 2D topological lasers and merit further investigation.  

\subsection{Future directions}

It is of interest to extend toplogical lasers to other material platforms and gain media. For example, lasing in 2D honeycomb and square lattices of plasmonic nanoparticles using organic dyes as the gain medium was recently demonstrated in Refs.~\onlinecite{Guo2019,Pourjamal2019}. Such class A lasers can avoid instabilities due to slow carrier dynamics. The radiative coupling present in such plasmonic systems means that the tight binding approximation is no longer valid, requiring re-examination of the results discussed above. 

The main selling point of topological lasers to date has been their potential for robust single mode continuous wave lasing, but for applications such as frequency comb generation or ultrashort pulse generation robust multimode emission is required. This can be achieved using lattices hosting multiple topological gaps and edge states~\cite{Pilozzi2017}, or by employing synthetic dimensions~\cite{YangCLEO}.

So far all topological lasers realized in experiment have been proofs of concept based on optically pumped gain media. Any real device applications will require electrical pumping, analysis of effects such as modulation bandwidth and how to avoid the instabilities discussed above~\cite{laser_book}, and most importantly, a ``killer application'' in which topological lasers outperform their conventional counterparts, such as tolerance to fabrication imperfections. As a step in this direction Suchomel et al.~\cite{Sucholmel2018} have implemented electrically-pumped polariton lasers in artificial honeycomb and square lattices, which can be readily generalized to topological lattices such as the time reversal-symmetric shrunken-expanded hexagon or valley Hall designs discussed in Sec~\ref{sec:lattices_intro}.

\section{Nonlinear nanophotonics} 
\label{sec:parametric} 

In the last five years, nanostructures made of high-index dielectric materials~\cite{Kuznetsov2016}, with judiciously designed resonant elements and lattice arrangements, have shown special promise for practical implementations of nonlinear topological photonics~\cite{Barik2018NN,Barik2018,Gorlach2018,Shalaev2018NatNano,He2018,Kruk2018,Smirnova2019}. This approach bridges the fundamental physics of topological phases with resonant nanophotonics and multipolar electrodynamics~\cite{Smirnova2016}.

The high-index dielectric nanostructures typically employed for topological nanophotonics possess strong optical nonlinearities enhanced by Mie-type resonances. In particular, silicon has a strong bulk third-order optical susceptibility~\cite{Shcherbakov2014,Smirnova2016ACS}, while III-V noncentrosymmetric  semiconductors are favorable for efficient second-order nonlinear applications due to a large volume quadratic nonlinearity~\cite{CamachoMorales2016,Frizyuk2019}. The resonant near-field enhancement associated with excitation of multipolar Mie modes in high-permittivty dielectric nanostructures further facilitate the nonlinear processes at the nanoscale.

To date, topological nanostructures that support subwavelength edge states and convert infrared radiation into visible light have been proposed and experimentally verified~\cite{Kruk2018,Smirnova2019}. The dynamic tunability of the topological properties in such nanostructures is approached via all-optical and thermo-optical tuning. Due to compactness and robustness to fabrication imperfections, topological nanophotonics is also being pursued for quantum information transport in integrated photonic platforms~\cite{Barik2018}.  

\subsection{Zigzag arrays} 

\begin{figure} [t]
\centerline{\includegraphics*[width=0.9\linewidth]{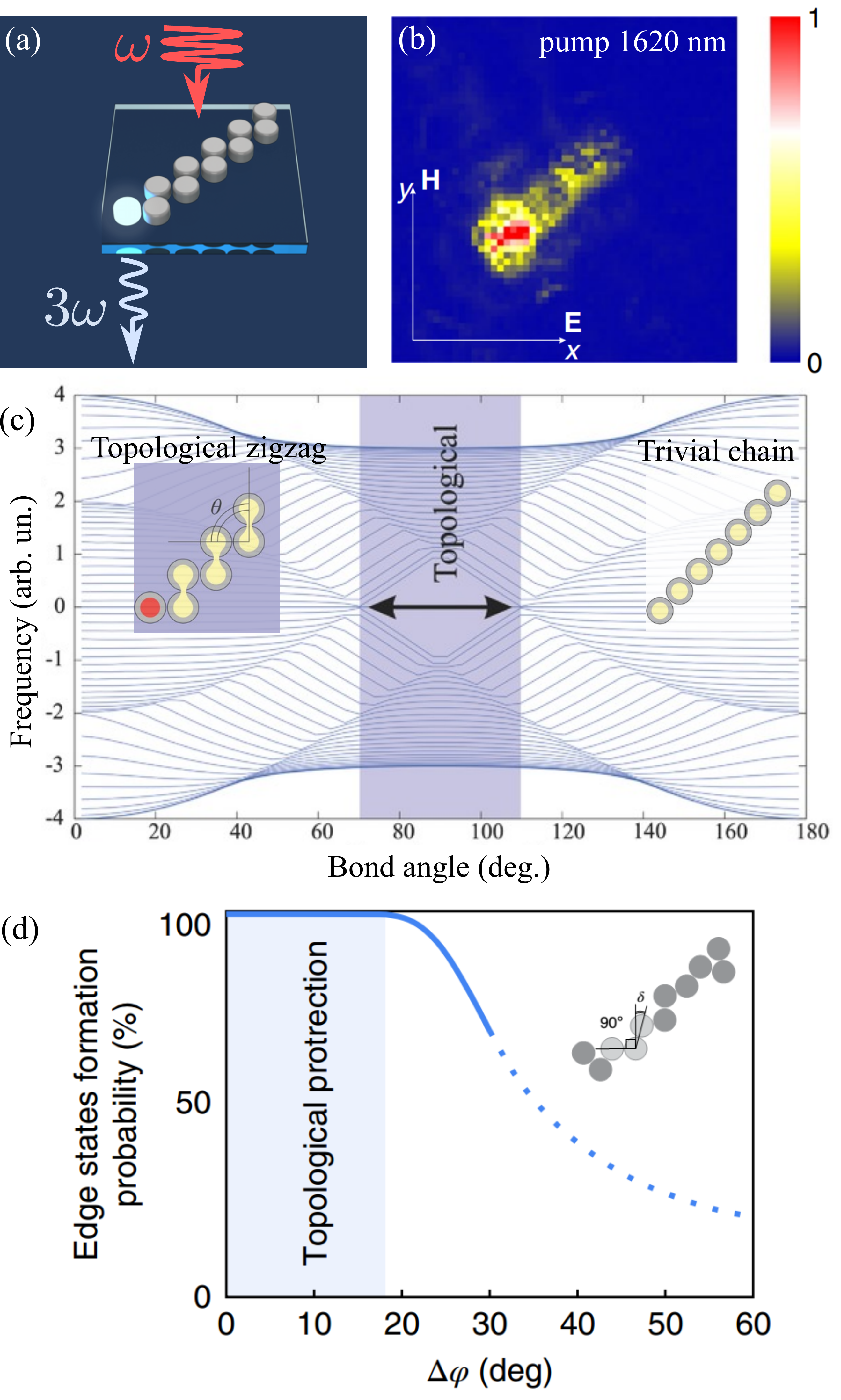}}
\caption{Nonlinear generation of light from a topological nanostructure~\cite{Kruk2018}. (a) Concept of THG in a zigzag array of nanoresonators: third-harmonic light (frequency $3\omega$) is generated by the topological edge state. Inset: SEM image of the sample. (b) Measured distribution of the third-harmonic field in 11-nanodisk zigzag array of Mie-resonant dielectric nanodisks. (c) Spectrum of the zigzag array calculated as the function of the bond angle. Shaded is the region where the topological edge states can exist~\cite{Kruk2017zigzag}.  (d) Robustness of the topological state to the disorder: edge state formation persists up to the disorder angle of $20^{\circ}$. \label{fig:zigzag}}%
\end{figure}

A simple yet fundamental topological model based on resonant nanoparticles is a zigzag array, originally proposed in 2015 for thin plasmonic nanodisks~\cite{Poddubny2014majorana}. Later, it was generalized to the case of dielectric particles with Mie resonances, followed by experimental studies for both microwaves~\cite{Slobozhanyuk2015} and optics~\cite{Slobozhanyuk2016LPR}. 
It can be described by a polarization-enriched generalized Su–Schrieffer–Heeger (SSH)-type model, with the nontrivial topological properties essentially captured in the framework the coupled dipole approximation.  Similar to the original SSH model, the alternation of strong and weak dipole-dipole couplings in zigzag geometries leads to the formation a boundary state at each boundary where the last coupling is weak. The straight chain is topologically trivial as it has a vanishing parity of the winding number, but for the zigzag chain the topological invariant is nonzero. The topological phase transition can be illustrated by plotting the energy spectrum of the finite chain as a function of the zigzag angle. The zigzag geometry engenders a chiral symmetric energy spectrum (see Fig.~\ref{fig:zigzag}). The structure exhibits a topological transition when the chain geometry changes from a line to a zigzag (the angle between three consecutive nanoparticles varies from 0 to $90^{\circ}$). The zigzag is topologically nontrivial in a range of bond angles near $90^{\circ}$.  When the angle lies within the shaded range, the system becomes gapped with two degenerate topological edge states in the middle of the gap.  

Very recently, unusual nonlinear properties due to topological phases in such arrays have been revealed in a third-harmonic generation experiment~\cite{Kruk2018}. Topologically nontrivial zigzag arrays of silicon nanodisks were fabricated on a glass substrate. Due to intrinsic nonlinearity of silicon, the topological edge state facilitates resonant generation of third-harmonic radiation. The topology-driven third-harmonic signal was shown to be robust against coupling disorder. A number of arrays with randomly generated bond angles between the disks was fabricated. In full agreement with theory, for disorder angle less than a critical value of $20^{\circ}$, edge states were observed in all cases. Remarkably, the observed third-harmonic radiation switched from one edge of the array to the other one, depending whether the system was illumination from the substrate or from air. This asymmetric harmonic generation is a type of nonreciprocal behavior and has potential applications for nanoscale topological optical diodes.

\subsection{Topological metasurfaces}

\begin{figure} [t!]
\centerline{\includegraphics*[width=0.9\linewidth]{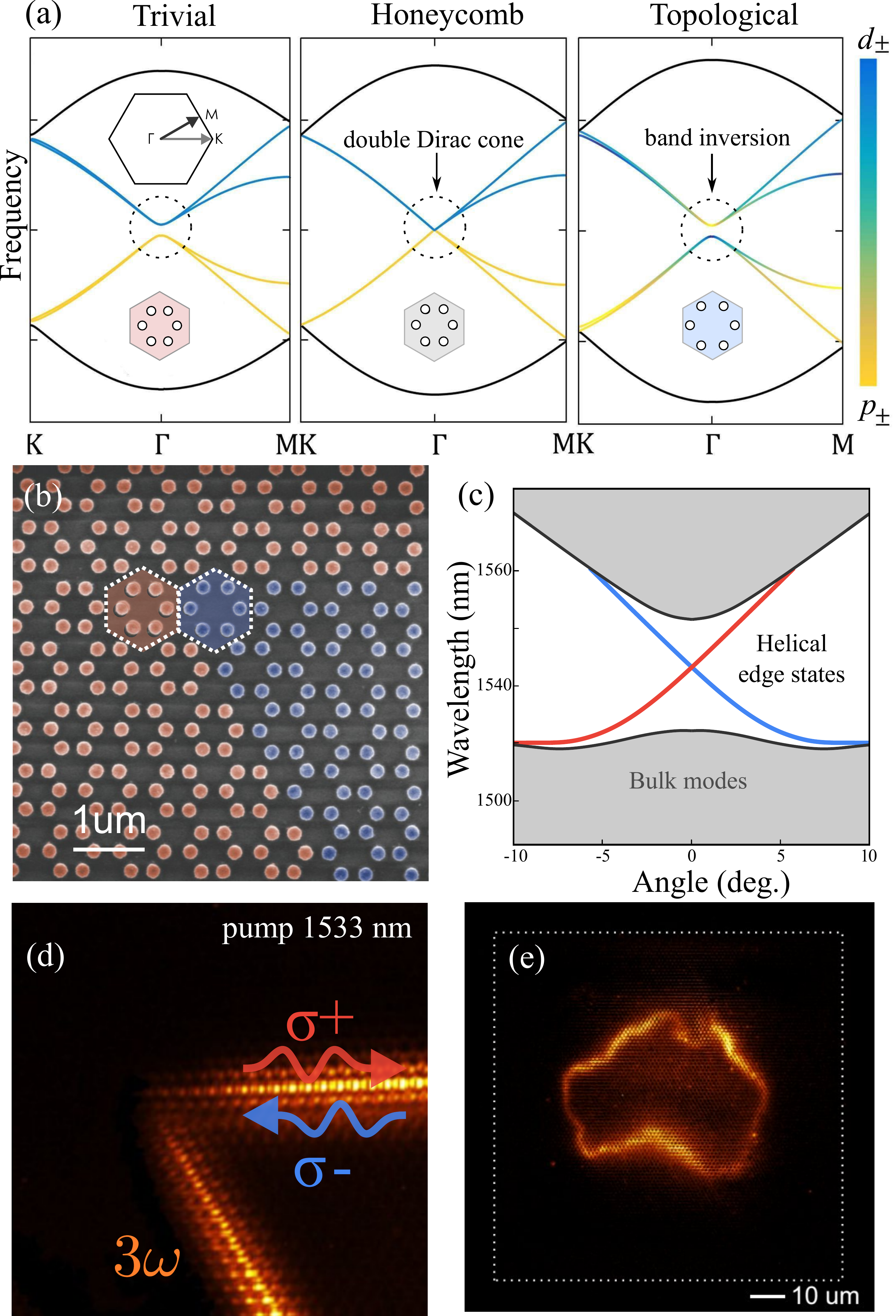}}
\caption{Third-harmonic generation from nanoscale helical edge states~\cite{Smirnova2019}. (a) Topological transition due to clustering the hexagonal unit cell. Band structures for shrunken (left), unperturbed (middle) and expanded (right) lattices of hexamers. Color of the bands encodes polarization ranging from pure c.p. dipolar ${p}_{\pm}$ to pure c.p. quadrupolar ${d}_{\pm}$ states. (b) SEM of the fabricated metasurface. (c) Calculated band diagram featuring gapped bands of bulk modes and Dirac-like crossing for the edge states. (d,e) Experimental images of third-harmonic generation by the edge states at the sharp-corner domain wall (d) and Australia-shaped contour (e). \label{fig:hexamers}}%
\end{figure}

Going to 2D, topology-controlled nonlinear light generation was demonstrated in a nanostructured metasurface with the domain wall supporting two counter-propagating spin-polarized edge waves (see Fig.~\ref{fig:hexamers}). Similar to the earlier theoretical proposal~\cite{Wu2015}, the topological metasurface was composed of hexamers of silicon nanoparticles. The nontrivial topological properties in QSHE phase are achieved by deforming a honeycomb lattice of silicon pillars into a triangular lattice of cylinder hexamers, as described in Sec.~\ref{sec:topo_intro}.

Figure~\ref{fig:hexamers}(c) shows the numerically computed bulk band diagram of the structure and the characteristic Dirac-like dispersion of the spin-momentum locked edge states residing in the band gap. Nonlinear imaging was employed to make the first direct observation of nanoscale helical edge states passing sharp corners~\cite{Smirnova2019}. The two pump polarizations couple to the edge modes with the opposite helicity values $\sigma_{+}$ and $\sigma_{-}$. The metasurface was excited by a tunable pulsed laser, and the third-harmonic signal was imaged onto a camera. The waveguiding domain wall in the geometry-independent photonic topological cavity was then clearly visualized via the third-harmonic field contour, as shown in Fig.~\ref{fig:hexamers}. 

For many practical applications, reconfigurability and dynamic tunability of photonic topological insulators are essential. In Ref.~\onlinecite{Shalaev2018} the position of the topological band gap in a pillar photonic crystal was proposed to be tuned by modifying the refractive index of a liquid crystals background medium with external electric field. Later, control over 
the spectral position of edge
states was implemented using pump-induced carrier generation in a topological photonic crystal slab~\cite{Shalaev2019}. Two theoretical proposals were made by Shalaev's group for ring resonators to realize switchable topological phase transitions, based on thermal tuning~\cite{Kudyshev2019termal} and integration with transparent conducting oxides~\cite{Kudyshev2019oxides}. 

Topological protection is of special interest for quantum photonic systems. The most common approach for generation of quantum light, being single photons, entangled photon pairs and correlated biphotons, relies on nonlinear frequency conversion, such as spontaneous parametric down-conversion (SPDC) and spontaneous four-wave mixing (SFWM) in nonlinear media. Nanophotonic structures with toplogical robustness of the spectrum of the supported photonic states against fabrication disorder and scattering losses can be potentially used to engineer robust quantum light sources and circuits. 

Topologically protected biphoton~\cite{BlancoRedondo2018} and entangled~\cite{Wang2019entangled} states were experimentally studied in the SSH-model-based array of coupled silicon nanowaveguides~\cite{blanco2016topological}. A biphoton correlation map resistant against coupling disorder was first reported in Ref.~\onlinecite{BlancoRedondo2018} for the waveguide array with a single long-long topological defect pumped at the infrared wavelength $1550$~nm. Due to the high third-order nonlinearity of silicon, the photon pairs generated via SFWM overlapped strongly with the topological defect mode localized at one sublattice with the topologically protected propagation constant. Subsequently, strong spatial entanglement between two topological states was revealed in the SSH geometry incorporating two coupled topological defects~\cite{Wang2019entangled}.

The use of ultra-thin metasurfaces helps to circumvent the restrictions associated with bulk phase-matching. Analytical analysis of the classical-quantum correspondence between sum frequency generation and SPDC from single particles and metasurfaces can be utilized to predict the generation yield~\cite{Marino2019}. 
Topological robustness has already been demonstrated for quantum transport of single photons in a perforated GaAs slab metasurface~\cite{Barik2018}. In the presence of the out-of-plane magnetic field, single photons with opposite circular polarizations were selectively generated by weakly pumped InAs quantum dots grown within the GaAs slab. Right- and left- handed single photons were observed to couple to oppositely propagating topological modes and propagate without backscattering at the bends. 

\subsection{Future prospects}

The above experiments have established topological dielectric nanostructures as a promising platform for robust generation and guiding of photons at the nanoscale. This topological all-dielectric platform can be used to build tunable and active topological photonic devices with integrated functional elements for advanced photonic circuitry (unidirectional waveguides, miniature topological cavities, low-power nanoscale lasers, etc.). Topological nanophotonic cavities may also be employed as nonlinear light sources, with near and far field characteristics tuned by topological phase transitions, and protected against fabrication imperfections. This has exciting prospects for singular optics and harmonic generation applications.

\section{Conclusions and Outlook}
\label{sec:outlook}

We have reviewed the basic physics and practical implementations of photonic systems that combine the studies of topological phases with nonlinear optics. Such systems can be modelled by nonlinear tight-binding models or nonlinear continuous-wave equations. Currently, there is a plethora of theoretical predictions of nonlinear phenomena in topological photonic structures, including solitons, modulational instability, frequency conversion, and optical switching. Many of these are now starting to be realized in experiments: the past two years have seen the first experimental demonstrations of lasing, harmonic generation, and nonlinearly-induced topological edge states.

While the study of electronic topological states has a long history, topological photonics is a comparatively young field of research. A pressing question now is how to harness this newly discovered degree of freedom in optical devices, for example to design and fabricate disorder-immune components for high-speed information transfer and processing. Topological photonic metasurfaces could form the basis for a new class of ultrathin devices with functionalities based on novel physical principles. As with conventional optical components, understanding and exploiting nonlinear effects offers many new opportunities, such as:

\begin{itemize}
\item Nonlinearities provide a straightforward way to reconfigure or otherwise manipulate topological lattices, and they are in particular essential for achieving ultra-fast modulation~\cite{Leykam2018}.

\item Parametric frequency conversion processes are technologically important. Feedback suppression enabled by certain topological edge states may be useful for stabilizing travelling wave amplifiers~\cite{Peano2016,Peano2016b}. Spontaneous wave mixing processes are an important source of entangled photon pairs for integrated quantum photonics applications.

\item Lasers are ubiquitous, and they become inherently nonlinear devices above threshold due to gain saturation, as well as they are always non-Hermitian. Topological edge states may be useful for the mode stabilization enabling high-power single-mode operation, although the extent to which this stabilization may hold in realistic devices is still under debates.

\item At a more fundamental level, nonlinear topological photonics provides a playground for exploring novel nonlinear wave equations originally derived in the context of high energy physics, and potentially realizing them in tabletop experiments. These models can support novel mechanisms for soliton formation (e.g. topological solitons, embedded solitons).
\end{itemize}

We envision nonlinear topological photonics to provide a fertile playground for not only studying interesting theoretical problems at the borderland between nonlinear dynamics and topology, but also as a route towards novel designs for disorder-robust photonic device applications, such as high-speed routing and switching, nanoscale lasers, and quantum light sources.

\begin{acknowledgments}
This work was supported by the Australian Research Council (grants DE190100430 and DP200101168) and the Institute for Basic Science in Korea (grant IBS-R024-Y1). Y.K. acknowledges a support from the Strategic Fund of the Australian National University. 
\end{acknowledgments}


\begin{thebibliography}{176}%
\makeatletter
\providecommand \@ifxundefined [1]{%
 \@ifx{#1\undefined}
}%
\providecommand \@ifnum [1]{%
 \ifnum #1\expandafter \@firstoftwo
 \else \expandafter \@secondoftwo
 \fi
}%
\providecommand \@ifx [1]{%
 \ifx #1\expandafter \@firstoftwo
 \else \expandafter \@secondoftwo
 \fi
}%
\providecommand \natexlab [1]{#1}%
\providecommand \enquote  [1]{``#1''}%
\providecommand \bibnamefont  [1]{#1}%
\providecommand \bibfnamefont [1]{#1}%
\providecommand \citenamefont [1]{#1}%
\providecommand \href@noop [0]{\@secondoftwo}%
\providecommand \href [0]{\begingroup \@sanitize@url \@href}%
\providecommand \@href[1]{\@@startlink{#1}\@@href}%
\providecommand \@@href[1]{\endgroup#1\@@endlink}%
\providecommand \@sanitize@url [0]{\catcode `\\12\catcode `\$12\catcode
  `\&12\catcode `\#12\catcode `\^12\catcode `\_12\catcode `\%12\relax}%
\providecommand \@@startlink[1]{}%
\providecommand \@@endlink[0]{}%
\providecommand \url  [0]{\begingroup\@sanitize@url \@url }%
\providecommand \@url [1]{\endgroup\@href {#1}{\urlprefix }}%
\providecommand \urlprefix  [0]{URL }%
\providecommand \Eprint [0]{\href }%
\providecommand \doibase [0]{http://dx.doi.org/}%
\providecommand \selectlanguage [0]{\@gobble}%
\providecommand \bibinfo  [0]{\@secondoftwo}%
\providecommand \bibfield  [0]{\@secondoftwo}%
\providecommand \translation [1]{[#1]}%
\providecommand \BibitemOpen [0]{}%
\providecommand \bibitemStop [0]{}%
\providecommand \bibitemNoStop [0]{.\EOS\space}%
\providecommand \EOS [0]{\spacefactor3000\relax}%
\providecommand \BibitemShut  [1]{\csname bibitem#1\endcsname}%
\let\auto@bib@innerbib\@empty
\bibitem [{\citenamefont {Lu}, \citenamefont {Joannopoulos},\ and\
  \citenamefont {Solja{\v{c}}i{\'{c}}}(2016)}]{Lu2016}%
  \BibitemOpen
  \bibfield  {author} {\bibinfo {author} {\bibfnamefont {L.}~\bibnamefont
  {Lu}}, \bibinfo {author} {\bibfnamefont {J.~D.}\ \bibnamefont
  {Joannopoulos}}, \ and\ \bibinfo {author} {\bibfnamefont {M.}~\bibnamefont
  {Solja{\v{c}}i{\'{c}}}},\ }\bibfield  {title} {\enquote {\bibinfo {title}
  {Topological states in photonic systems},}\ }\href {\doibase
  10.1038/nphys3796} {\bibfield  {journal} {\bibinfo  {journal} {Nature
  Physics}\ }\textbf {\bibinfo {volume} {12}},\ \bibinfo {pages} {626--629}
  (\bibinfo {year} {2016})}\BibitemShut {NoStop}%
\bibitem [{\citenamefont {Ozawa}\ \emph {et~al.}(2019)\citenamefont {Ozawa},
  \citenamefont {Price}, \citenamefont {Amo}, \citenamefont {Goldman},
  \citenamefont {Hafezi}, \citenamefont {Lu}, \citenamefont {Rechtsman},
  \citenamefont {Schuster}, \citenamefont {Simon}, \citenamefont {Zilberberg},\
  and\ \citenamefont {Carusotto}}]{Ozawa2019}%
  \BibitemOpen
  \bibfield  {author} {\bibinfo {author} {\bibfnamefont {T.}~\bibnamefont
  {Ozawa}}, \bibinfo {author} {\bibfnamefont {H.~M.}\ \bibnamefont {Price}},
  \bibinfo {author} {\bibfnamefont {A.}~\bibnamefont {Amo}}, \bibinfo {author}
  {\bibfnamefont {N.}~\bibnamefont {Goldman}}, \bibinfo {author} {\bibfnamefont
  {M.}~\bibnamefont {Hafezi}}, \bibinfo {author} {\bibfnamefont
  {L.}~\bibnamefont {Lu}}, \bibinfo {author} {\bibfnamefont {M.~C.}\
  \bibnamefont {Rechtsman}}, \bibinfo {author} {\bibfnamefont {D.}~\bibnamefont
  {Schuster}}, \bibinfo {author} {\bibfnamefont {J.}~\bibnamefont {Simon}},
  \bibinfo {author} {\bibfnamefont {O.}~\bibnamefont {Zilberberg}}, \ and\
  \bibinfo {author} {\bibfnamefont {I.}~\bibnamefont {Carusotto}},\ }\bibfield
  {title} {\enquote {\bibinfo {title} {Topological photonics},}\ }\href
  {\doibase 10.1103/RevModPhys.91.015006} {\bibfield  {journal} {\bibinfo
  {journal} {Rev. Mod. Phys.}\ }\textbf {\bibinfo {volume} {91}},\ \bibinfo
  {pages} {015006} (\bibinfo {year} {2019})}\BibitemShut {NoStop}%
\bibitem [{\citenamefont {Xie}\ \emph {et~al.}(2018)\citenamefont {Xie},
  \citenamefont {Wang}, \citenamefont {Zhu}, \citenamefont {Lu}, \citenamefont
  {Wang},\ and\ \citenamefont {Chen}}]{Xie2018}%
  \BibitemOpen
  \bibfield  {author} {\bibinfo {author} {\bibfnamefont {B.-Y.}\ \bibnamefont
  {Xie}}, \bibinfo {author} {\bibfnamefont {H.-F.}\ \bibnamefont {Wang}},
  \bibinfo {author} {\bibfnamefont {X.-Y.}\ \bibnamefont {Zhu}}, \bibinfo
  {author} {\bibfnamefont {M.-H.}\ \bibnamefont {Lu}}, \bibinfo {author}
  {\bibfnamefont {Z.~D.}\ \bibnamefont {Wang}}, \ and\ \bibinfo {author}
  {\bibfnamefont {Y.-F.}\ \bibnamefont {Chen}},\ }\bibfield  {title} {\enquote
  {\bibinfo {title} {Photonics meets topology},}\ }\href {\doibase
  10.1364/oe.26.024531} {\bibfield  {journal} {\bibinfo  {journal} {Optics
  Express}\ }\textbf {\bibinfo {volume} {26}},\ \bibinfo {pages} {24531}
  (\bibinfo {year} {2018})}\BibitemShut {NoStop}%
\bibitem [{\citenamefont {Longhi}(2017)}]{Longhi_2017}%
  \BibitemOpen
  \bibfield  {author} {\bibinfo {author} {\bibfnamefont {S.}~\bibnamefont
  {Longhi}},\ }\bibfield  {title} {\enquote {\bibinfo {title} {Parity-time
  symmetry meets photonics: A new twist in non-hermitian optics},}\ }\href
  {\doibase 10.1209/0295-5075/120/64001} {\bibfield  {journal} {\bibinfo
  {journal} {{EPL} (Europhysics Letters)}\ }\textbf {\bibinfo {volume} {120}},\
  \bibinfo {pages} {64001} (\bibinfo {year} {2017})}\BibitemShut {NoStop}%
\bibitem [{\citenamefont {Zhou}\ \emph {et~al.}(2017)\citenamefont {Zhou},
  \citenamefont {Wang}, \citenamefont {Leykam},\ and\ \citenamefont
  {Chong}}]{Zhou2017}%
  \BibitemOpen
  \bibfield  {author} {\bibinfo {author} {\bibfnamefont {X.}~\bibnamefont
  {Zhou}}, \bibinfo {author} {\bibfnamefont {Y.}~\bibnamefont {Wang}}, \bibinfo
  {author} {\bibfnamefont {D.}~\bibnamefont {Leykam}}, \ and\ \bibinfo {author}
  {\bibfnamefont {Y.~D.}\ \bibnamefont {Chong}},\ }\bibfield  {title} {\enquote
  {\bibinfo {title} {Optical isolation with nonlinear topological photonics},}\
  }\href {\doibase 10.1088/1367-2630/aa7cb5} {\bibfield  {journal} {\bibinfo
  {journal} {New J. Phys.}\ }\textbf {\bibinfo {volume} {19}},\ \bibinfo
  {pages} {095002} (\bibinfo {year} {2017})}\BibitemShut {NoStop}%
\bibitem [{\citenamefont {Kartashov}\ and\ \citenamefont
  {Skryabin}(2017)}]{Kartashov2017}%
  \BibitemOpen
  \bibfield  {author} {\bibinfo {author} {\bibfnamefont {Y.~V.}\ \bibnamefont
  {Kartashov}}\ and\ \bibinfo {author} {\bibfnamefont {D.~V.}\ \bibnamefont
  {Skryabin}},\ }\bibfield  {title} {\enquote {\bibinfo {title} {Bistable
  topological insulator with exciton-polaritons},}\ }\href {\doibase
  10.1103/PhysRevLett.119.253904} {\bibfield  {journal} {\bibinfo  {journal}
  {Phys. Rev. Lett.}\ }\textbf {\bibinfo {volume} {119}},\ \bibinfo {pages}
  {253904} (\bibinfo {year} {2017})}\BibitemShut {NoStop}%
\bibitem [{\citenamefont {Bandres}\ \emph {et~al.}(2018)\citenamefont
  {Bandres}, \citenamefont {Wittek}, \citenamefont {Harari}, \citenamefont
  {Parto}, \citenamefont {Ren}, \citenamefont {Segev}, \citenamefont
  {Christodoulides},\ and\ \citenamefont {Khajavikhan}}]{Segev2018b}%
  \BibitemOpen
  \bibfield  {author} {\bibinfo {author} {\bibfnamefont {M.~A.}\ \bibnamefont
  {Bandres}}, \bibinfo {author} {\bibfnamefont {S.}~\bibnamefont {Wittek}},
  \bibinfo {author} {\bibfnamefont {G.}~\bibnamefont {Harari}}, \bibinfo
  {author} {\bibfnamefont {M.}~\bibnamefont {Parto}}, \bibinfo {author}
  {\bibfnamefont {J.}~\bibnamefont {Ren}}, \bibinfo {author} {\bibfnamefont
  {M.}~\bibnamefont {Segev}}, \bibinfo {author} {\bibfnamefont {D.~N.}\
  \bibnamefont {Christodoulides}}, \ and\ \bibinfo {author} {\bibfnamefont
  {M.}~\bibnamefont {Khajavikhan}},\ }\bibfield  {title} {\enquote {\bibinfo
  {title} {Topological insulator laser: Experiments},}\ }\href {\doibase
  10.1126/science.aar4005} {\bibfield  {journal} {\bibinfo  {journal}
  {Science}\ }\textbf {\bibinfo {volume} {359}},\ \bibinfo {pages} {eaar4005}
  (\bibinfo {year} {2018})}\BibitemShut {NoStop}%
\bibitem [{\citenamefont {Chen}\ \emph {et~al.}(2018)\citenamefont {Chen},
  \citenamefont {Leykam}, \citenamefont {Chong},\ and\ \citenamefont
  {Yang}}]{Chen2018}%
  \BibitemOpen
  \bibfield  {author} {\bibinfo {author} {\bibfnamefont {W.}~\bibnamefont
  {Chen}}, \bibinfo {author} {\bibfnamefont {D.}~\bibnamefont {Leykam}},
  \bibinfo {author} {\bibfnamefont {Y.}~\bibnamefont {Chong}}, \ and\ \bibinfo
  {author} {\bibfnamefont {L.}~\bibnamefont {Yang}},\ }\bibfield  {title}
  {\enquote {\bibinfo {title} {Nonreciprocity in synthetic photonic materials
  with nonlinearity},}\ }\href {\doibase 10.1557/mrs.2018.124} {\bibfield
  {journal} {\bibinfo  {journal} {{MRS} Bulletin}\ }\textbf {\bibinfo {volume}
  {43}},\ \bibinfo {pages} {443--451} (\bibinfo {year} {2018})}\BibitemShut
  {NoStop}%
\bibitem [{\citenamefont {Mittal}, \citenamefont {Goldschmidt},\ and\
  \citenamefont {Hafezi}(2018)}]{Mittal2018}%
  \BibitemOpen
  \bibfield  {author} {\bibinfo {author} {\bibfnamefont {S.}~\bibnamefont
  {Mittal}}, \bibinfo {author} {\bibfnamefont {E.~A.}\ \bibnamefont
  {Goldschmidt}}, \ and\ \bibinfo {author} {\bibfnamefont {M.}~\bibnamefont
  {Hafezi}},\ }\bibfield  {title} {\enquote {\bibinfo {title} {A topological
  source of quantum light},}\ }\href {\doibase 10.1038/s41586-018-0478-3}
  {\bibfield  {journal} {\bibinfo  {journal} {Nature}\ }\textbf {\bibinfo
  {volume} {561}},\ \bibinfo {pages} {502--506} (\bibinfo {year}
  {2018})}\BibitemShut {NoStop}%
\bibitem [{\citenamefont {Kruk}\ \emph {et~al.}(2019)\citenamefont {Kruk},
  \citenamefont {Poddubny}, \citenamefont {Smirnova}, \citenamefont {Wang},
  \citenamefont {Slobozhanyuk}, \citenamefont {Shorokhov}, \citenamefont
  {Kravchenko}, \citenamefont {Luther-Davies},\ and\ \citenamefont
  {Kivshar}}]{Kruk2018}%
  \BibitemOpen
  \bibfield  {author} {\bibinfo {author} {\bibfnamefont {S.}~\bibnamefont
  {Kruk}}, \bibinfo {author} {\bibfnamefont {A.}~\bibnamefont {Poddubny}},
  \bibinfo {author} {\bibfnamefont {D.}~\bibnamefont {Smirnova}}, \bibinfo
  {author} {\bibfnamefont {L.}~\bibnamefont {Wang}}, \bibinfo {author}
  {\bibfnamefont {A.}~\bibnamefont {Slobozhanyuk}}, \bibinfo {author}
  {\bibfnamefont {A.}~\bibnamefont {Shorokhov}}, \bibinfo {author}
  {\bibfnamefont {I.}~\bibnamefont {Kravchenko}}, \bibinfo {author}
  {\bibfnamefont {B.}~\bibnamefont {Luther-Davies}}, \ and\ \bibinfo {author}
  {\bibfnamefont {Y.}~\bibnamefont {Kivshar}},\ }\bibfield  {title} {\enquote
  {\bibinfo {title} {Nonlinear light generation in topological
  nanostructures},}\ }\href {\doibase 10.1038/s41565-018-0324-7} {\bibfield
  {journal} {\bibinfo  {journal} {Nature Nanotechnology}\ }\textbf {\bibinfo
  {volume} {14}},\ \bibinfo {pages} {126--130} (\bibinfo {year}
  {2019})}\BibitemShut {NoStop}%
\bibitem [{\citenamefont {Leykam}\ \emph {et~al.}(2018)\citenamefont {Leykam},
  \citenamefont {Mittal}, \citenamefont {Hafezi},\ and\ \citenamefont
  {Chong}}]{Leykam2018}%
  \BibitemOpen
  \bibfield  {author} {\bibinfo {author} {\bibfnamefont {D.}~\bibnamefont
  {Leykam}}, \bibinfo {author} {\bibfnamefont {S.}~\bibnamefont {Mittal}},
  \bibinfo {author} {\bibfnamefont {M.}~\bibnamefont {Hafezi}}, \ and\ \bibinfo
  {author} {\bibfnamefont {Y.~D.}\ \bibnamefont {Chong}},\ }\bibfield  {title}
  {\enquote {\bibinfo {title} {Reconfigurable topological phases in
  next-nearest-neighbor coupled resonator lattices},}\ }\href {\doibase
  10.1103/PhysRevLett.121.023901} {\bibfield  {journal} {\bibinfo  {journal}
  {Phys. Rev. Lett.}\ }\textbf {\bibinfo {volume} {121}},\ \bibinfo {pages}
  {023901} (\bibinfo {year} {2018})}\BibitemShut {NoStop}%
\bibitem [{\citenamefont {Amo}(2018)}]{Amo2018}%
  \BibitemOpen
  \bibfield  {author} {\bibinfo {author} {\bibfnamefont {A.}~\bibnamefont
  {Amo}},\ }\bibfield  {title} {\enquote {\bibinfo {title} {When quantum optics
  meets topology},}\ }\href {\doibase 10.1126/science.aar7396} {\bibfield
  {journal} {\bibinfo  {journal} {Science}\ }\textbf {\bibinfo {volume}
  {359}},\ \bibinfo {pages} {638--639} (\bibinfo {year} {2018})}\BibitemShut
  {NoStop}%
\bibitem [{\citenamefont {Wang}\ \emph
  {et~al.}(2019{\natexlab{a}})\citenamefont {Wang}, \citenamefont {Lang},
  \citenamefont {Lee}, \citenamefont {Zhang},\ and\ \citenamefont
  {Chong}}]{Wang2019}%
  \BibitemOpen
  \bibfield  {author} {\bibinfo {author} {\bibfnamefont {Y.}~\bibnamefont
  {Wang}}, \bibinfo {author} {\bibfnamefont {L.-J.}\ \bibnamefont {Lang}},
  \bibinfo {author} {\bibfnamefont {C.~H.}\ \bibnamefont {Lee}}, \bibinfo
  {author} {\bibfnamefont {B.}~\bibnamefont {Zhang}}, \ and\ \bibinfo {author}
  {\bibfnamefont {Y.~D.}\ \bibnamefont {Chong}},\ }\bibfield  {title} {\enquote
  {\bibinfo {title} {Topologically enhanced harmonic generation in a nonlinear
  transmission line metamaterial},}\ }\href {\doibase
  10.1038/s41467-019-08966-9} {\bibfield  {journal} {\bibinfo  {journal}
  {Nature Communications}\ }\textbf {\bibinfo {volume} {10}},\ \bibinfo {pages}
  {1102} (\bibinfo {year} {2019}{\natexlab{a}})}\BibitemShut {NoStop}%
\bibitem [{\citenamefont {Barik}\ and\ \citenamefont
  {Hafezi}(2018)}]{Barik2018NN}%
  \BibitemOpen
  \bibfield  {author} {\bibinfo {author} {\bibfnamefont {S.}~\bibnamefont
  {Barik}}\ and\ \bibinfo {author} {\bibfnamefont {M.}~\bibnamefont {Hafezi}},\
  }\bibfield  {title} {\enquote {\bibinfo {title} {Robust and compact
  waveguides},}\ }\href {\doibase 10.1038/s41565-018-0314-9} {\bibfield
  {journal} {\bibinfo  {journal} {Nature Nanotechnology}\ }\textbf {\bibinfo
  {volume} {14}},\ \bibinfo {pages} {8--9} (\bibinfo {year}
  {2018})}\BibitemShut {NoStop}%
\bibitem [{\citenamefont {Shalaev}\ \emph
  {et~al.}(2018{\natexlab{a}})\citenamefont {Shalaev}, \citenamefont {Walasik},
  \citenamefont {Tsukernik}, \citenamefont {Xu},\ and\ \citenamefont
  {Litchinitser}}]{Shalaev2018NatNano}%
  \BibitemOpen
  \bibfield  {author} {\bibinfo {author} {\bibfnamefont {M.~I.}\ \bibnamefont
  {Shalaev}}, \bibinfo {author} {\bibfnamefont {W.}~\bibnamefont {Walasik}},
  \bibinfo {author} {\bibfnamefont {A.}~\bibnamefont {Tsukernik}}, \bibinfo
  {author} {\bibfnamefont {Y.}~\bibnamefont {Xu}}, \ and\ \bibinfo {author}
  {\bibfnamefont {N.~M.}\ \bibnamefont {Litchinitser}},\ }\bibfield  {title}
  {\enquote {\bibinfo {title} {Robust topologically protected transport in
  photonic crystals at telecommunication wavelengths},}\ }\href {\doibase
  10.1038/s41565-018-0297-6} {\bibfield  {journal} {\bibinfo  {journal} {Nature
  Nanotechnology}\ }\textbf {\bibinfo {volume} {14}},\ \bibinfo {pages}
  {31--34} (\bibinfo {year} {2018}{\natexlab{a}})}\BibitemShut {NoStop}%
\bibitem [{\citenamefont {Husko}\ \emph {et~al.}(2009)\citenamefont {Husko},
  \citenamefont {De~Rossi}, \citenamefont {Combrié}, \citenamefont {Tran},
  \citenamefont {Raineri},\ and\ \citenamefont {Wong}}]{Husko2009}%
  \BibitemOpen
  \bibfield  {author} {\bibinfo {author} {\bibfnamefont {C.}~\bibnamefont
  {Husko}}, \bibinfo {author} {\bibfnamefont {A.}~\bibnamefont {De~Rossi}},
  \bibinfo {author} {\bibfnamefont {S.}~\bibnamefont {Combrié}}, \bibinfo
  {author} {\bibfnamefont {Q.~V.}\ \bibnamefont {Tran}}, \bibinfo {author}
  {\bibfnamefont {F.}~\bibnamefont {Raineri}}, \ and\ \bibinfo {author}
  {\bibfnamefont {C.~W.}\ \bibnamefont {Wong}},\ }\bibfield  {title} {\enquote
  {\bibinfo {title} {Ultrafast all-optical modulation in gaas photonic crystal
  cavities},}\ }\href {\doibase 10.1063/1.3068755} {\bibfield  {journal}
  {\bibinfo  {journal} {Appl. Phys. Lett.}\ }\textbf {\bibinfo {volume} {94}},\
  \bibinfo {pages} {021111} (\bibinfo {year} {2009})}\BibitemShut {NoStop}%
\bibitem [{\citenamefont {Eggleton}\ \emph {et~al.}(2012)\citenamefont
  {Eggleton}, \citenamefont {Vo}, \citenamefont {Pant}, \citenamefont {Schr},
  \citenamefont {Pelusi}, \citenamefont {Choi}, \citenamefont {Madden},\ and\
  \citenamefont {Luther-Davies}}]{Eggleton2012}%
  \BibitemOpen
  \bibfield  {author} {\bibinfo {author} {\bibfnamefont {B.}~\bibnamefont
  {Eggleton}}, \bibinfo {author} {\bibfnamefont {T.}~\bibnamefont {Vo}},
  \bibinfo {author} {\bibfnamefont {R.}~\bibnamefont {Pant}}, \bibinfo {author}
  {\bibfnamefont {J.}~\bibnamefont {Schr}}, \bibinfo {author} {\bibfnamefont
  {M.}~\bibnamefont {Pelusi}}, \bibinfo {author} {\bibfnamefont {D.~Y.}\
  \bibnamefont {Choi}}, \bibinfo {author} {\bibfnamefont {S.}~\bibnamefont
  {Madden}}, \ and\ \bibinfo {author} {\bibfnamefont {B.}~\bibnamefont
  {Luther-Davies}},\ }\bibfield  {title} {\enquote {\bibinfo {title} {Photonic
  chip based ultrafast optical processing based on high nonlinearity dispersion
  engineered chalcogenide waveguides},}\ }\href {\doibase
  10.1002/lpor.201100024} {\bibfield  {journal} {\bibinfo  {journal} {Laser \&
  Photonics Reviews}\ }\textbf {\bibinfo {volume} {6}},\ \bibinfo {pages}
  {97--114} (\bibinfo {year} {2012})}\BibitemShut {NoStop}%
\bibitem [{\citenamefont {Silveirinha}(2016)}]{Silveirinha2016}%
  \BibitemOpen
  \bibfield  {author} {\bibinfo {author} {\bibfnamefont {M.~G.}\ \bibnamefont
  {Silveirinha}},\ }\bibfield  {title} {\enquote {\bibinfo {title} {Bulk-edge
  correspondence for topological photonic continua},}\ }\href {\doibase
  10.1103/PhysRevB.94.205105} {\bibfield  {journal} {\bibinfo  {journal} {Phys.
  Rev. B}\ }\textbf {\bibinfo {volume} {94}},\ \bibinfo {pages} {205105}
  (\bibinfo {year} {2016})}\BibitemShut {NoStop}%
\bibitem [{\citenamefont {Bliokh}\ \emph {et~al.}(2019)\citenamefont {Bliokh},
  \citenamefont {Leykam}, \citenamefont {Lein},\ and\ \citenamefont
  {Nori}}]{Bliokh2019}%
  \BibitemOpen
  \bibfield  {author} {\bibinfo {author} {\bibfnamefont {K.~Y.}\ \bibnamefont
  {Bliokh}}, \bibinfo {author} {\bibfnamefont {D.}~\bibnamefont {Leykam}},
  \bibinfo {author} {\bibfnamefont {M.}~\bibnamefont {Lein}}, \ and\ \bibinfo
  {author} {\bibfnamefont {F.}~\bibnamefont {Nori}},\ }\bibfield  {title}
  {\enquote {\bibinfo {title} {Topological non-hermitian origin of surface
  maxwell waves},}\ }\href {\doibase 10.1038/s41467-019-08397-6} {\bibfield
  {journal} {\bibinfo  {journal} {Nature Communications}\ }\textbf {\bibinfo
  {volume} {10}},\ \bibinfo {pages} {580} (\bibinfo {year} {2019})}\BibitemShut
  {NoStop}%
\bibitem [{\citenamefont {Hsieh}\ \emph {et~al.}(2011)\citenamefont {Hsieh},
  \citenamefont {McIver}, \citenamefont {Torchinsky}, \citenamefont {Gardner},
  \citenamefont {Lee},\ and\ \citenamefont {Gedik}}]{Hsieh2011}%
  \BibitemOpen
  \bibfield  {author} {\bibinfo {author} {\bibfnamefont {D.}~\bibnamefont
  {Hsieh}}, \bibinfo {author} {\bibfnamefont {J.~W.}\ \bibnamefont {McIver}},
  \bibinfo {author} {\bibfnamefont {D.~H.}\ \bibnamefont {Torchinsky}},
  \bibinfo {author} {\bibfnamefont {D.~R.}\ \bibnamefont {Gardner}}, \bibinfo
  {author} {\bibfnamefont {Y.~S.}\ \bibnamefont {Lee}}, \ and\ \bibinfo
  {author} {\bibfnamefont {N.}~\bibnamefont {Gedik}},\ }\bibfield  {title}
  {\enquote {\bibinfo {title} {Nonlinear optical probe of tunable surface
  electrons on a topological insulator},}\ }\href {\doibase
  10.1103/PhysRevLett.106.057401} {\bibfield  {journal} {\bibinfo  {journal}
  {Phys. Rev. Lett.}\ }\textbf {\bibinfo {volume} {106}},\ \bibinfo {pages}
  {057401} (\bibinfo {year} {2011})}\BibitemShut {NoStop}%
\bibitem [{\citenamefont {Zhao}\ \emph {et~al.}(2012)\citenamefont {Zhao},
  \citenamefont {Zhang}, \citenamefont {Qi}, \citenamefont {Chen},
  \citenamefont {Wang}, \citenamefont {Wen},\ and\ \citenamefont
  {Tang}}]{Zhao2012}%
  \BibitemOpen
  \bibfield  {author} {\bibinfo {author} {\bibfnamefont {C.}~\bibnamefont
  {Zhao}}, \bibinfo {author} {\bibfnamefont {H.}~\bibnamefont {Zhang}},
  \bibinfo {author} {\bibfnamefont {X.}~\bibnamefont {Qi}}, \bibinfo {author}
  {\bibfnamefont {Y.}~\bibnamefont {Chen}}, \bibinfo {author} {\bibfnamefont
  {Z.}~\bibnamefont {Wang}}, \bibinfo {author} {\bibfnamefont {S.}~\bibnamefont
  {Wen}}, \ and\ \bibinfo {author} {\bibfnamefont {D.}~\bibnamefont {Tang}},\
  }\bibfield  {title} {\enquote {\bibinfo {title} {Ultra-short pulse generation
  by a topological insulator based saturable absorber},}\ }\href {\doibase
  10.1063/1.4767919} {\bibfield  {journal} {\bibinfo  {journal} {Appl. Phys.
  Lett.}\ }\textbf {\bibinfo {volume} {101}},\ \bibinfo {pages} {211106}
  (\bibinfo {year} {2012})}\BibitemShut {NoStop}%
\bibitem [{\citenamefont {Chen}\ \emph {et~al.}(2014)\citenamefont {Chen},
  \citenamefont {Zhao}, \citenamefont {Li}, \citenamefont {Huang},
  \citenamefont {Lu}, \citenamefont {Zhang},\ and\ \citenamefont
  {Wen}}]{Chen2014}%
  \BibitemOpen
  \bibfield  {author} {\bibinfo {author} {\bibfnamefont {S.}~\bibnamefont
  {Chen}}, \bibinfo {author} {\bibfnamefont {C.}~\bibnamefont {Zhao}}, \bibinfo
  {author} {\bibfnamefont {Y.}~\bibnamefont {Li}}, \bibinfo {author}
  {\bibfnamefont {H.}~\bibnamefont {Huang}}, \bibinfo {author} {\bibfnamefont
  {S.}~\bibnamefont {Lu}}, \bibinfo {author} {\bibfnamefont {H.}~\bibnamefont
  {Zhang}}, \ and\ \bibinfo {author} {\bibfnamefont {S.}~\bibnamefont {Wen}},\
  }\bibfield  {title} {\enquote {\bibinfo {title} {Broadband optical and
  microwave nonlinear response in topological insulator},}\ }\href {\doibase
  10.1364/OME.4.000587} {\bibfield  {journal} {\bibinfo  {journal} {Opt. Mater.
  Express}\ }\textbf {\bibinfo {volume} {4}},\ \bibinfo {pages} {587--596}
  (\bibinfo {year} {2014})}\BibitemShut {NoStop}%
\bibitem [{\citenamefont {Bernevig}\ and\ \citenamefont
  {Hughes}(2013)}]{bernevig2013}%
  \BibitemOpen
  \bibfield  {author} {\bibinfo {author} {\bibfnamefont {B.}~\bibnamefont
  {Bernevig}}\ and\ \bibinfo {author} {\bibfnamefont {T.}~\bibnamefont
  {Hughes}},\ }\href {https://books.google.ru/books?id=wOn7JHSSxrsC} {\emph
  {\bibinfo {title} {Topological Insulators and Topological Superconductors}}}\
  (\bibinfo  {publisher} {Princeton University Press},\ \bibinfo {year}
  {2013})\BibitemShut {NoStop}%
\bibitem [{\citenamefont {Thouless}\ \emph {et~al.}(1982)\citenamefont
  {Thouless}, \citenamefont {Kohmoto}, \citenamefont {Nightingale},\ and\
  \citenamefont {den Nijs}}]{Thouless1982}%
  \BibitemOpen
  \bibfield  {author} {\bibinfo {author} {\bibfnamefont {D.~J.}\ \bibnamefont
  {Thouless}}, \bibinfo {author} {\bibfnamefont {M.}~\bibnamefont {Kohmoto}},
  \bibinfo {author} {\bibfnamefont {M.~P.}\ \bibnamefont {Nightingale}}, \ and\
  \bibinfo {author} {\bibfnamefont {M.}~\bibnamefont {den Nijs}},\ }\bibfield
  {title} {\enquote {\bibinfo {title} {Quantized hall conductance in a
  two-dimensional periodic potential},}\ }\href {\doibase
  10.1103/PhysRevLett.49.405} {\bibfield  {journal} {\bibinfo  {journal} {Phys.
  Rev. Lett.}\ }\textbf {\bibinfo {volume} {49}},\ \bibinfo {pages} {405--408}
  (\bibinfo {year} {1982})}\BibitemShut {NoStop}%
\bibitem [{\citenamefont {Kane}\ and\ \citenamefont {Mele}(2005)}]{Kane2005}%
  \BibitemOpen
  \bibfield  {author} {\bibinfo {author} {\bibfnamefont {C.~L.}\ \bibnamefont
  {Kane}}\ and\ \bibinfo {author} {\bibfnamefont {E.~J.}\ \bibnamefont
  {Mele}},\ }\bibfield  {title} {\enquote {\bibinfo {title} {${Z}_{2}$
  topological order and the quantum spin hall effect},}\ }\href {\doibase
  10.1103/PhysRevLett.95.146802} {\bibfield  {journal} {\bibinfo  {journal}
  {Phys. Rev. Lett.}\ }\textbf {\bibinfo {volume} {95}},\ \bibinfo {pages}
  {146802} (\bibinfo {year} {2005})}\BibitemShut {NoStop}%
\bibitem [{\citenamefont {Bernevig}, \citenamefont {Hughes},\ and\
  \citenamefont {Zhang}(2006)}]{Bernevig2006}%
  \BibitemOpen
  \bibfield  {author} {\bibinfo {author} {\bibfnamefont {B.~A.}\ \bibnamefont
  {Bernevig}}, \bibinfo {author} {\bibfnamefont {T.~L.}\ \bibnamefont
  {Hughes}}, \ and\ \bibinfo {author} {\bibfnamefont {S.-C.}\ \bibnamefont
  {Zhang}},\ }\bibfield  {title} {\enquote {\bibinfo {title} {Quantum spin
  {H}all effect and topological phase transition in {HgTe} quantum wells},}\
  }\href {\doibase 10.1126/science.1133734} {\bibfield  {journal} {\bibinfo
  {journal} {Science}\ }\textbf {\bibinfo {volume} {314}},\ \bibinfo {pages}
  {1757--1761} (\bibinfo {year} {2006})}\BibitemShut {NoStop}%
\bibitem [{\citenamefont {K{\"o}nig}\ \emph {et~al.}(2007)\citenamefont
  {K{\"o}nig}, \citenamefont {Wiedmann}, \citenamefont {Br{\"u}ne},
  \citenamefont {Roth}, \citenamefont {Buhmann}, \citenamefont {Molenkamp},
  \citenamefont {Qi},\ and\ \citenamefont {Zhang}}]{Konig2007}%
  \BibitemOpen
  \bibfield  {author} {\bibinfo {author} {\bibfnamefont {M.}~\bibnamefont
  {K{\"o}nig}}, \bibinfo {author} {\bibfnamefont {S.}~\bibnamefont {Wiedmann}},
  \bibinfo {author} {\bibfnamefont {C.}~\bibnamefont {Br{\"u}ne}}, \bibinfo
  {author} {\bibfnamefont {A.}~\bibnamefont {Roth}}, \bibinfo {author}
  {\bibfnamefont {H.}~\bibnamefont {Buhmann}}, \bibinfo {author} {\bibfnamefont
  {L.~W.}\ \bibnamefont {Molenkamp}}, \bibinfo {author} {\bibfnamefont {X.-L.}\
  \bibnamefont {Qi}}, \ and\ \bibinfo {author} {\bibfnamefont {S.-C.}\
  \bibnamefont {Zhang}},\ }\bibfield  {title} {\enquote {\bibinfo {title}
  {Quantum spin hall insulator state in hgte quantum wells},}\ }\href {\doibase
  10.1126/science.1148047} {\bibfield  {journal} {\bibinfo  {journal}
  {Science}\ }\textbf {\bibinfo {volume} {318}},\ \bibinfo {pages} {766--770}
  (\bibinfo {year} {2007})}\BibitemShut {NoStop}%
\bibitem [{\citenamefont {Volkov}\ and\ \citenamefont
  {Pankratov}(1985)}]{Volkov1985}%
  \BibitemOpen
  \bibfield  {author} {\bibinfo {author} {\bibfnamefont {B.~A.}\ \bibnamefont
  {Volkov}}\ and\ \bibinfo {author} {\bibfnamefont {O.~A.}\ \bibnamefont
  {Pankratov}},\ }\bibfield  {title} {\enquote {\bibinfo {title}
  {Two-dimensional massless electrons in an inverted contact},}\ }\href
  {http://www.jetpletters.ac.ru/ps/1420/article_21570.shtml} {\bibfield
  {journal} {\bibinfo  {journal} {JETP Lett.}\ }\textbf {\bibinfo {volume}
  {42}},\ \bibinfo {pages} {178} (\bibinfo {year} {1985})}\BibitemShut
  {NoStop}%
\bibitem [{\citenamefont {Gerchikov}\ and\ \citenamefont
  {Subashiev}(1990)}]{Gerchikov1990}%
  \BibitemOpen
  \bibfield  {author} {\bibinfo {author} {\bibfnamefont {L.~G.}\ \bibnamefont
  {Gerchikov}}\ and\ \bibinfo {author} {\bibfnamefont {A.~V.}\ \bibnamefont
  {Subashiev}},\ }\bibfield  {title} {\enquote {\bibinfo {title} {Interface
  states in subband structure of semiconductor quantum wells},}\ }\href
  {\doibase 10.1002/pssb.2221600207} {\bibfield  {journal} {\bibinfo  {journal}
  {physica status solidi (b)}\ }\textbf {\bibinfo {volume} {160}},\ \bibinfo
  {pages} {443--457} (\bibinfo {year} {1990})}\BibitemShut {NoStop}%
\bibitem [{\citenamefont {Fu}, \citenamefont {Kane},\ and\ \citenamefont
  {Mele}(2007)}]{Fu2007}%
  \BibitemOpen
  \bibfield  {author} {\bibinfo {author} {\bibfnamefont {L.}~\bibnamefont
  {Fu}}, \bibinfo {author} {\bibfnamefont {C.~L.}\ \bibnamefont {Kane}}, \ and\
  \bibinfo {author} {\bibfnamefont {E.~J.}\ \bibnamefont {Mele}},\ }\bibfield
  {title} {\enquote {\bibinfo {title} {Topological insulators in three
  dimensions},}\ }\href {\doibase 10.1103/PhysRevLett.98.106803} {\bibfield
  {journal} {\bibinfo  {journal} {Phys. Rev. Lett.}\ }\textbf {\bibinfo
  {volume} {98}},\ \bibinfo {pages} {106803} (\bibinfo {year}
  {2007})}\BibitemShut {NoStop}%
\bibitem [{\citenamefont {Manoharan}(2010)}]{Manoharan2010}%
  \BibitemOpen
  \bibfield  {author} {\bibinfo {author} {\bibfnamefont {H.~C.}\ \bibnamefont
  {Manoharan}},\ }\bibfield  {title} {\enquote {\bibinfo {title} {A romance
  with many dimensions},}\ }\href {\doibase 10.1038/nnano.2010.138} {\bibfield
  {journal} {\bibinfo  {journal} {Nature Nanotechnology}\ }\textbf {\bibinfo
  {volume} {5}},\ \bibinfo {pages} {477--479} (\bibinfo {year}
  {2010})}\BibitemShut {NoStop}%
\bibitem [{\citenamefont {Wu}\ and\ \citenamefont {Hu}(2015)}]{Wu2015}%
  \BibitemOpen
  \bibfield  {author} {\bibinfo {author} {\bibfnamefont {L.-H.}\ \bibnamefont
  {Wu}}\ and\ \bibinfo {author} {\bibfnamefont {X.}~\bibnamefont {Hu}},\
  }\bibfield  {title} {\enquote {\bibinfo {title} {Scheme for achieving a
  topological photonic crystal by using dielectric material},}\ }\href
  {\doibase 10.1103/PhysRevLett.114.223901} {\bibfield  {journal} {\bibinfo
  {journal} {Phys. Rev. Lett.}\ }\textbf {\bibinfo {volume} {114}},\ \bibinfo
  {pages} {223901} (\bibinfo {year} {2015})}\BibitemShut {NoStop}%
\bibitem [{\citenamefont {Semenoff}, \citenamefont {Semenoff},\ and\
  \citenamefont {Zhou}(2008)}]{Semenoff2008}%
  \BibitemOpen
  \bibfield  {author} {\bibinfo {author} {\bibfnamefont {G.~W.}\ \bibnamefont
  {Semenoff}}, \bibinfo {author} {\bibfnamefont {V.}~\bibnamefont {Semenoff}},
  \ and\ \bibinfo {author} {\bibfnamefont {F.}~\bibnamefont {Zhou}},\
  }\bibfield  {title} {\enquote {\bibinfo {title} {Domain walls in gapped
  graphene},}\ }\href {\doibase 10.1103/PhysRevLett.101.087204} {\bibfield
  {journal} {\bibinfo  {journal} {Phys. Rev. Lett.}\ }\textbf {\bibinfo
  {volume} {101}},\ \bibinfo {pages} {087204} (\bibinfo {year}
  {2008})}\BibitemShut {NoStop}%
\bibitem [{\citenamefont {Yao}, \citenamefont {Yang},\ and\ \citenamefont
  {Niu}(2009)}]{Yao2009}%
  \BibitemOpen
  \bibfield  {author} {\bibinfo {author} {\bibfnamefont {W.}~\bibnamefont
  {Yao}}, \bibinfo {author} {\bibfnamefont {S.~A.}\ \bibnamefont {Yang}}, \
  and\ \bibinfo {author} {\bibfnamefont {Q.}~\bibnamefont {Niu}},\ }\bibfield
  {title} {\enquote {\bibinfo {title} {Edge states in graphene: From gapped
  flat-band to gapless chiral modes},}\ }\href {\doibase
  10.1103/PhysRevLett.102.096801} {\bibfield  {journal} {\bibinfo  {journal}
  {Phys. Rev. Lett.}\ }\textbf {\bibinfo {volume} {102}},\ \bibinfo {pages}
  {096801} (\bibinfo {year} {2009})}\BibitemShut {NoStop}%
\bibitem [{\citenamefont {Wang}\ \emph {et~al.}(2008)\citenamefont {Wang},
  \citenamefont {Chong}, \citenamefont {Joannopoulos},\ and\ \citenamefont
  {Solja\ifmmode \check{c}\else \v{c}\fi{}i\ifmmode~\acute{c}\else
  \'{c}\fi{}}}]{Wang2008}%
  \BibitemOpen
  \bibfield  {author} {\bibinfo {author} {\bibfnamefont {Z.}~\bibnamefont
  {Wang}}, \bibinfo {author} {\bibfnamefont {Y.~D.}\ \bibnamefont {Chong}},
  \bibinfo {author} {\bibfnamefont {J.~D.}\ \bibnamefont {Joannopoulos}}, \
  and\ \bibinfo {author} {\bibfnamefont {M.}~\bibnamefont {Solja\ifmmode
  \check{c}\else \v{c}\fi{}i\ifmmode~\acute{c}\else \'{c}\fi{}}},\ }\bibfield
  {title} {\enquote {\bibinfo {title} {Reflection-free one-way edge modes in a
  gyromagnetic photonic crystal},}\ }\href {\doibase
  10.1103/PhysRevLett.100.013905} {\bibfield  {journal} {\bibinfo  {journal}
  {Phys. Rev. Lett.}\ }\textbf {\bibinfo {volume} {100}},\ \bibinfo {pages}
  {013905} (\bibinfo {year} {2008})}\BibitemShut {NoStop}%
\bibitem [{\citenamefont {Rechtsman}\ \emph {et~al.}(2013)\citenamefont
  {Rechtsman}, \citenamefont {Zeuner}, \citenamefont {Plotnik}, \citenamefont
  {Lumer}, \citenamefont {Podolsky}, \citenamefont {Dreisow}, \citenamefont
  {Nolte}, \citenamefont {Segev},\ and\ \citenamefont
  {Szameit}}]{rechtsman2013photonic}%
  \BibitemOpen
  \bibfield  {author} {\bibinfo {author} {\bibfnamefont {M.~C.}\ \bibnamefont
  {Rechtsman}}, \bibinfo {author} {\bibfnamefont {J.~M.}\ \bibnamefont
  {Zeuner}}, \bibinfo {author} {\bibfnamefont {Y.}~\bibnamefont {Plotnik}},
  \bibinfo {author} {\bibfnamefont {Y.}~\bibnamefont {Lumer}}, \bibinfo
  {author} {\bibfnamefont {D.}~\bibnamefont {Podolsky}}, \bibinfo {author}
  {\bibfnamefont {F.}~\bibnamefont {Dreisow}}, \bibinfo {author} {\bibfnamefont
  {S.}~\bibnamefont {Nolte}}, \bibinfo {author} {\bibfnamefont
  {M.}~\bibnamefont {Segev}}, \ and\ \bibinfo {author} {\bibfnamefont
  {A.}~\bibnamefont {Szameit}},\ }\bibfield  {title} {\enquote {\bibinfo
  {title} {Photonic floquet topological insulators},}\ }\href
  {https://doi.org/10.1038/nature12066} {\bibfield  {journal} {\bibinfo
  {journal} {Nature}\ }\textbf {\bibinfo {volume} {496}},\ \bibinfo {pages}
  {196} (\bibinfo {year} {2013})}\BibitemShut {NoStop}%
\bibitem [{\citenamefont {Slobozhanyuk}\ \emph {et~al.}(2019)\citenamefont
  {Slobozhanyuk}, \citenamefont {Shchelokova}, \citenamefont {Ni},
  \citenamefont {Hossein~Mousavi}, \citenamefont {Smirnova}, \citenamefont
  {Belov}, \citenamefont {Alù}, \citenamefont {Kivshar},\ and\ \citenamefont
  {Khanikaev}}]{Slobozhanyuk2019}%
  \BibitemOpen
  \bibfield  {author} {\bibinfo {author} {\bibfnamefont {A.}~\bibnamefont
  {Slobozhanyuk}}, \bibinfo {author} {\bibfnamefont {A.~V.}\ \bibnamefont
  {Shchelokova}}, \bibinfo {author} {\bibfnamefont {X.}~\bibnamefont {Ni}},
  \bibinfo {author} {\bibfnamefont {S.}~\bibnamefont {Hossein~Mousavi}},
  \bibinfo {author} {\bibfnamefont {D.~A.}\ \bibnamefont {Smirnova}}, \bibinfo
  {author} {\bibfnamefont {P.~A.}\ \bibnamefont {Belov}}, \bibinfo {author}
  {\bibfnamefont {A.}~\bibnamefont {Alù}}, \bibinfo {author} {\bibfnamefont
  {Y.~S.}\ \bibnamefont {Kivshar}}, \ and\ \bibinfo {author} {\bibfnamefont
  {A.~B.}\ \bibnamefont {Khanikaev}},\ }\bibfield  {title} {\enquote {\bibinfo
  {title} {Near-field imaging of spin-locked edge states in all-dielectric
  topological metasurfaces},}\ }\href {\doibase 10.1063/1.5055601} {\bibfield
  {journal} {\bibinfo  {journal} {Appl. Phys. Lett.}\ }\textbf {\bibinfo
  {volume} {114}},\ \bibinfo {pages} {031103} (\bibinfo {year}
  {2019})}\BibitemShut {NoStop}%
\bibitem [{\citenamefont {Bahari}\ \emph {et~al.}(2017)\citenamefont {Bahari},
  \citenamefont {Ndao}, \citenamefont {Vallini}, \citenamefont {El~Amili},
  \citenamefont {Fainman},\ and\ \citenamefont {Kant{\'e}}}]{Bahari2017}%
  \BibitemOpen
  \bibfield  {author} {\bibinfo {author} {\bibfnamefont {B.}~\bibnamefont
  {Bahari}}, \bibinfo {author} {\bibfnamefont {A.}~\bibnamefont {Ndao}},
  \bibinfo {author} {\bibfnamefont {F.}~\bibnamefont {Vallini}}, \bibinfo
  {author} {\bibfnamefont {A.}~\bibnamefont {El~Amili}}, \bibinfo {author}
  {\bibfnamefont {Y.}~\bibnamefont {Fainman}}, \ and\ \bibinfo {author}
  {\bibfnamefont {B.}~\bibnamefont {Kant{\'e}}},\ }\bibfield  {title} {\enquote
  {\bibinfo {title} {Nonreciprocal lasing in topological cavities of arbitrary
  geometries},}\ }\href {\doibase 10.1126/science.aao4551} {\bibfield
  {journal} {\bibinfo  {journal} {Science}\ }\textbf {\bibinfo {volume}
  {358}},\ \bibinfo {pages} {636--640} (\bibinfo {year} {2017})}\BibitemShut
  {NoStop}%
\bibitem [{\citenamefont {Benalcazar}, \citenamefont {Bernevig},\ and\
  \citenamefont {Hughes}(2017)}]{Benalcazar2017}%
  \BibitemOpen
  \bibfield  {author} {\bibinfo {author} {\bibfnamefont {W.~A.}\ \bibnamefont
  {Benalcazar}}, \bibinfo {author} {\bibfnamefont {B.~A.}\ \bibnamefont
  {Bernevig}}, \ and\ \bibinfo {author} {\bibfnamefont {T.~L.}\ \bibnamefont
  {Hughes}},\ }\bibfield  {title} {\enquote {\bibinfo {title} {Quantized
  electric multipole insulators},}\ }\href {\doibase 10.1126/science.aah6442}
  {\bibfield  {journal} {\bibinfo  {journal} {Science}\ }\textbf {\bibinfo
  {volume} {357}},\ \bibinfo {pages} {61--66} (\bibinfo {year}
  {2017})}\BibitemShut {NoStop}%
\bibitem [{\citenamefont {{Lu}}, \citenamefont {{Joannopoulos}},\ and\
  \citenamefont {{Solja{\v c}i{\'c}}}(2014)}]{Lu2014}%
  \BibitemOpen
  \bibfield  {author} {\bibinfo {author} {\bibfnamefont {L.}~\bibnamefont
  {{Lu}}}, \bibinfo {author} {\bibfnamefont {J.~D.}\ \bibnamefont
  {{Joannopoulos}}}, \ and\ \bibinfo {author} {\bibfnamefont {M.}~\bibnamefont
  {{Solja{\v c}i{\'c}}}},\ }\bibfield  {title} {\enquote {\bibinfo {title}
  {{Topological photonics}},}\ }\href {\doibase 10.1038/nphoton.2014.248}
  {\bibfield  {journal} {\bibinfo  {journal} {Nature Photonics}\ }\textbf
  {\bibinfo {volume} {8}},\ \bibinfo {pages} {821--829} (\bibinfo {year}
  {2014})}\BibitemShut {NoStop}%
\bibitem [{\citenamefont {Khanikaev}\ and\ \citenamefont
  {Shvets}(2017)}]{Khanikaev2017}%
  \BibitemOpen
  \bibfield  {author} {\bibinfo {author} {\bibfnamefont {A.~B.}\ \bibnamefont
  {Khanikaev}}\ and\ \bibinfo {author} {\bibfnamefont {G.}~\bibnamefont
  {Shvets}},\ }\bibfield  {title} {\enquote {\bibinfo {title} {Two-dimensional
  topological photonics},}\ }\href {\doibase 10.1038/s41566-017-0048-5}
  {\bibfield  {journal} {\bibinfo  {journal} {Nature Photonics}\ }\textbf
  {\bibinfo {volume} {11}},\ \bibinfo {pages} {763--773} (\bibinfo {year}
  {2017})}\BibitemShut {NoStop}%
\bibitem [{\citenamefont {Sun}\ \emph {et~al.}(2017)\citenamefont {Sun},
  \citenamefont {He}, \citenamefont {Liu}, \citenamefont {Lu}, \citenamefont
  {Zhu},\ and\ \citenamefont {Chen}}]{Sun2017}%
  \BibitemOpen
  \bibfield  {author} {\bibinfo {author} {\bibfnamefont {X.-C.}\ \bibnamefont
  {Sun}}, \bibinfo {author} {\bibfnamefont {C.}~\bibnamefont {He}}, \bibinfo
  {author} {\bibfnamefont {X.-P.}\ \bibnamefont {Liu}}, \bibinfo {author}
  {\bibfnamefont {M.-H.}\ \bibnamefont {Lu}}, \bibinfo {author} {\bibfnamefont
  {S.-N.}\ \bibnamefont {Zhu}}, \ and\ \bibinfo {author} {\bibfnamefont
  {Y.-F.}\ \bibnamefont {Chen}},\ }\bibfield  {title} {\enquote {\bibinfo
  {title} {Two-dimensional topological photonic systems},}\ }\href {\doibase
  10.1016/j.pquantelec.2017.07.004} {\bibfield  {journal} {\bibinfo  {journal}
  {Progress in Quantum Electronics}\ }\textbf {\bibinfo {volume} {55}},\
  \bibinfo {pages} {52--73} (\bibinfo {year} {2017})}\BibitemShut {NoStop}%
\bibitem [{\citenamefont {Rider}\ \emph {et~al.}(2019)\citenamefont {Rider},
  \citenamefont {Palmer}, \citenamefont {Pocock}, \citenamefont {Xiao},
  \citenamefont {Huidobro},\ and\ \citenamefont {Giannini}}]{Rider2019}%
  \BibitemOpen
  \bibfield  {author} {\bibinfo {author} {\bibfnamefont {M.~S.}\ \bibnamefont
  {Rider}}, \bibinfo {author} {\bibfnamefont {S.~J.}\ \bibnamefont {Palmer}},
  \bibinfo {author} {\bibfnamefont {S.~R.}\ \bibnamefont {Pocock}}, \bibinfo
  {author} {\bibfnamefont {X.}~\bibnamefont {Xiao}}, \bibinfo {author}
  {\bibfnamefont {P.~A.}\ \bibnamefont {Huidobro}}, \ and\ \bibinfo {author}
  {\bibfnamefont {V.}~\bibnamefont {Giannini}},\ }\bibfield  {title} {\enquote
  {\bibinfo {title} {A perspective on topological nanophotonics: Current status
  and future challenges},}\ }\href {\doibase 10.1063/1.5086433} {\bibfield
  {journal} {\bibinfo  {journal} {Journal of Applied Physics}\ }\textbf
  {\bibinfo {volume} {125}},\ \bibinfo {pages} {120901} (\bibinfo {year}
  {2019})}\BibitemShut {NoStop}%
\bibitem [{\citenamefont {Shen}(2013)}]{Shen2013}%
  \BibitemOpen
  \bibfield  {author} {\bibinfo {author} {\bibfnamefont {S.-Q.}\ \bibnamefont
  {Shen}},\ }\href@noop {} {\emph {\bibinfo {title} {{T}opological
  {I}nsulators. {D}irac {E}quation in {C}ondensed {M}atters}}},\ Springer
  Series in {S}olid-{S}tate {S}ciences\ (\bibinfo  {publisher} {Springer},\
  \bibinfo {address} {Heidelberg},\ \bibinfo {year} {2013})\BibitemShut
  {NoStop}%
\bibitem [{\citenamefont {{Wang}}\ \emph {et~al.}(2009)\citenamefont {{Wang}},
  \citenamefont {{Chong}}, \citenamefont {{Joannopoulos}},\ and\ \citenamefont
  {{Solja{\v c}i{\'c}}}}]{Wang2009}%
  \BibitemOpen
  \bibfield  {author} {\bibinfo {author} {\bibfnamefont {Z.}~\bibnamefont
  {{Wang}}}, \bibinfo {author} {\bibfnamefont {Y.}~\bibnamefont {{Chong}}},
  \bibinfo {author} {\bibfnamefont {J.~D.}\ \bibnamefont {{Joannopoulos}}}, \
  and\ \bibinfo {author} {\bibfnamefont {M.}~\bibnamefont {{Solja{\v
  c}i{\'c}}}},\ }\bibfield  {title} {\enquote {\bibinfo {title} {{Observation
  of unidirectional backscattering-immune topological electromagnetic
  states}},}\ }\href {\doibase 10.1038/nature08293} {\bibfield  {journal}
  {\bibinfo  {journal} {Nature}\ }\textbf {\bibinfo {volume} {461}},\ \bibinfo
  {pages} {772--775} (\bibinfo {year} {2009})}\BibitemShut {NoStop}%
\bibitem [{\citenamefont {Haldane}\ and\ \citenamefont
  {Raghu}(2008)}]{Haldane2008}%
  \BibitemOpen
  \bibfield  {author} {\bibinfo {author} {\bibfnamefont {F.~D.~M.}\
  \bibnamefont {Haldane}}\ and\ \bibinfo {author} {\bibfnamefont
  {S.}~\bibnamefont {Raghu}},\ }\bibfield  {title} {\enquote {\bibinfo {title}
  {Possible realization of directional optical waveguides in photonic crystals
  with broken time-reversal symmetry},}\ }\href {\doibase
  10.1103/PhysRevLett.100.013904} {\bibfield  {journal} {\bibinfo  {journal}
  {Phys. Rev. Lett.}\ }\textbf {\bibinfo {volume} {100}},\ \bibinfo {pages}
  {013904} (\bibinfo {year} {2008})}\BibitemShut {NoStop}%
\bibitem [{\citenamefont {Raghu}\ and\ \citenamefont
  {Haldane}(2008)}]{Raghu2008}%
  \BibitemOpen
  \bibfield  {author} {\bibinfo {author} {\bibfnamefont {S.}~\bibnamefont
  {Raghu}}\ and\ \bibinfo {author} {\bibfnamefont {F.~D.~M.}\ \bibnamefont
  {Haldane}},\ }\bibfield  {title} {\enquote {\bibinfo {title} {Analogs of
  quantum-hall-effect edge states in photonic crystals},}\ }\href {\doibase
  10.1103/PhysRevA.78.033834} {\bibfield  {journal} {\bibinfo  {journal} {Phys.
  Rev. A}\ }\textbf {\bibinfo {volume} {78}},\ \bibinfo {pages} {033834}
  (\bibinfo {year} {2008})}\BibitemShut {NoStop}%
\bibitem [{\citenamefont {Hafezi}\ \emph {et~al.}(2013)\citenamefont {Hafezi},
  \citenamefont {Mittal}, \citenamefont {Fan}, \citenamefont {Migdall},\ and\
  \citenamefont {Taylor}}]{hafezi2013imaging}%
  \BibitemOpen
  \bibfield  {author} {\bibinfo {author} {\bibfnamefont {M.}~\bibnamefont
  {Hafezi}}, \bibinfo {author} {\bibfnamefont {S.}~\bibnamefont {Mittal}},
  \bibinfo {author} {\bibfnamefont {J.}~\bibnamefont {Fan}}, \bibinfo {author}
  {\bibfnamefont {A.}~\bibnamefont {Migdall}}, \ and\ \bibinfo {author}
  {\bibfnamefont {J.~M.}\ \bibnamefont {Taylor}},\ }\bibfield  {title}
  {\enquote {\bibinfo {title} {Imaging topological edge states in silicon
  photonics},}\ }\href {https://doi.org/10.1038/nphoton.2013.274} {\bibfield
  {journal} {\bibinfo  {journal} {Nature Photonics}\ }\textbf {\bibinfo
  {volume} {7}},\ \bibinfo {pages} {1001} (\bibinfo {year} {2013})}\BibitemShut
  {NoStop}%
\bibitem [{\citenamefont {Kraus}\ \emph {et~al.}(2012)\citenamefont {Kraus},
  \citenamefont {Lahini}, \citenamefont {Ringel}, \citenamefont {Verbin},\ and\
  \citenamefont {Zilberberg}}]{Kraus2012}%
  \BibitemOpen
  \bibfield  {author} {\bibinfo {author} {\bibfnamefont {Y.~E.}\ \bibnamefont
  {Kraus}}, \bibinfo {author} {\bibfnamefont {Y.}~\bibnamefont {Lahini}},
  \bibinfo {author} {\bibfnamefont {Z.}~\bibnamefont {Ringel}}, \bibinfo
  {author} {\bibfnamefont {M.}~\bibnamefont {Verbin}}, \ and\ \bibinfo {author}
  {\bibfnamefont {O.}~\bibnamefont {Zilberberg}},\ }\bibfield  {title}
  {\enquote {\bibinfo {title} {Topological states and adiabatic pumping in
  quasicrystals},}\ }\href {\doibase 10.1103/PhysRevLett.109.106402} {\bibfield
   {journal} {\bibinfo  {journal} {Phys. Rev. Lett.}\ }\textbf {\bibinfo
  {volume} {109}},\ \bibinfo {pages} {106402} (\bibinfo {year}
  {2012})}\BibitemShut {NoStop}%
\bibitem [{\citenamefont {Verbin}\ \emph {et~al.}(2013)\citenamefont {Verbin},
  \citenamefont {Zilberberg}, \citenamefont {Kraus}, \citenamefont {Lahini},\
  and\ \citenamefont {Silberberg}}]{Verbin2013}%
  \BibitemOpen
  \bibfield  {author} {\bibinfo {author} {\bibfnamefont {M.}~\bibnamefont
  {Verbin}}, \bibinfo {author} {\bibfnamefont {O.}~\bibnamefont {Zilberberg}},
  \bibinfo {author} {\bibfnamefont {Y.~E.}\ \bibnamefont {Kraus}}, \bibinfo
  {author} {\bibfnamefont {Y.}~\bibnamefont {Lahini}}, \ and\ \bibinfo {author}
  {\bibfnamefont {Y.}~\bibnamefont {Silberberg}},\ }\bibfield  {title}
  {\enquote {\bibinfo {title} {Observation of topological phase transitions in
  photonic quasicrystals},}\ }\href {\doibase 10.1103/PhysRevLett.110.076403}
  {\bibfield  {journal} {\bibinfo  {journal} {Phys. Rev. Lett.}\ }\textbf
  {\bibinfo {volume} {110}},\ \bibinfo {pages} {076403} (\bibinfo {year}
  {2013})}\BibitemShut {NoStop}%
\bibitem [{\citenamefont {Slobozhanyuk}\ \emph
  {et~al.}(2016{\natexlab{a}})\citenamefont {Slobozhanyuk}, \citenamefont
  {Mousavi}, \citenamefont {Ni}, \citenamefont {Smirnova}, \citenamefont
  {Kivshar},\ and\ \citenamefont {Khanikaev}}]{Slobozhanyuk2016}%
  \BibitemOpen
  \bibfield  {author} {\bibinfo {author} {\bibfnamefont {A.}~\bibnamefont
  {Slobozhanyuk}}, \bibinfo {author} {\bibfnamefont {S.~H.}\ \bibnamefont
  {Mousavi}}, \bibinfo {author} {\bibfnamefont {X.}~\bibnamefont {Ni}},
  \bibinfo {author} {\bibfnamefont {D.}~\bibnamefont {Smirnova}}, \bibinfo
  {author} {\bibfnamefont {Y.~S.}\ \bibnamefont {Kivshar}}, \ and\ \bibinfo
  {author} {\bibfnamefont {A.~B.}\ \bibnamefont {Khanikaev}},\ }\bibfield
  {title} {\enquote {\bibinfo {title} {Three-dimensional all-dielectric
  photonic topological insulator},}\ }\href {\doibase 10.1038/nphoton.2016.253}
  {\bibfield  {journal} {\bibinfo  {journal} {Nature Photonics}\ }\textbf
  {\bibinfo {volume} {11}},\ \bibinfo {pages} {130--136} (\bibinfo {year}
  {2016}{\natexlab{a}})}\BibitemShut {NoStop}%
\bibitem [{\citenamefont {Hafezi}\ \emph {et~al.}(2011)\citenamefont {Hafezi},
  \citenamefont {Demler}, \citenamefont {Lukin},\ and\ \citenamefont
  {Taylor}}]{Hafezi2011}%
  \BibitemOpen
  \bibfield  {author} {\bibinfo {author} {\bibfnamefont {M.}~\bibnamefont
  {Hafezi}}, \bibinfo {author} {\bibfnamefont {E.~A.}\ \bibnamefont {Demler}},
  \bibinfo {author} {\bibfnamefont {M.~D.}\ \bibnamefont {Lukin}}, \ and\
  \bibinfo {author} {\bibfnamefont {J.~M.}\ \bibnamefont {Taylor}},\ }\bibfield
   {title} {\enquote {\bibinfo {title} {Robust optical delay lines with
  topological protection},}\ }\href {\doibase 10.1038/nphys2063} {\bibfield
  {journal} {\bibinfo  {journal} {Nature Physics}\ }\textbf {\bibinfo {volume}
  {7}},\ \bibinfo {pages} {907--912} (\bibinfo {year} {2011})}\BibitemShut
  {NoStop}%
\bibitem [{\citenamefont {{Khanikaev}}\ \emph {et~al.}(2013)\citenamefont
  {{Khanikaev}}, \citenamefont {{Hossein Mousavi}}, \citenamefont {{Tse}},
  \citenamefont {{Kargarian}}, \citenamefont {{MacDonald}},\ and\ \citenamefont
  {{Shvets}}}]{Khanikaev2013}%
  \BibitemOpen
  \bibfield  {author} {\bibinfo {author} {\bibfnamefont {A.~B.}\ \bibnamefont
  {{Khanikaev}}}, \bibinfo {author} {\bibfnamefont {S.}~\bibnamefont {{Hossein
  Mousavi}}}, \bibinfo {author} {\bibfnamefont {W.-K.}\ \bibnamefont {{Tse}}},
  \bibinfo {author} {\bibfnamefont {M.}~\bibnamefont {{Kargarian}}}, \bibinfo
  {author} {\bibfnamefont {A.~H.}\ \bibnamefont {{MacDonald}}}, \ and\ \bibinfo
  {author} {\bibfnamefont {G.}~\bibnamefont {{Shvets}}},\ }\bibfield  {title}
  {\enquote {\bibinfo {title} {{Photonic topological insulators}},}\ }\href
  {\doibase 10.1038/nmat3520} {\bibfield  {journal} {\bibinfo  {journal}
  {Nature Materials}\ }\textbf {\bibinfo {volume} {12}},\ \bibinfo {pages}
  {233--239} (\bibinfo {year} {2013})}\BibitemShut {NoStop}%
\bibitem [{\citenamefont {Ma}\ \emph {et~al.}(2015)\citenamefont {Ma},
  \citenamefont {Khanikaev}, \citenamefont {Mousavi},\ and\ \citenamefont
  {Shvets}}]{Ma2015}%
  \BibitemOpen
  \bibfield  {author} {\bibinfo {author} {\bibfnamefont {T.}~\bibnamefont
  {Ma}}, \bibinfo {author} {\bibfnamefont {A.~B.}\ \bibnamefont {Khanikaev}},
  \bibinfo {author} {\bibfnamefont {S.~H.}\ \bibnamefont {Mousavi}}, \ and\
  \bibinfo {author} {\bibfnamefont {G.}~\bibnamefont {Shvets}},\ }\bibfield
  {title} {\enquote {\bibinfo {title} {Guiding electromagnetic waves around
  sharp corners: Topologically protected photonic transport in
  metawaveguides},}\ }\href {\doibase 10.1103/PhysRevLett.114.127401}
  {\bibfield  {journal} {\bibinfo  {journal} {Phys. Rev. Lett.}\ }\textbf
  {\bibinfo {volume} {114}},\ \bibinfo {pages} {127401} (\bibinfo {year}
  {2015})}\BibitemShut {NoStop}%
\bibitem [{\citenamefont {Slobozhanyuk}\ \emph
  {et~al.}(2016{\natexlab{b}})\citenamefont {Slobozhanyuk}, \citenamefont
  {Khanikaev}, \citenamefont {Filonov}, \citenamefont {Smirnova}, \citenamefont
  {Miroshnichenko},\ and\ \citenamefont {Kivshar}}]{Slobozhanyuk2016SciRep}%
  \BibitemOpen
  \bibfield  {author} {\bibinfo {author} {\bibfnamefont {A.~P.}\ \bibnamefont
  {Slobozhanyuk}}, \bibinfo {author} {\bibfnamefont {A.~B.}\ \bibnamefont
  {Khanikaev}}, \bibinfo {author} {\bibfnamefont {D.~S.}\ \bibnamefont
  {Filonov}}, \bibinfo {author} {\bibfnamefont {D.~A.}\ \bibnamefont
  {Smirnova}}, \bibinfo {author} {\bibfnamefont {A.~E.}\ \bibnamefont
  {Miroshnichenko}}, \ and\ \bibinfo {author} {\bibfnamefont {Y.~S.}\
  \bibnamefont {Kivshar}},\ }\bibfield  {title} {\enquote {\bibinfo {title}
  {Experimental demonstration of topological effects in bianisotropic
  metamaterials},}\ }\href {\doibase 10.1038/srep22270} {\bibfield  {journal}
  {\bibinfo  {journal} {Scientific Reports}\ }\textbf {\bibinfo {volume} {6}}
  (\bibinfo {year} {2016}{\natexlab{b}}),\ 10.1038/srep22270}\BibitemShut
  {NoStop}%
\bibitem [{\citenamefont {Ma}\ and\ \citenamefont {Shvets}(2016)}]{Ma2016}%
  \BibitemOpen
  \bibfield  {author} {\bibinfo {author} {\bibfnamefont {T.}~\bibnamefont
  {Ma}}\ and\ \bibinfo {author} {\bibfnamefont {G.}~\bibnamefont {Shvets}},\
  }\bibfield  {title} {\enquote {\bibinfo {title} {All-{S}i valley-{H}all
  photonic topological insulator},}\ }\href
  {http://stacks.iop.org/1367-2630/18/i=2/a=025012} {\bibfield  {journal}
  {\bibinfo  {journal} {New Journal of Physics}\ }\textbf {\bibinfo {volume}
  {18}},\ \bibinfo {pages} {025012} (\bibinfo {year} {2016})}\BibitemShut
  {NoStop}%
\bibitem [{\citenamefont {Gorlach}\ \emph {et~al.}(2018)\citenamefont
  {Gorlach}, \citenamefont {Ni}, \citenamefont {Smirnova}, \citenamefont
  {Korobkin}, \citenamefont {Zhirihin}, \citenamefont {Slobozhanyuk},
  \citenamefont {Belov}, \citenamefont {Al{\`{u}}},\ and\ \citenamefont
  {Khanikaev}}]{Gorlach2018}%
  \BibitemOpen
  \bibfield  {author} {\bibinfo {author} {\bibfnamefont {M.~A.}\ \bibnamefont
  {Gorlach}}, \bibinfo {author} {\bibfnamefont {X.}~\bibnamefont {Ni}},
  \bibinfo {author} {\bibfnamefont {D.~A.}\ \bibnamefont {Smirnova}}, \bibinfo
  {author} {\bibfnamefont {D.}~\bibnamefont {Korobkin}}, \bibinfo {author}
  {\bibfnamefont {D.}~\bibnamefont {Zhirihin}}, \bibinfo {author}
  {\bibfnamefont {A.~P.}\ \bibnamefont {Slobozhanyuk}}, \bibinfo {author}
  {\bibfnamefont {P.~A.}\ \bibnamefont {Belov}}, \bibinfo {author}
  {\bibfnamefont {A.}~\bibnamefont {Al{\`{u}}}}, \ and\ \bibinfo {author}
  {\bibfnamefont {A.~B.}\ \bibnamefont {Khanikaev}},\ }\bibfield  {title}
  {\enquote {\bibinfo {title} {Far-field probing of leaky topological states in
  all-dielectric metasurfaces},}\ }\href {\doibase 10.1038/s41467-018-03330-9}
  {\bibfield  {journal} {\bibinfo  {journal} {Nature Communications}\ }\textbf
  {\bibinfo {volume} {9}},\ \bibinfo {pages} {909} (\bibinfo {year}
  {2018})}\BibitemShut {NoStop}%
\bibitem [{\citenamefont {Barik}\ \emph {et~al.}(2018)\citenamefont {Barik},
  \citenamefont {Karasahin}, \citenamefont {Flower}, \citenamefont {Cai},
  \citenamefont {Miyake}, \citenamefont {DeGottardi}, \citenamefont {Hafezi},\
  and\ \citenamefont {Waks}}]{Barik2018}%
  \BibitemOpen
  \bibfield  {author} {\bibinfo {author} {\bibfnamefont {S.}~\bibnamefont
  {Barik}}, \bibinfo {author} {\bibfnamefont {A.}~\bibnamefont {Karasahin}},
  \bibinfo {author} {\bibfnamefont {C.}~\bibnamefont {Flower}}, \bibinfo
  {author} {\bibfnamefont {T.}~\bibnamefont {Cai}}, \bibinfo {author}
  {\bibfnamefont {H.}~\bibnamefont {Miyake}}, \bibinfo {author} {\bibfnamefont
  {W.}~\bibnamefont {DeGottardi}}, \bibinfo {author} {\bibfnamefont
  {M.}~\bibnamefont {Hafezi}}, \ and\ \bibinfo {author} {\bibfnamefont
  {E.}~\bibnamefont {Waks}},\ }\bibfield  {title} {\enquote {\bibinfo {title}
  {A topological quantum optics interface},}\ }\href {\doibase
  10.1126/science.aaq0327} {\bibfield  {journal} {\bibinfo  {journal}
  {Science}\ }\textbf {\bibinfo {volume} {359}},\ \bibinfo {pages} {666}
  (\bibinfo {year} {2018})}\BibitemShut {NoStop}%
\bibitem [{\citenamefont {He}\ \emph {et~al.}(2019)\citenamefont {He},
  \citenamefont {Liang}, \citenamefont {Yuan}, \citenamefont {Qiu},
  \citenamefont {Chen}, \citenamefont {Zhao},\ and\ \citenamefont
  {Dong}}]{He2018}%
  \BibitemOpen
  \bibfield  {author} {\bibinfo {author} {\bibfnamefont {X.-T.}\ \bibnamefont
  {He}}, \bibinfo {author} {\bibfnamefont {E.-T.}\ \bibnamefont {Liang}},
  \bibinfo {author} {\bibfnamefont {J.-J.}\ \bibnamefont {Yuan}}, \bibinfo
  {author} {\bibfnamefont {H.-Y.}\ \bibnamefont {Qiu}}, \bibinfo {author}
  {\bibfnamefont {X.-D.}\ \bibnamefont {Chen}}, \bibinfo {author}
  {\bibfnamefont {F.-L.}\ \bibnamefont {Zhao}}, \ and\ \bibinfo {author}
  {\bibfnamefont {J.-W.}\ \bibnamefont {Dong}},\ }\bibfield  {title} {\enquote
  {\bibinfo {title} {A silicon-on-insulator slab for topological valley
  transport},}\ }\href {\doibase 10.1038/s41467-019-08881-z} {\bibfield
  {journal} {\bibinfo  {journal} {Nature Communications}\ }\textbf {\bibinfo
  {volume} {10}},\ \bibinfo {pages} {872} (\bibinfo {year} {2019})}\BibitemShut
  {NoStop}%
\bibitem [{\citenamefont {Peng}\ \emph {et~al.}(2019)\citenamefont {Peng},
  \citenamefont {Schilder}, \citenamefont {Ni}, \citenamefont {van~de Groep},
  \citenamefont {Brongersma}, \citenamefont {Al\`u}, \citenamefont {Khanikaev},
  \citenamefont {Atwater},\ and\ \citenamefont {Polman}}]{Peng2018}%
  \BibitemOpen
  \bibfield  {author} {\bibinfo {author} {\bibfnamefont {S.}~\bibnamefont
  {Peng}}, \bibinfo {author} {\bibfnamefont {N.~J.}\ \bibnamefont {Schilder}},
  \bibinfo {author} {\bibfnamefont {X.}~\bibnamefont {Ni}}, \bibinfo {author}
  {\bibfnamefont {J.}~\bibnamefont {van~de Groep}}, \bibinfo {author}
  {\bibfnamefont {M.~L.}\ \bibnamefont {Brongersma}}, \bibinfo {author}
  {\bibfnamefont {A.}~\bibnamefont {Al\`u}}, \bibinfo {author} {\bibfnamefont
  {A.~B.}\ \bibnamefont {Khanikaev}}, \bibinfo {author} {\bibfnamefont {H.~A.}\
  \bibnamefont {Atwater}}, \ and\ \bibinfo {author} {\bibfnamefont
  {A.}~\bibnamefont {Polman}},\ }\bibfield  {title} {\enquote {\bibinfo {title}
  {Probing the band structure of topological silicon photonic lattices in the
  visible spectrum},}\ }\href {\doibase 10.1103/PhysRevLett.122.117401}
  {\bibfield  {journal} {\bibinfo  {journal} {Phys. Rev. Lett.}\ }\textbf
  {\bibinfo {volume} {122}},\ \bibinfo {pages} {117401} (\bibinfo {year}
  {2019})}\BibitemShut {NoStop}%
\bibitem [{\citenamefont {Smirnova}, \citenamefont {Padmanabhan},\ and\
  \citenamefont {Leykam}(2019)}]{Smirnova2019}%
  \BibitemOpen
  \bibfield  {author} {\bibinfo {author} {\bibfnamefont {D.~A.}\ \bibnamefont
  {Smirnova}}, \bibinfo {author} {\bibfnamefont {P.}~\bibnamefont
  {Padmanabhan}}, \ and\ \bibinfo {author} {\bibfnamefont {D.}~\bibnamefont
  {Leykam}},\ }\bibfield  {title} {\enquote {\bibinfo {title} {Parity anomaly
  laser},}\ }\href {\doibase 10.1364/OL.44.001120} {\bibfield  {journal}
  {\bibinfo  {journal} {Opt. Lett.}\ }\textbf {\bibinfo {volume} {44}},\
  \bibinfo {pages} {1120--1123} (\bibinfo {year} {2019})}\BibitemShut {NoStop}%
\bibitem [{\citenamefont {Kuznetsov}\ \emph {et~al.}(2016)\citenamefont
  {Kuznetsov}, \citenamefont {Miroshnichenko}, \citenamefont {Brongersma},
  \citenamefont {Kivshar},\ and\ \citenamefont {Luk'yanchuk}}]{Kuznetsov2016}%
  \BibitemOpen
  \bibfield  {author} {\bibinfo {author} {\bibfnamefont {A.~I.}\ \bibnamefont
  {Kuznetsov}}, \bibinfo {author} {\bibfnamefont {A.~E.}\ \bibnamefont
  {Miroshnichenko}}, \bibinfo {author} {\bibfnamefont {M.~L.}\ \bibnamefont
  {Brongersma}}, \bibinfo {author} {\bibfnamefont {Y.~S.}\ \bibnamefont
  {Kivshar}}, \ and\ \bibinfo {author} {\bibfnamefont {B.}~\bibnamefont
  {Luk'yanchuk}},\ }\bibfield  {title} {\enquote {\bibinfo {title} {Optically
  resonant dielectric nanostructures},}\ }\href {\doibase
  10.1126/science.aag2472} {\bibfield  {journal} {\bibinfo  {journal}
  {Science}\ }\textbf {\bibinfo {volume} {354}},\ \bibinfo {pages} {aag2472}
  (\bibinfo {year} {2016})}\BibitemShut {NoStop}%
\bibitem [{\citenamefont {Berry}(1984)}]{Berry1984}%
  \BibitemOpen
  \bibfield  {author} {\bibinfo {author} {\bibfnamefont {M.~V.}\ \bibnamefont
  {Berry}},\ }\bibfield  {title} {\enquote {\bibinfo {title} {Quantal phase
  factors accompanying adiabatic changes},}\ }\href {\doibase
  10.1098/rspa.1984.0023} {\bibfield  {journal} {\bibinfo  {journal}
  {Proceedings of the Royal Society A: Mathematical, Physical and Engineering
  Sciences}\ }\textbf {\bibinfo {volume} {392}},\ \bibinfo {pages} {45--57}
  (\bibinfo {year} {1984})}\BibitemShut {NoStop}%
\bibitem [{\citenamefont {Leykam}\ and\ \citenamefont
  {Desyatnikov}(2016)}]{conical_review}%
  \BibitemOpen
  \bibfield  {author} {\bibinfo {author} {\bibfnamefont {D.}~\bibnamefont
  {Leykam}}\ and\ \bibinfo {author} {\bibfnamefont {A.~S.}\ \bibnamefont
  {Desyatnikov}},\ }\bibfield  {title} {\enquote {\bibinfo {title} {Conical
  intersections for light and matter waves},}\ }\href {\doibase
  10.1080/23746149.2016.1144482} {\bibfield  {journal} {\bibinfo  {journal}
  {Advances in Physics: X}\ }\textbf {\bibinfo {volume} {1}},\ \bibinfo {pages}
  {101--113} (\bibinfo {year} {2016})}\BibitemShut {NoStop}%
\bibitem [{\citenamefont {Zak}(1989)}]{Zak1989}%
  \BibitemOpen
  \bibfield  {author} {\bibinfo {author} {\bibfnamefont {J.}~\bibnamefont
  {Zak}},\ }\bibfield  {title} {\enquote {\bibinfo {title} {Berry's phase for
  energy bands in solids},}\ }\href {\doibase 10.1103/PhysRevLett.62.2747}
  {\bibfield  {journal} {\bibinfo  {journal} {Phys. Rev. Lett.}\ }\textbf
  {\bibinfo {volume} {62}},\ \bibinfo {pages} {2747--2750} (\bibinfo {year}
  {1989})}\BibitemShut {NoStop}%
\bibitem [{\citenamefont {Peleg}\ \emph {et~al.}(2007)\citenamefont {Peleg},
  \citenamefont {Bartal}, \citenamefont {Freedman}, \citenamefont {Manela},
  \citenamefont {Segev},\ and\ \citenamefont {Christodoulides}}]{Peleg2007}%
  \BibitemOpen
  \bibfield  {author} {\bibinfo {author} {\bibfnamefont {O.}~\bibnamefont
  {Peleg}}, \bibinfo {author} {\bibfnamefont {G.}~\bibnamefont {Bartal}},
  \bibinfo {author} {\bibfnamefont {B.}~\bibnamefont {Freedman}}, \bibinfo
  {author} {\bibfnamefont {O.}~\bibnamefont {Manela}}, \bibinfo {author}
  {\bibfnamefont {M.}~\bibnamefont {Segev}}, \ and\ \bibinfo {author}
  {\bibfnamefont {D.~N.}\ \bibnamefont {Christodoulides}},\ }\bibfield  {title}
  {\enquote {\bibinfo {title} {Conical diffraction and gap solitons in
  honeycomb photonic lattices},}\ }\href {\doibase
  10.1103/PhysRevLett.98.103901} {\bibfield  {journal} {\bibinfo  {journal}
  {Phys. Rev. Lett.}\ }\textbf {\bibinfo {volume} {98}},\ \bibinfo {pages}
  {103901} (\bibinfo {year} {2007})}\BibitemShut {NoStop}%
\bibitem [{\citenamefont {Lumer}\ \emph {et~al.}(2013)\citenamefont {Lumer},
  \citenamefont {Plotnik}, \citenamefont {Rechtsman},\ and\ \citenamefont
  {Segev}}]{Plotnik2013}%
  \BibitemOpen
  \bibfield  {author} {\bibinfo {author} {\bibfnamefont {Y.}~\bibnamefont
  {Lumer}}, \bibinfo {author} {\bibfnamefont {Y.}~\bibnamefont {Plotnik}},
  \bibinfo {author} {\bibfnamefont {M.~C.}\ \bibnamefont {Rechtsman}}, \ and\
  \bibinfo {author} {\bibfnamefont {M.}~\bibnamefont {Segev}},\ }\bibfield
  {title} {\enquote {\bibinfo {title} {Self-localized states in photonic
  topological insulators},}\ }\href {\doibase 10.1103/PhysRevLett.111.243905}
  {\bibfield  {journal} {\bibinfo  {journal} {Phys. Rev. Lett.}\ }\textbf
  {\bibinfo {volume} {111}},\ \bibinfo {pages} {243905} (\bibinfo {year}
  {2013})}\BibitemShut {NoStop}%
\bibitem [{\citenamefont {Haldane}(1988)}]{Haldane1988}%
  \BibitemOpen
  \bibfield  {author} {\bibinfo {author} {\bibfnamefont {F.~D.~M.}\
  \bibnamefont {Haldane}},\ }\bibfield  {title} {\enquote {\bibinfo {title}
  {Model for a quantum hall effect without landau levels: Condensed-matter
  realization of the "parity anomaly"},}\ }\href {\doibase
  10.1103/PhysRevLett.61.2015} {\bibfield  {journal} {\bibinfo  {journal}
  {Phys. Rev. Lett.}\ }\textbf {\bibinfo {volume} {61}},\ \bibinfo {pages}
  {2015--2018} (\bibinfo {year} {1988})}\BibitemShut {NoStop}%
\bibitem [{\citenamefont {Kane}\ and\ \citenamefont
  {Lubensky}(2013)}]{kane2014topological}%
  \BibitemOpen
  \bibfield  {author} {\bibinfo {author} {\bibfnamefont {C.~L.}\ \bibnamefont
  {Kane}}\ and\ \bibinfo {author} {\bibfnamefont {T.~C.}\ \bibnamefont
  {Lubensky}},\ }\bibfield  {title} {\enquote {\bibinfo {title} {Topological
  boundary modes in isostatic lattices},}\ }\href
  {https://doi.org/10.1038/nphys2835} {\bibfield  {journal} {\bibinfo
  {journal} {Nature Physics}\ }\textbf {\bibinfo {volume} {10}},\ \bibinfo
  {pages} {39} (\bibinfo {year} {2013})}\BibitemShut {NoStop}%
\bibitem [{\citenamefont {Dyakov}\ \emph {et~al.}(2012)\citenamefont {Dyakov},
  \citenamefont {Baldycheva}, \citenamefont {Perova}, \citenamefont {Li},
  \citenamefont {Astrova}, \citenamefont {Gippius},\ and\ \citenamefont
  {Tikhodeev}}]{Dyakov2012}%
  \BibitemOpen
  \bibfield  {author} {\bibinfo {author} {\bibfnamefont {S.~A.}\ \bibnamefont
  {Dyakov}}, \bibinfo {author} {\bibfnamefont {A.}~\bibnamefont {Baldycheva}},
  \bibinfo {author} {\bibfnamefont {T.~S.}\ \bibnamefont {Perova}}, \bibinfo
  {author} {\bibfnamefont {G.~V.}\ \bibnamefont {Li}}, \bibinfo {author}
  {\bibfnamefont {E.~V.}\ \bibnamefont {Astrova}}, \bibinfo {author}
  {\bibfnamefont {N.~A.}\ \bibnamefont {Gippius}}, \ and\ \bibinfo {author}
  {\bibfnamefont {S.~G.}\ \bibnamefont {Tikhodeev}},\ }\bibfield  {title}
  {\enquote {\bibinfo {title} {Surface states in the optical spectra of
  two-dimensional photonic crystals with various surface terminations},}\
  }\href {\doibase 10.1103/PhysRevB.86.115126} {\bibfield  {journal} {\bibinfo
  {journal} {Phys. Rev. B}\ }\textbf {\bibinfo {volume} {86}},\ \bibinfo
  {pages} {115126} (\bibinfo {year} {2012})}\BibitemShut {NoStop}%
\bibitem [{\citenamefont {Kivshar}(2008)}]{Kivshar2008}%
  \BibitemOpen
  \bibfield  {author} {\bibinfo {author} {\bibfnamefont {Y.~S.}\ \bibnamefont
  {Kivshar}},\ }\bibfield  {title} {\enquote {\bibinfo {title} {Nonlinear tamm
  states and surface effects in periodic photonic structures},}\ }\href
  {\doibase 10.1002/lapl.200810062} {\bibfield  {journal} {\bibinfo  {journal}
  {Laser Physics Letters}\ }\textbf {\bibinfo {volume} {5}},\ \bibinfo {pages}
  {703--713} (\bibinfo {year} {2008})}\BibitemShut {NoStop}%
\bibitem [{\citenamefont {Eisenberg}\ \emph {et~al.}(1998)\citenamefont
  {Eisenberg}, \citenamefont {Silberberg}, \citenamefont {Morandotti},
  \citenamefont {Boyd},\ and\ \citenamefont {Aitchison}}]{Eisenberg1998}%
  \BibitemOpen
  \bibfield  {author} {\bibinfo {author} {\bibfnamefont {H.~S.}\ \bibnamefont
  {Eisenberg}}, \bibinfo {author} {\bibfnamefont {Y.}~\bibnamefont
  {Silberberg}}, \bibinfo {author} {\bibfnamefont {R.}~\bibnamefont
  {Morandotti}}, \bibinfo {author} {\bibfnamefont {A.~R.}\ \bibnamefont
  {Boyd}}, \ and\ \bibinfo {author} {\bibfnamefont {J.~S.}\ \bibnamefont
  {Aitchison}},\ }\bibfield  {title} {\enquote {\bibinfo {title} {Discrete
  spatial optical solitons in waveguide arrays},}\ }\href {\doibase
  10.1103/PhysRevLett.81.3383} {\bibfield  {journal} {\bibinfo  {journal}
  {Phys. Rev. Lett.}\ }\textbf {\bibinfo {volume} {81}},\ \bibinfo {pages}
  {3383--3386} (\bibinfo {year} {1998})}\BibitemShut {NoStop}%
\bibitem [{\citenamefont {Szameit}\ \emph {et~al.}(2005)\citenamefont
  {Szameit}, \citenamefont {Bl\"{o}mer}, \citenamefont {Burghoff},
  \citenamefont {Schreiber}, \citenamefont {Pertsch}, \citenamefont {Nolte},
  \citenamefont {T\"{u}nnermann},\ and\ \citenamefont {Lederer}}]{Szameit2005}%
  \BibitemOpen
  \bibfield  {author} {\bibinfo {author} {\bibfnamefont {A.}~\bibnamefont
  {Szameit}}, \bibinfo {author} {\bibfnamefont {D.}~\bibnamefont {Bl\"{o}mer}},
  \bibinfo {author} {\bibfnamefont {J.}~\bibnamefont {Burghoff}}, \bibinfo
  {author} {\bibfnamefont {T.}~\bibnamefont {Schreiber}}, \bibinfo {author}
  {\bibfnamefont {T.}~\bibnamefont {Pertsch}}, \bibinfo {author} {\bibfnamefont
  {S.}~\bibnamefont {Nolte}}, \bibinfo {author} {\bibfnamefont
  {A.}~\bibnamefont {T\"{u}nnermann}}, \ and\ \bibinfo {author} {\bibfnamefont
  {F.}~\bibnamefont {Lederer}},\ }\bibfield  {title} {\enquote {\bibinfo
  {title} {Discrete nonlinear localization in femtosecond laser written
  waveguides in fused silica},}\ }\href {\doibase 10.1364/OPEX.13.010552}
  {\bibfield  {journal} {\bibinfo  {journal} {Opt. Express}\ }\textbf {\bibinfo
  {volume} {13}},\ \bibinfo {pages} {10552--10557} (\bibinfo {year}
  {2005})}\BibitemShut {NoStop}%
\bibitem [{\citenamefont {Fleischer}\ \emph {et~al.}(2003)\citenamefont
  {Fleischer}, \citenamefont {Carmon}, \citenamefont {Segev}, \citenamefont
  {Efremidis},\ and\ \citenamefont {Christodoulides}}]{Fleischer2003}%
  \BibitemOpen
  \bibfield  {author} {\bibinfo {author} {\bibfnamefont {J.~W.}\ \bibnamefont
  {Fleischer}}, \bibinfo {author} {\bibfnamefont {T.}~\bibnamefont {Carmon}},
  \bibinfo {author} {\bibfnamefont {M.}~\bibnamefont {Segev}}, \bibinfo
  {author} {\bibfnamefont {N.~K.}\ \bibnamefont {Efremidis}}, \ and\ \bibinfo
  {author} {\bibfnamefont {D.~N.}\ \bibnamefont {Christodoulides}},\ }\bibfield
   {title} {\enquote {\bibinfo {title} {Observation of discrete solitons in
  optically induced real time waveguide arrays},}\ }\href {\doibase
  10.1103/PhysRevLett.90.023902} {\bibfield  {journal} {\bibinfo  {journal}
  {Phys. Rev. Lett.}\ }\textbf {\bibinfo {volume} {90}},\ \bibinfo {pages}
  {023902} (\bibinfo {year} {2003})}\BibitemShut {NoStop}%
\bibitem [{\citenamefont {Iwanow}\ \emph {et~al.}(2004)\citenamefont {Iwanow},
  \citenamefont {Schiek}, \citenamefont {Stegeman}, \citenamefont {Pertsch},
  \citenamefont {Lederer}, \citenamefont {Min},\ and\ \citenamefont
  {Sohler}}]{Iwanow2004}%
  \BibitemOpen
  \bibfield  {author} {\bibinfo {author} {\bibfnamefont {R.}~\bibnamefont
  {Iwanow}}, \bibinfo {author} {\bibfnamefont {R.}~\bibnamefont {Schiek}},
  \bibinfo {author} {\bibfnamefont {G.~I.}\ \bibnamefont {Stegeman}}, \bibinfo
  {author} {\bibfnamefont {T.}~\bibnamefont {Pertsch}}, \bibinfo {author}
  {\bibfnamefont {F.}~\bibnamefont {Lederer}}, \bibinfo {author} {\bibfnamefont
  {Y.}~\bibnamefont {Min}}, \ and\ \bibinfo {author} {\bibfnamefont
  {W.}~\bibnamefont {Sohler}},\ }\bibfield  {title} {\enquote {\bibinfo {title}
  {Observation of discrete quadratic solitons},}\ }\href {\doibase
  10.1103/PhysRevLett.93.113902} {\bibfield  {journal} {\bibinfo  {journal}
  {Phys. Rev. Lett.}\ }\textbf {\bibinfo {volume} {93}},\ \bibinfo {pages}
  {113902} (\bibinfo {year} {2004})}\BibitemShut {NoStop}%
\bibitem [{\citenamefont {Li}\ \emph {et~al.}(2014)\citenamefont {Li},
  \citenamefont {Huang}, \citenamefont {Wang}, \citenamefont {Petek},\ and\
  \citenamefont {Chen}}]{Li2014}%
  \BibitemOpen
  \bibfield  {author} {\bibinfo {author} {\bibfnamefont {M.}~\bibnamefont
  {Li}}, \bibinfo {author} {\bibfnamefont {S.}~\bibnamefont {Huang}}, \bibinfo
  {author} {\bibfnamefont {Q.}~\bibnamefont {Wang}}, \bibinfo {author}
  {\bibfnamefont {H.}~\bibnamefont {Petek}}, \ and\ \bibinfo {author}
  {\bibfnamefont {K.~P.}\ \bibnamefont {Chen}},\ }\bibfield  {title} {\enquote
  {\bibinfo {title} {Nonlinear optical localization in embedded chalcogenide
  waveguide arrays},}\ }\href {\doibase 10.1063/1.4879619} {\bibfield
  {journal} {\bibinfo  {journal} {AIP Advances}\ }\textbf {\bibinfo {volume}
  {4}},\ \bibinfo {pages} {057120} (\bibinfo {year} {2014})}\BibitemShut
  {NoStop}%
\bibitem [{\citenamefont {Fan}\ \emph {et~al.}(2012)\citenamefont {Fan},
  \citenamefont {Wang}, \citenamefont {Varghese}, \citenamefont {Shen},
  \citenamefont {Niu}, \citenamefont {Xuan}, \citenamefont {Weiner},\ and\
  \citenamefont {Qi}}]{Fan2012}%
  \BibitemOpen
  \bibfield  {author} {\bibinfo {author} {\bibfnamefont {L.}~\bibnamefont
  {Fan}}, \bibinfo {author} {\bibfnamefont {J.}~\bibnamefont {Wang}}, \bibinfo
  {author} {\bibfnamefont {L.~T.}\ \bibnamefont {Varghese}}, \bibinfo {author}
  {\bibfnamefont {H.}~\bibnamefont {Shen}}, \bibinfo {author} {\bibfnamefont
  {B.}~\bibnamefont {Niu}}, \bibinfo {author} {\bibfnamefont {Y.}~\bibnamefont
  {Xuan}}, \bibinfo {author} {\bibfnamefont {A.~M.}\ \bibnamefont {Weiner}}, \
  and\ \bibinfo {author} {\bibfnamefont {M.}~\bibnamefont {Qi}},\ }\bibfield
  {title} {\enquote {\bibinfo {title} {An all-silicon passive optical diode},}\
  }\href {\doibase 10.1126/science.1214383} {\bibfield  {journal} {\bibinfo
  {journal} {Science}\ }\textbf {\bibinfo {volume} {335}},\ \bibinfo {pages}
  {447--450} (\bibinfo {year} {2012})}\BibitemShut {NoStop}%
\bibitem [{\citenamefont {Yu}\ \emph {et~al.}(2015)\citenamefont {Yu},
  \citenamefont {Chen}, \citenamefont {Hu}, \citenamefont {Xue}, \citenamefont
  {Yvind},\ and\ \citenamefont {Mork}}]{Yu2015}%
  \BibitemOpen
  \bibfield  {author} {\bibinfo {author} {\bibfnamefont {Y.}~\bibnamefont
  {Yu}}, \bibinfo {author} {\bibfnamefont {Y.}~\bibnamefont {Chen}}, \bibinfo
  {author} {\bibfnamefont {H.}~\bibnamefont {Hu}}, \bibinfo {author}
  {\bibfnamefont {W.}~\bibnamefont {Xue}}, \bibinfo {author} {\bibfnamefont
  {K.}~\bibnamefont {Yvind}}, \ and\ \bibinfo {author} {\bibfnamefont
  {J.}~\bibnamefont {Mork}},\ }\bibfield  {title} {\enquote {\bibinfo {title}
  {Nonreciprocal transmission in a nonlinear photonic-crystal {F}ano structure
  with broken symmetry},}\ }\href {\doibase 10.1002/lpor.201400207} {\bibfield
  {journal} {\bibinfo  {journal} {Laser \& Photonics Reviews}\ }\textbf
  {\bibinfo {volume} {9}},\ \bibinfo {pages} {241--247} (\bibinfo {year}
  {2015})}\BibitemShut {NoStop}%
\bibitem [{\citenamefont {Ota}\ \emph {et~al.}(2018)\citenamefont {Ota},
  \citenamefont {Katsumi}, \citenamefont {Watanabe}, \citenamefont {Iwamoto},\
  and\ \citenamefont {Arakawa}}]{Ota2018}%
  \BibitemOpen
  \bibfield  {author} {\bibinfo {author} {\bibfnamefont {Y.}~\bibnamefont
  {Ota}}, \bibinfo {author} {\bibfnamefont {R.}~\bibnamefont {Katsumi}},
  \bibinfo {author} {\bibfnamefont {K.}~\bibnamefont {Watanabe}}, \bibinfo
  {author} {\bibfnamefont {S.}~\bibnamefont {Iwamoto}}, \ and\ \bibinfo
  {author} {\bibfnamefont {Y.}~\bibnamefont {Arakawa}},\ }\bibfield  {title}
  {\enquote {\bibinfo {title} {Topological photonic crystal nanocavity
  laser},}\ }\href {\doibase 10.1038/s42005-018-0083-7} {\bibfield  {journal}
  {\bibinfo  {journal} {Communications Physics}\ }\textbf {\bibinfo {volume}
  {1}},\ \bibinfo {pages} {86} (\bibinfo {year} {2018})}\BibitemShut {NoStop}%
\bibitem [{\citenamefont {Smirnova}\ \emph {et~al.}(2019)\citenamefont
  {Smirnova}, \citenamefont {Kruk}, \citenamefont {Leykam}, \citenamefont
  {Melik-Gaykazyan}, \citenamefont {Choi},\ and\ \citenamefont
  {Kivshar}}]{Smirnova2018}%
  \BibitemOpen
  \bibfield  {author} {\bibinfo {author} {\bibfnamefont {D.}~\bibnamefont
  {Smirnova}}, \bibinfo {author} {\bibfnamefont {S.}~\bibnamefont {Kruk}},
  \bibinfo {author} {\bibfnamefont {D.}~\bibnamefont {Leykam}}, \bibinfo
  {author} {\bibfnamefont {E.}~\bibnamefont {Melik-Gaykazyan}}, \bibinfo
  {author} {\bibfnamefont {D.-Y.}\ \bibnamefont {Choi}}, \ and\ \bibinfo
  {author} {\bibfnamefont {Y.}~\bibnamefont {Kivshar}},\ }\bibfield  {title}
  {\enquote {\bibinfo {title} {Third-harmonic generation in photonic
  topological metasurfaces},}\ }\href {\doibase 10.1103/PhysRevLett.123.103901}
  {\bibfield  {journal} {\bibinfo  {journal} {Phys. Rev. Lett.}\ }\textbf
  {\bibinfo {volume} {123}},\ \bibinfo {pages} {103901} (\bibinfo {year}
  {2019})}\BibitemShut {NoStop}%
\bibitem [{\citenamefont {Wimmer}\ \emph {et~al.}(2015)\citenamefont {Wimmer},
  \citenamefont {Regensburger}, \citenamefont {Miri}, \citenamefont {Bersch},
  \citenamefont {Christodoulides},\ and\ \citenamefont {Peschel}}]{Wimmer2015}%
  \BibitemOpen
  \bibfield  {author} {\bibinfo {author} {\bibfnamefont {M.}~\bibnamefont
  {Wimmer}}, \bibinfo {author} {\bibfnamefont {A.}~\bibnamefont
  {Regensburger}}, \bibinfo {author} {\bibfnamefont {M.-A.}\ \bibnamefont
  {Miri}}, \bibinfo {author} {\bibfnamefont {C.}~\bibnamefont {Bersch}},
  \bibinfo {author} {\bibfnamefont {D.~N.}\ \bibnamefont {Christodoulides}}, \
  and\ \bibinfo {author} {\bibfnamefont {U.}~\bibnamefont {Peschel}},\
  }\bibfield  {title} {\enquote {\bibinfo {title} {Observation of optical
  solitons in pt-symmetric lattices},}\ }\href
  {https://doi.org/10.1038/ncomms8782} {\bibfield  {journal} {\bibinfo
  {journal} {Nature Communications}\ }\textbf {\bibinfo {volume} {6}},\
  \bibinfo {pages} {7782} (\bibinfo {year} {2015})}\BibitemShut {NoStop}%
\bibitem [{\citenamefont {Tanese}\ \emph {et~al.}(2013)\citenamefont {Tanese},
  \citenamefont {Flayac}, \citenamefont {Solnyshkov}, \citenamefont {Amo},
  \citenamefont {Lema{\^i}tre}, \citenamefont {Galopin}, \citenamefont
  {Braive}, \citenamefont {Senellart}, \citenamefont {Sagnes}, \citenamefont
  {Malpuech},\ and\ \citenamefont {Bloch}}]{Tanese2013}%
  \BibitemOpen
  \bibfield  {author} {\bibinfo {author} {\bibfnamefont {D.}~\bibnamefont
  {Tanese}}, \bibinfo {author} {\bibfnamefont {H.}~\bibnamefont {Flayac}},
  \bibinfo {author} {\bibfnamefont {D.}~\bibnamefont {Solnyshkov}}, \bibinfo
  {author} {\bibfnamefont {A.}~\bibnamefont {Amo}}, \bibinfo {author}
  {\bibfnamefont {A.}~\bibnamefont {Lema{\^i}tre}}, \bibinfo {author}
  {\bibfnamefont {E.}~\bibnamefont {Galopin}}, \bibinfo {author} {\bibfnamefont
  {R.}~\bibnamefont {Braive}}, \bibinfo {author} {\bibfnamefont
  {P.}~\bibnamefont {Senellart}}, \bibinfo {author} {\bibfnamefont
  {I.}~\bibnamefont {Sagnes}}, \bibinfo {author} {\bibfnamefont
  {G.}~\bibnamefont {Malpuech}}, \ and\ \bibinfo {author} {\bibfnamefont
  {J.}~\bibnamefont {Bloch}},\ }\bibfield  {title} {\enquote {\bibinfo {title}
  {Polariton condensation in solitonic gap states in a one-dimensional periodic
  potential},}\ }\href {https://doi.org/10.1038/ncomms2760} {\bibfield
  {journal} {\bibinfo  {journal} {Nature Communications}\ }\textbf {\bibinfo
  {volume} {4}},\ \bibinfo {pages} {1749} (\bibinfo {year} {2013})},\ \bibinfo
  {note} {article}\BibitemShut {NoStop}%
\bibitem [{\citenamefont {Dobrykh}\ \emph {et~al.}(2018)\citenamefont
  {Dobrykh}, \citenamefont {Yulin}, \citenamefont {Slobozhanyuk}, \citenamefont
  {Poddubny},\ and\ \citenamefont {Kivshar}}]{Dobrykh2018}%
  \BibitemOpen
  \bibfield  {author} {\bibinfo {author} {\bibfnamefont {D.~A.}\ \bibnamefont
  {Dobrykh}}, \bibinfo {author} {\bibfnamefont {A.~V.}\ \bibnamefont {Yulin}},
  \bibinfo {author} {\bibfnamefont {A.~P.}\ \bibnamefont {Slobozhanyuk}},
  \bibinfo {author} {\bibfnamefont {A.~N.}\ \bibnamefont {Poddubny}}, \ and\
  \bibinfo {author} {\bibfnamefont {Y.~S.}\ \bibnamefont {Kivshar}},\
  }\bibfield  {title} {\enquote {\bibinfo {title} {Nonlinear control of
  electromagnetic topological edge states},}\ }\href {\doibase
  10.1103/PhysRevLett.121.163901} {\bibfield  {journal} {\bibinfo  {journal}
  {Phys. Rev. Lett.}\ }\textbf {\bibinfo {volume} {121}},\ \bibinfo {pages}
  {163901} (\bibinfo {year} {2018})}\BibitemShut {NoStop}%
\bibitem [{\citenamefont {Ma}\ \emph {et~al.}(2019)\citenamefont {Ma},
  \citenamefont {Saxberg}, \citenamefont {Owens}, \citenamefont {Leung},
  \citenamefont {Lu}, \citenamefont {Simon},\ and\ \citenamefont
  {Schuster}}]{Ma2019}%
  \BibitemOpen
  \bibfield  {author} {\bibinfo {author} {\bibfnamefont {R.}~\bibnamefont
  {Ma}}, \bibinfo {author} {\bibfnamefont {B.}~\bibnamefont {Saxberg}},
  \bibinfo {author} {\bibfnamefont {C.}~\bibnamefont {Owens}}, \bibinfo
  {author} {\bibfnamefont {N.}~\bibnamefont {Leung}}, \bibinfo {author}
  {\bibfnamefont {Y.}~\bibnamefont {Lu}}, \bibinfo {author} {\bibfnamefont
  {J.}~\bibnamefont {Simon}}, \ and\ \bibinfo {author} {\bibfnamefont {D.~I.}\
  \bibnamefont {Schuster}},\ }\bibfield  {title} {\enquote {\bibinfo {title} {A
  dissipatively stabilized mott insulator of photons},}\ }\href {\doibase
  10.1038/s41586-019-0897-9} {\bibfield  {journal} {\bibinfo  {journal}
  {Nature}\ }\textbf {\bibinfo {volume} {566}},\ \bibinfo {pages} {51--57}
  (\bibinfo {year} {2019})}\BibitemShut {NoStop}%
\bibitem [{\citenamefont {Hadad}\ \emph
  {et~al.}(2018{\natexlab{a}})\citenamefont {Hadad}, \citenamefont {Soric},
  \citenamefont {Khanikaev},\ and\ \citenamefont {Al{\`u}}}]{hadad2018self}%
  \BibitemOpen
  \bibfield  {author} {\bibinfo {author} {\bibfnamefont {Y.}~\bibnamefont
  {Hadad}}, \bibinfo {author} {\bibfnamefont {J.~C.}\ \bibnamefont {Soric}},
  \bibinfo {author} {\bibfnamefont {A.~B.}\ \bibnamefont {Khanikaev}}, \ and\
  \bibinfo {author} {\bibfnamefont {A.}~\bibnamefont {Al{\`u}}},\ }\bibfield
  {title} {\enquote {\bibinfo {title} {Self-induced topological protection in
  nonlinear circuit arrays},}\ }\href {\doibase 10.1038/s41928-018-0042-z}
  {\bibfield  {journal} {\bibinfo  {journal} {Nature Electronics}\ }\textbf
  {\bibinfo {volume} {1}},\ \bibinfo {pages} {178--182} (\bibinfo {year}
  {2018}{\natexlab{a}})}\BibitemShut {NoStop}%
\bibitem [{\citenamefont {Christodoulides}\ and\ \citenamefont
  {Joseph}(1988)}]{Christodoulides1988}%
  \BibitemOpen
  \bibfield  {author} {\bibinfo {author} {\bibfnamefont {D.~N.}\ \bibnamefont
  {Christodoulides}}\ and\ \bibinfo {author} {\bibfnamefont {R.~I.}\
  \bibnamefont {Joseph}},\ }\bibfield  {title} {\enquote {\bibinfo {title}
  {Discrete self-focusing in nonlinear arrays of coupled waveguides},}\ }\href
  {\doibase 10.1364/OL.13.000794} {\bibfield  {journal} {\bibinfo  {journal}
  {Opt. Lett.}\ }\textbf {\bibinfo {volume} {13}},\ \bibinfo {pages} {794--796}
  (\bibinfo {year} {1988})}\BibitemShut {NoStop}%
\bibitem [{\citenamefont {Lederer}\ \emph {et~al.}(2008)\citenamefont
  {Lederer}, \citenamefont {Stegeman}, \citenamefont {Christodoulides},
  \citenamefont {Assanto}, \citenamefont {Segev},\ and\ \citenamefont
  {Silberberg}}]{Lederer_review}%
  \BibitemOpen
  \bibfield  {author} {\bibinfo {author} {\bibfnamefont {F.}~\bibnamefont
  {Lederer}}, \bibinfo {author} {\bibfnamefont {G.~I.}\ \bibnamefont
  {Stegeman}}, \bibinfo {author} {\bibfnamefont {D.~N.}\ \bibnamefont
  {Christodoulides}}, \bibinfo {author} {\bibfnamefont {G.}~\bibnamefont
  {Assanto}}, \bibinfo {author} {\bibfnamefont {M.}~\bibnamefont {Segev}}, \
  and\ \bibinfo {author} {\bibfnamefont {Y.}~\bibnamefont {Silberberg}},\
  }\bibfield  {title} {\enquote {\bibinfo {title} {Discrete solitons in
  optics},}\ }\href {\doibase https://doi.org/10.1016/j.physrep.2008.04.004}
  {\bibfield  {journal} {\bibinfo  {journal} {Physics Reports}\ }\textbf
  {\bibinfo {volume} {463}},\ \bibinfo {pages} {1 -- 126} (\bibinfo {year}
  {2008})}\BibitemShut {NoStop}%
\bibitem [{\citenamefont {Denz}\ \emph {et~al.}(2010)\citenamefont {Denz},
  \citenamefont {Flach}, \citenamefont {Kivshar} \emph {et~al.}}]{Denz_review}%
  \BibitemOpen
  \bibfield  {author} {\bibinfo {author} {\bibfnamefont {C.}~\bibnamefont
  {Denz}}, \bibinfo {author} {\bibfnamefont {S.}~\bibnamefont {Flach}},
  \bibinfo {author} {\bibfnamefont {Y.~S.}\ \bibnamefont {Kivshar}},  \emph
  {et~al.},\ }\href@noop {} {\emph {\bibinfo {title} {Nonlinearities in
  periodic structures and metamaterials}}},\ Vol.\ \bibinfo {volume} {150}\
  (\bibinfo  {publisher} {Springer},\ \bibinfo {year} {2010})\BibitemShut
  {NoStop}%
\bibitem [{\citenamefont {Chen}, \citenamefont {Segev},\ and\ \citenamefont
  {Christodoulides}(2012)}]{Chen_review}%
  \BibitemOpen
  \bibfield  {author} {\bibinfo {author} {\bibfnamefont {Z.}~\bibnamefont
  {Chen}}, \bibinfo {author} {\bibfnamefont {M.}~\bibnamefont {Segev}}, \ and\
  \bibinfo {author} {\bibfnamefont {D.~N.}\ \bibnamefont {Christodoulides}},\
  }\bibfield  {title} {\enquote {\bibinfo {title} {Optical spatial solitons:
  historical overview and recent advances},}\ }\href {\doibase
  10.1088/0034-4885/75/8/086401} {\bibfield  {journal} {\bibinfo  {journal}
  {Reports on Progress in Physics}\ }\textbf {\bibinfo {volume} {75}},\
  \bibinfo {pages} {086401} (\bibinfo {year} {2012})}\BibitemShut {NoStop}%
\bibitem [{\citenamefont {Lahini}\ \emph {et~al.}(2007)\citenamefont {Lahini},
  \citenamefont {Frumker}, \citenamefont {Silberberg}, \citenamefont
  {Droulias}, \citenamefont {Hizanidis}, \citenamefont {Morandotti},\ and\
  \citenamefont {Christodoulides}}]{Lahini2007}%
  \BibitemOpen
  \bibfield  {author} {\bibinfo {author} {\bibfnamefont {Y.}~\bibnamefont
  {Lahini}}, \bibinfo {author} {\bibfnamefont {E.}~\bibnamefont {Frumker}},
  \bibinfo {author} {\bibfnamefont {Y.}~\bibnamefont {Silberberg}}, \bibinfo
  {author} {\bibfnamefont {S.}~\bibnamefont {Droulias}}, \bibinfo {author}
  {\bibfnamefont {K.}~\bibnamefont {Hizanidis}}, \bibinfo {author}
  {\bibfnamefont {R.}~\bibnamefont {Morandotti}}, \ and\ \bibinfo {author}
  {\bibfnamefont {D.~N.}\ \bibnamefont {Christodoulides}},\ }\bibfield  {title}
  {\enquote {\bibinfo {title} {Discrete $x$-wave formation in nonlinear
  waveguide arrays},}\ }\href {\doibase 10.1103/PhysRevLett.98.023901}
  {\bibfield  {journal} {\bibinfo  {journal} {Phys. Rev. Lett.}\ }\textbf
  {\bibinfo {volume} {98}},\ \bibinfo {pages} {023901} (\bibinfo {year}
  {2007})}\BibitemShut {NoStop}%
\bibitem [{\citenamefont {Bisianov}\ \emph {et~al.}(2018)\citenamefont
  {Bisianov}, \citenamefont {Kremer}, \citenamefont {Szameit},\ and\
  \citenamefont {Peschel}}]{Bisianov2018}%
  \BibitemOpen
  \bibfield  {author} {\bibinfo {author} {\bibfnamefont {A.}~\bibnamefont
  {Bisianov}}, \bibinfo {author} {\bibfnamefont {M.}~\bibnamefont {Kremer}},
  \bibinfo {author} {\bibfnamefont {A.}~\bibnamefont {Szameit}}, \ and\
  \bibinfo {author} {\bibfnamefont {U.}~\bibnamefont {Peschel}},\ }\bibfield
  {title} {\enquote {\bibinfo {title} {Experimental observation of the coupling
  of a nonlinear wave to a topological edge state},}\ }in\ \href {\doibase
  10.1364/CLEO_QELS.2018.FM1E.1} {\emph {\bibinfo {booktitle} {Conference on
  Lasers and Electro-Optics}}}\ (\bibinfo  {publisher} {Optical Society of
  America},\ \bibinfo {year} {2018})\ p.\ \bibinfo {pages} {FM1E.1}\BibitemShut
  {NoStop}%
\bibitem [{\citenamefont {Anderson}\ \emph {et~al.}(2016)\citenamefont
  {Anderson}, \citenamefont {Ma}, \citenamefont {Owens}, \citenamefont
  {Schuster},\ and\ \citenamefont {Simon}}]{Anderson2016}%
  \BibitemOpen
  \bibfield  {author} {\bibinfo {author} {\bibfnamefont {B.~M.}\ \bibnamefont
  {Anderson}}, \bibinfo {author} {\bibfnamefont {R.}~\bibnamefont {Ma}},
  \bibinfo {author} {\bibfnamefont {C.}~\bibnamefont {Owens}}, \bibinfo
  {author} {\bibfnamefont {D.~I.}\ \bibnamefont {Schuster}}, \ and\ \bibinfo
  {author} {\bibfnamefont {J.}~\bibnamefont {Simon}},\ }\bibfield  {title}
  {\enquote {\bibinfo {title} {Engineering topological many-body materials in
  microwave cavity arrays},}\ }\href {\doibase 10.1103/PhysRevX.6.041043}
  {\bibfield  {journal} {\bibinfo  {journal} {Phys. Rev. X}\ }\textbf {\bibinfo
  {volume} {6}},\ \bibinfo {pages} {041043} (\bibinfo {year}
  {2016})}\BibitemShut {NoStop}%
\bibitem [{\citenamefont {Owens}\ \emph {et~al.}(2018)\citenamefont {Owens},
  \citenamefont {LaChapelle}, \citenamefont {Saxberg}, \citenamefont
  {Anderson}, \citenamefont {Ma}, \citenamefont {Simon},\ and\ \citenamefont
  {Schuster}}]{Owens2018}%
  \BibitemOpen
  \bibfield  {author} {\bibinfo {author} {\bibfnamefont {C.}~\bibnamefont
  {Owens}}, \bibinfo {author} {\bibfnamefont {A.}~\bibnamefont {LaChapelle}},
  \bibinfo {author} {\bibfnamefont {B.}~\bibnamefont {Saxberg}}, \bibinfo
  {author} {\bibfnamefont {B.~M.}\ \bibnamefont {Anderson}}, \bibinfo {author}
  {\bibfnamefont {R.}~\bibnamefont {Ma}}, \bibinfo {author} {\bibfnamefont
  {J.}~\bibnamefont {Simon}}, \ and\ \bibinfo {author} {\bibfnamefont {D.~I.}\
  \bibnamefont {Schuster}},\ }\bibfield  {title} {\enquote {\bibinfo {title}
  {Quarter-flux hofstadter lattice in a qubit-compatible microwave cavity
  array},}\ }\href {\doibase 10.1103/PhysRevA.97.013818} {\bibfield  {journal}
  {\bibinfo  {journal} {Phys. Rev. A}\ }\textbf {\bibinfo {volume} {97}},\
  \bibinfo {pages} {013818} (\bibinfo {year} {2018})}\BibitemShut {NoStop}%
\bibitem [{\citenamefont {S\"{u}sstrunk}\ and\ \citenamefont
  {Huber}(2015)}]{Susstrunk2015}%
  \BibitemOpen
  \bibfield  {author} {\bibinfo {author} {\bibfnamefont {R.}~\bibnamefont
  {S\"{u}sstrunk}}\ and\ \bibinfo {author} {\bibfnamefont {S.~D.}\ \bibnamefont
  {Huber}},\ }\bibfield  {title} {\enquote {\bibinfo {title} {Observation of
  phononic helical edge states in a mechanical topological insulator},}\ }\href
  {\doibase 10.1126/science.aab0239} {\bibfield  {journal} {\bibinfo  {journal}
  {Science}\ }\textbf {\bibinfo {volume} {349}},\ \bibinfo {pages} {47--50}
  (\bibinfo {year} {2015})}\BibitemShut {NoStop}%
\bibitem [{\citenamefont {Snee}\ and\ \citenamefont {Ma}(2019)}]{Snee2018}%
  \BibitemOpen
  \bibfield  {author} {\bibinfo {author} {\bibfnamefont {D.~D.}\ \bibnamefont
  {Snee}}\ and\ \bibinfo {author} {\bibfnamefont {Y.-P.}\ \bibnamefont {Ma}},\
  }\bibfield  {title} {\enquote {\bibinfo {title} {Edge solitons in a nonlinear
  mechanical topological insulator},}\ }\href {\doibase
  https://doi.org/10.1016/j.eml.2019.100487} {\bibfield  {journal} {\bibinfo
  {journal} {Extreme Mechanics Letters}\ }\textbf {\bibinfo {volume} {30}},\
  \bibinfo {pages} {100487} (\bibinfo {year} {2019})}\BibitemShut {NoStop}%
\bibitem [{\citenamefont {Chaunsali}\ and\ \citenamefont
  {Theocharis}(2019)}]{Chaunsali2019}%
  \BibitemOpen
  \bibfield  {author} {\bibinfo {author} {\bibfnamefont {R.}~\bibnamefont
  {Chaunsali}}\ and\ \bibinfo {author} {\bibfnamefont {G.}~\bibnamefont
  {Theocharis}},\ }\bibfield  {title} {\enquote {\bibinfo {title} {Self-induced
  topological transition in phononic crystals by nonlinearity management},}\
  }\href {\doibase 10.1103/PhysRevB.100.014302} {\bibfield  {journal} {\bibinfo
   {journal} {Phys. Rev. B}\ }\textbf {\bibinfo {volume} {100}},\ \bibinfo
  {pages} {014302} (\bibinfo {year} {2019})}\BibitemShut {NoStop}%
\bibitem [{\citenamefont {Lumer}\ \emph {et~al.}(2016)\citenamefont {Lumer},
  \citenamefont {Rechtsman}, \citenamefont {Plotnik},\ and\ \citenamefont
  {Segev}}]{Lumer2016}%
  \BibitemOpen
  \bibfield  {author} {\bibinfo {author} {\bibfnamefont {Y.}~\bibnamefont
  {Lumer}}, \bibinfo {author} {\bibfnamefont {M.~C.}\ \bibnamefont
  {Rechtsman}}, \bibinfo {author} {\bibfnamefont {Y.}~\bibnamefont {Plotnik}},
  \ and\ \bibinfo {author} {\bibfnamefont {M.}~\bibnamefont {Segev}},\
  }\bibfield  {title} {\enquote {\bibinfo {title} {Instability of bosonic
  topological edge states in the presence of interactions},}\ }\href {\doibase
  10.1103/PhysRevA.94.021801} {\bibfield  {journal} {\bibinfo  {journal} {Phys.
  Rev. A}\ }\textbf {\bibinfo {volume} {94}},\ \bibinfo {pages} {021801}
  (\bibinfo {year} {2016})}\BibitemShut {NoStop}%
\bibitem [{\citenamefont {Solnyshkov}\ \emph {et~al.}(2017)\citenamefont
  {Solnyshkov}, \citenamefont {Bleu}, \citenamefont {Teklu},\ and\
  \citenamefont {Malpuech}}]{Solnyshkov2017}%
  \BibitemOpen
  \bibfield  {author} {\bibinfo {author} {\bibfnamefont {D.}~\bibnamefont
  {Solnyshkov}}, \bibinfo {author} {\bibfnamefont {O.}~\bibnamefont {Bleu}},
  \bibinfo {author} {\bibfnamefont {B.}~\bibnamefont {Teklu}}, \ and\ \bibinfo
  {author} {\bibfnamefont {G.}~\bibnamefont {Malpuech}},\ }\bibfield  {title}
  {\enquote {\bibinfo {title} {Chirality of topological gap solitons in bosonic
  dimer chains},}\ }\href {\doibase 10.1103/physrevlett.118.023901} {\bibfield
  {journal} {\bibinfo  {journal} {Phys. Rev. Lett.}\ }\textbf {\bibinfo
  {volume} {118}},\ \bibinfo {pages} {023901} (\bibinfo {year}
  {2017})}\BibitemShut {NoStop}%
\bibitem [{\citenamefont {Smirnova}\ \emph {et~al.}()\citenamefont {Smirnova},
  \citenamefont {Smirnov}, \citenamefont {Leykam},\ and\ \citenamefont
  {Kivshar}}]{Smirnova2019LPR}%
  \BibitemOpen
  \bibfield  {author} {\bibinfo {author} {\bibfnamefont {D.~A.}\ \bibnamefont
  {Smirnova}}, \bibinfo {author} {\bibfnamefont {L.~A.}\ \bibnamefont
  {Smirnov}}, \bibinfo {author} {\bibfnamefont {D.}~\bibnamefont {Leykam}}, \
  and\ \bibinfo {author} {\bibfnamefont {Y.~S.}\ \bibnamefont {Kivshar}},\
  }\bibfield  {title} {\enquote {\bibinfo {title} {Topological edge states and
  gap solitons in the nonlinear dirac model},}\ }\href {\doibase
  10.1002/lpor.201900223} {\bibinfo  {journal} {Laser \& Photonics Reviews}\ ,\
  \bibinfo {pages} {1900223}}\BibitemShut {NoStop}%
\bibitem [{\citenamefont {Leykam}\ and\ \citenamefont
  {Chong}(2016)}]{Leykam2016}%
  \BibitemOpen
\bibfield  {journal} {  }\bibfield  {author} {\bibinfo {author} {\bibfnamefont
  {D.}~\bibnamefont {Leykam}}\ and\ \bibinfo {author} {\bibfnamefont {Y.~D.}\
  \bibnamefont {Chong}},\ }\bibfield  {title} {\enquote {\bibinfo {title} {Edge
  solitons in nonlinear-photonic topological insulators},}\ }\href {\doibase
  10.1103/PhysRevLett.117.143901} {\bibfield  {journal} {\bibinfo  {journal}
  {Phys. Rev. Lett.}\ }\textbf {\bibinfo {volume} {117}},\ \bibinfo {pages}
  {143901} (\bibinfo {year} {2016})}\BibitemShut {NoStop}%
\bibitem [{\citenamefont {Poddubny}\ and\ \citenamefont
  {Smirnova}(2018)}]{RingSoliton2018}%
  \BibitemOpen
  \bibfield  {author} {\bibinfo {author} {\bibfnamefont {A.~N.}\ \bibnamefont
  {Poddubny}}\ and\ \bibinfo {author} {\bibfnamefont {D.~A.}\ \bibnamefont
  {Smirnova}},\ }\bibfield  {title} {\enquote {\bibinfo {title} {Ring {D}irac
  solitons in nonlinear topological systems},}\ }\href {\doibase
  10.1103/PhysRevA.98.013827} {\bibfield  {journal} {\bibinfo  {journal} {Phys.
  Rev. A}\ }\textbf {\bibinfo {volume} {98}},\ \bibinfo {pages} {013827}
  (\bibinfo {year} {2018})}\BibitemShut {NoStop}%
\bibitem [{\citenamefont {Liu}\ and\ \citenamefont
  {Fu}(2010)}]{BerryPhaseNL2010}%
  \BibitemOpen
  \bibfield  {author} {\bibinfo {author} {\bibfnamefont {J.}~\bibnamefont
  {Liu}}\ and\ \bibinfo {author} {\bibfnamefont {L.~B.}\ \bibnamefont {Fu}},\
  }\bibfield  {title} {\enquote {\bibinfo {title} {Berry phase in nonlinear
  systems},}\ }\href {\doibase 10.1103/PhysRevA.81.052112} {\bibfield
  {journal} {\bibinfo  {journal} {Phys. Rev. A}\ }\textbf {\bibinfo {volume}
  {81}},\ \bibinfo {pages} {052112} (\bibinfo {year} {2010})}\BibitemShut
  {NoStop}%
\bibitem [{\citenamefont {Bomantara}\ \emph {et~al.}(2017)\citenamefont
  {Bomantara}, \citenamefont {Zhao}, \citenamefont {Zhou},\ and\ \citenamefont
  {Gong}}]{Bomantara2017}%
  \BibitemOpen
  \bibfield  {author} {\bibinfo {author} {\bibfnamefont {R.~W.}\ \bibnamefont
  {Bomantara}}, \bibinfo {author} {\bibfnamefont {W.}~\bibnamefont {Zhao}},
  \bibinfo {author} {\bibfnamefont {L.}~\bibnamefont {Zhou}}, \ and\ \bibinfo
  {author} {\bibfnamefont {J.}~\bibnamefont {Gong}},\ }\bibfield  {title}
  {\enquote {\bibinfo {title} {Nonlinear {D}irac cones},}\ }\href {\doibase
  10.1103/PhysRevB.96.121406} {\bibfield  {journal} {\bibinfo  {journal} {Phys.
  Rev. B}\ }\textbf {\bibinfo {volume} {96}},\ \bibinfo {pages} {121406}
  (\bibinfo {year} {2017})}\BibitemShut {NoStop}%
\bibitem [{\citenamefont {Solnyshkov}, \citenamefont {Bleu},\ and\
  \citenamefont {Malpuech}(2018)}]{Solnyshkov2018}%
  \BibitemOpen
  \bibfield  {author} {\bibinfo {author} {\bibfnamefont {D.~D.}\ \bibnamefont
  {Solnyshkov}}, \bibinfo {author} {\bibfnamefont {O.}~\bibnamefont {Bleu}}, \
  and\ \bibinfo {author} {\bibfnamefont {G.}~\bibnamefont {Malpuech}},\
  }\bibfield  {title} {\enquote {\bibinfo {title} {Topological optical isolator
  based on polariton graphene},}\ }\href {\doibase 10.1063/1.5018902}
  {\bibfield  {journal} {\bibinfo  {journal} {Appl. Phys. Lett.}\ }\textbf
  {\bibinfo {volume} {112}},\ \bibinfo {pages} {031106} (\bibinfo {year}
  {2018})}\BibitemShut {NoStop}%
\bibitem [{\citenamefont {Hadad}, \citenamefont {Khanikaev},\ and\
  \citenamefont {Al\`u}(2016)}]{Hadad2016}%
  \BibitemOpen
  \bibfield  {author} {\bibinfo {author} {\bibfnamefont {Y.}~\bibnamefont
  {Hadad}}, \bibinfo {author} {\bibfnamefont {A.~B.}\ \bibnamefont
  {Khanikaev}}, \ and\ \bibinfo {author} {\bibfnamefont {A.}~\bibnamefont
  {Al\`u}},\ }\bibfield  {title} {\enquote {\bibinfo {title} {Self-induced
  topological transitions and edge states supported by nonlinear staggered
  potentials},}\ }\href {\doibase 10.1103/PhysRevB.93.155112} {\bibfield
  {journal} {\bibinfo  {journal} {Phys. Rev. B}\ }\textbf {\bibinfo {volume}
  {93}},\ \bibinfo {pages} {155112} (\bibinfo {year} {2016})}\BibitemShut
  {NoStop}%
\bibitem [{\citenamefont {Gulevich}\ \emph {et~al.}(2017)\citenamefont
  {Gulevich}, \citenamefont {Yudin}, \citenamefont {Skryabin}, \citenamefont
  {Iorsh},\ and\ \citenamefont {Shelykh}}]{Gulevich2017}%
  \BibitemOpen
  \bibfield  {author} {\bibinfo {author} {\bibfnamefont {D.~R.}\ \bibnamefont
  {Gulevich}}, \bibinfo {author} {\bibfnamefont {D.}~\bibnamefont {Yudin}},
  \bibinfo {author} {\bibfnamefont {D.~V.}\ \bibnamefont {Skryabin}}, \bibinfo
  {author} {\bibfnamefont {I.~V.}\ \bibnamefont {Iorsh}}, \ and\ \bibinfo
  {author} {\bibfnamefont {I.~A.}\ \bibnamefont {Shelykh}},\ }\bibfield
  {title} {\enquote {\bibinfo {title} {Exploring nonlinear topological states
  of matter with exciton-polaritons: {E}dge solitons in {K}agome lattice},}\
  }\href {\doibase 10.1038/s41598-017-01646-y} {\bibfield  {journal} {\bibinfo
  {journal} {Sci. Rep.}\ }\textbf {\bibinfo {volume} {7}},\ \bibinfo {pages}
  {1780} (\bibinfo {year} {2017})}\BibitemShut {NoStop}%
\bibitem [{\citenamefont {Marzuola}\ \emph {et~al.}(2019)\citenamefont
  {Marzuola}, \citenamefont {Rechtsman}, \citenamefont {Osting},\ and\
  \citenamefont {Bandres}}]{Marzuola2019arxiv}%
  \BibitemOpen
  \bibfield  {author} {\bibinfo {author} {\bibfnamefont {J.~L.}\ \bibnamefont
  {Marzuola}}, \bibinfo {author} {\bibfnamefont {M.}~\bibnamefont {Rechtsman}},
  \bibinfo {author} {\bibfnamefont {B.}~\bibnamefont {Osting}}, \ and\ \bibinfo
  {author} {\bibfnamefont {M.}~\bibnamefont {Bandres}},\ }\bibfield  {title}
  {\enquote {\bibinfo {title} {Bulk soliton dynamics in bosonic topological
  insulators},}\ }\href {https://arxiv.org/abs/1904.10312} {\bibfield
  {journal} {\bibinfo  {journal} {arXiv:1904.10312}\ } (\bibinfo {year}
  {2019})}\BibitemShut {NoStop}%
\bibitem [{\citenamefont {Mukherjee}\ and\ \citenamefont
  {Rechtsman}(2019)}]{Mukherjee2019arxiv}%
  \BibitemOpen
  \bibfield  {author} {\bibinfo {author} {\bibfnamefont {S.}~\bibnamefont
  {Mukherjee}}\ and\ \bibinfo {author} {\bibfnamefont {M.~C.}\ \bibnamefont
  {Rechtsman}},\ }\bibfield  {title} {\enquote {\bibinfo {title} {Observation
  of topological band gap solitons},}\ }\href
  {https://arxiv.org/abs/1911.05260} {\bibfield  {journal} {\bibinfo  {journal}
  {arXiv}\ ,\ \bibinfo {pages} {1911.05260}} (\bibinfo {year}
  {2019})}\BibitemShut {NoStop}%
\bibitem [{\citenamefont {Ablowitz}, \citenamefont {Curtis},\ and\
  \citenamefont {Ma}(2014)}]{Ablowitz2014}%
  \BibitemOpen
  \bibfield  {author} {\bibinfo {author} {\bibfnamefont {M.~J.}\ \bibnamefont
  {Ablowitz}}, \bibinfo {author} {\bibfnamefont {C.~W.}\ \bibnamefont
  {Curtis}}, \ and\ \bibinfo {author} {\bibfnamefont {Y.-P.}\ \bibnamefont
  {Ma}},\ }\bibfield  {title} {\enquote {\bibinfo {title} {Linear and nonlinear
  traveling edge waves in optical honeycomb lattices},}\ }\href {\doibase
  10.1103/PhysRevA.90.023813} {\bibfield  {journal} {\bibinfo  {journal} {Phys.
  Rev. A}\ }\textbf {\bibinfo {volume} {90}},\ \bibinfo {pages} {023813}
  (\bibinfo {year} {2014})}\BibitemShut {NoStop}%
\bibitem [{\citenamefont {Kartashov}\ and\ \citenamefont
  {Skryabin}(2016)}]{Kartashov2016}%
  \BibitemOpen
  \bibfield  {author} {\bibinfo {author} {\bibfnamefont {Y.~V.}\ \bibnamefont
  {Kartashov}}\ and\ \bibinfo {author} {\bibfnamefont {D.~V.}\ \bibnamefont
  {Skryabin}},\ }\bibfield  {title} {\enquote {\bibinfo {title} {Modulational
  instability and solitary waves in polariton topological insulators},}\ }\href
  {\doibase 10.1364/optica.3.001228} {\bibfield  {journal} {\bibinfo  {journal}
  {Optica}\ }\textbf {\bibinfo {volume} {3}},\ \bibinfo {pages} {1228}
  (\bibinfo {year} {2016})}\BibitemShut {NoStop}%
\bibitem [{\citenamefont {Cuevas\char21{}Maraver}\ \emph
  {et~al.}(2016)\citenamefont {Cuevas\char21{}Maraver}, \citenamefont
  {Kevrekidis}, \citenamefont {Saxena}, \citenamefont {Comech},\ and\
  \citenamefont {Lan}}]{Maravero2016}%
  \BibitemOpen
  \bibfield  {author} {\bibinfo {author} {\bibfnamefont {J.}~\bibnamefont
  {Cuevas\char21{}Maraver}}, \bibinfo {author} {\bibfnamefont {P.~G.}\
  \bibnamefont {Kevrekidis}}, \bibinfo {author} {\bibfnamefont
  {A.}~\bibnamefont {Saxena}}, \bibinfo {author} {\bibfnamefont
  {A.}~\bibnamefont {Comech}}, \ and\ \bibinfo {author} {\bibfnamefont
  {R.}~\bibnamefont {Lan}},\ }\bibfield  {title} {\enquote {\bibinfo {title}
  {Stability of solitary waves and vortices in a 2{D} nonlinear {D}irac
  model},}\ }\href {\doibase 10.1103/PhysRevLett.116.214101} {\bibfield
  {journal} {\bibinfo  {journal} {Phys. Rev. Lett.}\ }\textbf {\bibinfo
  {volume} {116}},\ \bibinfo {pages} {214101} (\bibinfo {year}
  {2016})}\BibitemShut {NoStop}%
\bibitem [{\citenamefont {Cuevas-Maraver}\ \emph {et~al.}(2017)\citenamefont
  {Cuevas-Maraver}, \citenamefont {Kevrekidis}, \citenamefont {Aceves},\ and\
  \citenamefont {Saxena}}]{Maraver2017b}%
  \BibitemOpen
  \bibfield  {author} {\bibinfo {author} {\bibfnamefont {J.}~\bibnamefont
  {Cuevas-Maraver}}, \bibinfo {author} {\bibfnamefont {P.~G.}\ \bibnamefont
  {Kevrekidis}}, \bibinfo {author} {\bibfnamefont {A.~B.}\ \bibnamefont
  {Aceves}}, \ and\ \bibinfo {author} {\bibfnamefont {A.}~\bibnamefont
  {Saxena}},\ }\bibfield  {title} {\enquote {\bibinfo {title} {Solitary waves
  in a two-dimensional nonlinear {D}irac equation: from discrete to
  continuum},}\ }\href {\doibase 10.1088/1751-8121/aa8e36} {\bibfield
  {journal} {\bibinfo  {journal} {J. Phys. A}\ }\textbf {\bibinfo {volume}
  {50}},\ \bibinfo {pages} {495207} (\bibinfo {year} {2017})}\BibitemShut
  {NoStop}%
\bibitem [{\citenamefont {Cuevas-Maraver}\ \emph {et~al.}(2018)\citenamefont
  {Cuevas-Maraver}, \citenamefont {Boussaïd}, \citenamefont {Comech},
  \citenamefont {Lan}, \citenamefont {Kevrekidis},\ and\ \citenamefont
  {Saxena}}]{Maraver2018}%
  \BibitemOpen
  \bibfield  {author} {\bibinfo {author} {\bibfnamefont {J.}~\bibnamefont
  {Cuevas-Maraver}}, \bibinfo {author} {\bibfnamefont {N.}~\bibnamefont
  {Boussaïd}}, \bibinfo {author} {\bibfnamefont {A.}~\bibnamefont {Comech}},
  \bibinfo {author} {\bibfnamefont {R.}~\bibnamefont {Lan}}, \bibinfo {author}
  {\bibfnamefont {P.~G.}\ \bibnamefont {Kevrekidis}}, \ and\ \bibinfo {author}
  {\bibfnamefont {A.}~\bibnamefont {Saxena}},\ }\bibfield  {title} {\enquote
  {\bibinfo {title} {Solitary waves in the nonlinear dirac equation},}\ }in\
  \href {\doibase 10.1007/978-3-319-66766-9_4} {\emph {\bibinfo {booktitle}
  {Understanding Complex Systems}}}\ (\bibinfo  {publisher} {Springer
  International Publishing},\ \bibinfo {year} {2018})\ pp.\ \bibinfo {pages}
  {89--143}\BibitemShut {NoStop}%
\bibitem [{\citenamefont {Hadad}, \citenamefont {Vitelli},\ and\ \citenamefont
  {Alu}(2017)}]{Hadad2017}%
  \BibitemOpen
  \bibfield  {author} {\bibinfo {author} {\bibfnamefont {Y.}~\bibnamefont
  {Hadad}}, \bibinfo {author} {\bibfnamefont {V.}~\bibnamefont {Vitelli}}, \
  and\ \bibinfo {author} {\bibfnamefont {A.}~\bibnamefont {Alu}},\ }\bibfield
  {title} {\enquote {\bibinfo {title} {Solitons and propagating domain walls in
  topological resonator arrays},}\ }\href {\doibase
  10.1021/acsphotonics.7b00303} {\bibfield  {journal} {\bibinfo  {journal} {ACS
  Photonics}\ }\textbf {\bibinfo {volume} {4}},\ \bibinfo {pages} {1974--1979}
  (\bibinfo {year} {2017})}\BibitemShut {NoStop}%
\bibitem [{\citenamefont {Gerasimenko}, \citenamefont {Tarasinski},\ and\
  \citenamefont {Beenakker}(2016)}]{Gerasimenko2016}%
  \BibitemOpen
  \bibfield  {author} {\bibinfo {author} {\bibfnamefont {Y.}~\bibnamefont
  {Gerasimenko}}, \bibinfo {author} {\bibfnamefont {B.}~\bibnamefont
  {Tarasinski}}, \ and\ \bibinfo {author} {\bibfnamefont {C.~W.~J.}\
  \bibnamefont {Beenakker}},\ }\bibfield  {title} {\enquote {\bibinfo {title}
  {Attractor-repeller pair of topological zero modes in a nonlinear quantum
  walk},}\ }\href {\doibase 10.1103/PhysRevA.93.022329} {\bibfield  {journal}
  {\bibinfo  {journal} {Phys. Rev. A}\ }\textbf {\bibinfo {volume} {93}},\
  \bibinfo {pages} {022329} (\bibinfo {year} {2016})}\BibitemShut {NoStop}%
\bibitem [{\citenamefont {Hadad}\ \emph
  {et~al.}(2018{\natexlab{b}})\citenamefont {Hadad}, \citenamefont {Soric},
  \citenamefont {Khanikaev},\ and\ \citenamefont {Al{\`{u}}}}]{Hadad2018}%
  \BibitemOpen
  \bibfield  {author} {\bibinfo {author} {\bibfnamefont {Y.}~\bibnamefont
  {Hadad}}, \bibinfo {author} {\bibfnamefont {J.~C.}\ \bibnamefont {Soric}},
  \bibinfo {author} {\bibfnamefont {A.~B.}\ \bibnamefont {Khanikaev}}, \ and\
  \bibinfo {author} {\bibfnamefont {A.}~\bibnamefont {Al{\`{u}}}},\ }\bibfield
  {title} {\enquote {\bibinfo {title} {Self-induced topological protection in
  nonlinear circuit arrays},}\ }\href {\doibase 10.1038/s41928-018-0042-z}
  {\bibfield  {journal} {\bibinfo  {journal} {Nature Electronics}\ }\textbf
  {\bibinfo {volume} {1}},\ \bibinfo {pages} {178--182} (\bibinfo {year}
  {2018}{\natexlab{b}})}\BibitemShut {NoStop}%
\bibitem [{\citenamefont {Bleu}, \citenamefont {Solnyshkov},\ and\
  \citenamefont {Malpuech}(2016)}]{Bleu2016}%
  \BibitemOpen
  \bibfield  {author} {\bibinfo {author} {\bibfnamefont {O.}~\bibnamefont
  {Bleu}}, \bibinfo {author} {\bibfnamefont {D.~D.}\ \bibnamefont
  {Solnyshkov}}, \ and\ \bibinfo {author} {\bibfnamefont {G.}~\bibnamefont
  {Malpuech}},\ }\bibfield  {title} {\enquote {\bibinfo {title} {Interacting
  quantum fluid in a polariton {C}hern insulator},}\ }\href {\doibase
  10.1103/PhysRevB.93.085438} {\bibfield  {journal} {\bibinfo  {journal} {Phys.
  Rev. B}\ }\textbf {\bibinfo {volume} {93}},\ \bibinfo {pages} {085438}
  (\bibinfo {year} {2016})}\BibitemShut {NoStop}%
\bibitem [{\citenamefont {Li}\ \emph {et~al.}(2018{\natexlab{a}})\citenamefont
  {Li}, \citenamefont {Ye}, \citenamefont {Chen}, \citenamefont {Kartashov},
  \citenamefont {Ferrando}, \citenamefont {Torner},\ and\ \citenamefont
  {Skryabin}}]{Li2018}%
  \BibitemOpen
  \bibfield  {author} {\bibinfo {author} {\bibfnamefont {C.}~\bibnamefont
  {Li}}, \bibinfo {author} {\bibfnamefont {F.}~\bibnamefont {Ye}}, \bibinfo
  {author} {\bibfnamefont {X.}~\bibnamefont {Chen}}, \bibinfo {author}
  {\bibfnamefont {Y.~V.}\ \bibnamefont {Kartashov}}, \bibinfo {author}
  {\bibfnamefont {A.}~\bibnamefont {Ferrando}}, \bibinfo {author}
  {\bibfnamefont {L.}~\bibnamefont {Torner}}, \ and\ \bibinfo {author}
  {\bibfnamefont {D.~V.}\ \bibnamefont {Skryabin}},\ }\bibfield  {title}
  {\enquote {\bibinfo {title} {{L}ieb polariton topological insulators},}\
  }\href {\doibase 10.1103/PhysRevB.97.081103} {\bibfield  {journal} {\bibinfo
  {journal} {Phys. Rev. B}\ }\textbf {\bibinfo {volume} {97}},\ \bibinfo
  {pages} {081103} (\bibinfo {year} {2018}{\natexlab{a}})}\BibitemShut
  {NoStop}%
\bibitem [{\citenamefont {Ningyuan}\ \emph {et~al.}(2015)\citenamefont
  {Ningyuan}, \citenamefont {Owens}, \citenamefont {Sommer}, \citenamefont
  {Schuster},\ and\ \citenamefont {Simon}}]{Simon2015}%
  \BibitemOpen
  \bibfield  {author} {\bibinfo {author} {\bibfnamefont {J.}~\bibnamefont
  {Ningyuan}}, \bibinfo {author} {\bibfnamefont {C.}~\bibnamefont {Owens}},
  \bibinfo {author} {\bibfnamefont {A.}~\bibnamefont {Sommer}}, \bibinfo
  {author} {\bibfnamefont {D.}~\bibnamefont {Schuster}}, \ and\ \bibinfo
  {author} {\bibfnamefont {J.}~\bibnamefont {Simon}},\ }\bibfield  {title}
  {\enquote {\bibinfo {title} {Time- and site-resolved dynamics in a
  topological circuit},}\ }\href {\doibase 10.1103/PhysRevX.5.021031}
  {\bibfield  {journal} {\bibinfo  {journal} {Phys. Rev. X}\ }\textbf {\bibinfo
  {volume} {5}},\ \bibinfo {pages} {021031} (\bibinfo {year}
  {2015})}\BibitemShut {NoStop}%
\bibitem [{\citenamefont {Albert}, \citenamefont {Glazman},\ and\ \citenamefont
  {Jiang}(2015)}]{Albert2015}%
  \BibitemOpen
  \bibfield  {author} {\bibinfo {author} {\bibfnamefont {V.~V.}\ \bibnamefont
  {Albert}}, \bibinfo {author} {\bibfnamefont {L.~I.}\ \bibnamefont {Glazman}},
  \ and\ \bibinfo {author} {\bibfnamefont {L.}~\bibnamefont {Jiang}},\
  }\bibfield  {title} {\enquote {\bibinfo {title} {Topological properties of
  linear circuit lattices},}\ }\href {\doibase 10.1103/PhysRevLett.114.173902}
  {\bibfield  {journal} {\bibinfo  {journal} {Phys. Rev. Lett.}\ }\textbf
  {\bibinfo {volume} {114}},\ \bibinfo {pages} {173902} (\bibinfo {year}
  {2015})}\BibitemShut {NoStop}%
\bibitem [{\citenamefont {Rosenthal}\ \emph {et~al.}(2018)\citenamefont
  {Rosenthal}, \citenamefont {Ehrlich}, \citenamefont {Rudner}, \citenamefont
  {Higginbotham},\ and\ \citenamefont {Lehnert}}]{Rosenthal2018}%
  \BibitemOpen
  \bibfield  {author} {\bibinfo {author} {\bibfnamefont {E.~I.}\ \bibnamefont
  {Rosenthal}}, \bibinfo {author} {\bibfnamefont {N.~K.}\ \bibnamefont
  {Ehrlich}}, \bibinfo {author} {\bibfnamefont {M.~S.}\ \bibnamefont {Rudner}},
  \bibinfo {author} {\bibfnamefont {A.~P.}\ \bibnamefont {Higginbotham}}, \
  and\ \bibinfo {author} {\bibfnamefont {K.~W.}\ \bibnamefont {Lehnert}},\
  }\bibfield  {title} {\enquote {\bibinfo {title} {Topological phase transition
  measured in a dissipative metamaterial},}\ }\href {\doibase
  10.1103/PhysRevB.97.220301} {\bibfield  {journal} {\bibinfo  {journal} {Phys.
  Rev. B}\ }\textbf {\bibinfo {volume} {97}},\ \bibinfo {pages} {220301}
  (\bibinfo {year} {2018})}\BibitemShut {NoStop}%
\bibitem [{\citenamefont {Lee}\ \emph {et~al.}(2018)\citenamefont {Lee},
  \citenamefont {Imhof}, \citenamefont {Berger}, \citenamefont {Bayer},
  \citenamefont {Brehm}, \citenamefont {Molenkamp}, \citenamefont {Kiessling},\
  and\ \citenamefont {Thomale}}]{Lee2018}%
  \BibitemOpen
  \bibfield  {author} {\bibinfo {author} {\bibfnamefont {C.~H.}\ \bibnamefont
  {Lee}}, \bibinfo {author} {\bibfnamefont {S.}~\bibnamefont {Imhof}}, \bibinfo
  {author} {\bibfnamefont {C.}~\bibnamefont {Berger}}, \bibinfo {author}
  {\bibfnamefont {F.}~\bibnamefont {Bayer}}, \bibinfo {author} {\bibfnamefont
  {J.}~\bibnamefont {Brehm}}, \bibinfo {author} {\bibfnamefont {L.~W.}\
  \bibnamefont {Molenkamp}}, \bibinfo {author} {\bibfnamefont {T.}~\bibnamefont
  {Kiessling}}, \ and\ \bibinfo {author} {\bibfnamefont {R.}~\bibnamefont
  {Thomale}},\ }\bibfield  {title} {\enquote {\bibinfo {title} {Topolectrical
  circuits},}\ }\href {\doibase 10.1038/s42005-018-0035-2} {\bibfield
  {journal} {\bibinfo  {journal} {Communications Physics}\ }\textbf {\bibinfo
  {volume} {1}},\ \bibinfo {pages} {39} (\bibinfo {year} {2018})}\BibitemShut
  {NoStop}%
\bibitem [{\citenamefont {Imhof}\ \emph {et~al.}(2018)\citenamefont {Imhof},
  \citenamefont {Berger}, \citenamefont {Bayer}, \citenamefont {Brehm},
  \citenamefont {Molenkamp}, \citenamefont {Kiessling}, \citenamefont
  {Schindler}, \citenamefont {Lee}, \citenamefont {Greiter}, \citenamefont
  {Neupert},\ and\ \citenamefont {Thomale}}]{Imhof2018}%
  \BibitemOpen
  \bibfield  {author} {\bibinfo {author} {\bibfnamefont {S.}~\bibnamefont
  {Imhof}}, \bibinfo {author} {\bibfnamefont {C.}~\bibnamefont {Berger}},
  \bibinfo {author} {\bibfnamefont {F.}~\bibnamefont {Bayer}}, \bibinfo
  {author} {\bibfnamefont {J.}~\bibnamefont {Brehm}}, \bibinfo {author}
  {\bibfnamefont {L.~W.}\ \bibnamefont {Molenkamp}}, \bibinfo {author}
  {\bibfnamefont {T.}~\bibnamefont {Kiessling}}, \bibinfo {author}
  {\bibfnamefont {F.}~\bibnamefont {Schindler}}, \bibinfo {author}
  {\bibfnamefont {C.~H.}\ \bibnamefont {Lee}}, \bibinfo {author} {\bibfnamefont
  {M.}~\bibnamefont {Greiter}}, \bibinfo {author} {\bibfnamefont
  {T.}~\bibnamefont {Neupert}}, \ and\ \bibinfo {author} {\bibfnamefont
  {R.}~\bibnamefont {Thomale}},\ }\bibfield  {title} {\enquote {\bibinfo
  {title} {Topolectrical-circuit realization of topological corner modes},}\
  }\href {\doibase 10.1038/s41567-018-0246-1} {\bibfield  {journal} {\bibinfo
  {journal} {Nature Physics}\ }\textbf {\bibinfo {volume} {14}},\ \bibinfo
  {pages} {925--929} (\bibinfo {year} {2018})}\BibitemShut {NoStop}%
\bibitem [{\citenamefont {Li}\ \emph {et~al.}(2018{\natexlab{b}})\citenamefont
  {Li}, \citenamefont {Sun}, \citenamefont {Zhu}, \citenamefont {Guo},
  \citenamefont {Jiang}, \citenamefont {Kariyado}, \citenamefont {Chen},\ and\
  \citenamefont {Hu}}]{Li2018b}%
  \BibitemOpen
  \bibfield  {author} {\bibinfo {author} {\bibfnamefont {Y.}~\bibnamefont
  {Li}}, \bibinfo {author} {\bibfnamefont {Y.}~\bibnamefont {Sun}}, \bibinfo
  {author} {\bibfnamefont {W.}~\bibnamefont {Zhu}}, \bibinfo {author}
  {\bibfnamefont {Z.}~\bibnamefont {Guo}}, \bibinfo {author} {\bibfnamefont
  {J.}~\bibnamefont {Jiang}}, \bibinfo {author} {\bibfnamefont
  {T.}~\bibnamefont {Kariyado}}, \bibinfo {author} {\bibfnamefont
  {H.}~\bibnamefont {Chen}}, \ and\ \bibinfo {author} {\bibfnamefont
  {X.}~\bibnamefont {Hu}},\ }\bibfield  {title} {\enquote {\bibinfo {title}
  {Topological lc-circuits based on microstrips and observation of
  electromagnetic modes with orbital angular momentum},}\ }\href {\doibase
  10.1038/s41467-018-07084-2} {\bibfield  {journal} {\bibinfo  {journal}
  {Nature Communications}\ }\textbf {\bibinfo {volume} {9}},\ \bibinfo {pages}
  {4598} (\bibinfo {year} {2018}{\natexlab{b}})}\BibitemShut {NoStop}%
\bibitem [{\citenamefont {Liu}\ \emph {et~al.}(2019)\citenamefont {Liu},
  \citenamefont {Gao}, \citenamefont {Zhang}, \citenamefont {Ma}, \citenamefont
  {Zhang}, \citenamefont {Liu}, \citenamefont {Xiang}, \citenamefont {Cui},\
  and\ \citenamefont {Zhang}}]{Liu2019}%
  \BibitemOpen
  \bibfield  {author} {\bibinfo {author} {\bibfnamefont {S.}~\bibnamefont
  {Liu}}, \bibinfo {author} {\bibfnamefont {W.}~\bibnamefont {Gao}}, \bibinfo
  {author} {\bibfnamefont {Q.}~\bibnamefont {Zhang}}, \bibinfo {author}
  {\bibfnamefont {S.}~\bibnamefont {Ma}}, \bibinfo {author} {\bibfnamefont
  {L.}~\bibnamefont {Zhang}}, \bibinfo {author} {\bibfnamefont
  {C.}~\bibnamefont {Liu}}, \bibinfo {author} {\bibfnamefont {Y.~J.}\
  \bibnamefont {Xiang}}, \bibinfo {author} {\bibfnamefont {T.~J.}\ \bibnamefont
  {Cui}}, \ and\ \bibinfo {author} {\bibfnamefont {S.}~\bibnamefont {Zhang}},\
  }\bibfield  {title} {\enquote {\bibinfo {title} {Topologically protected edge
  state in two-dimensional {S}u--{S}chrieffer--{H}eeger circuit},}\ }\href
  {\doibase 10.34133/2019/8609875} {\bibfield  {journal} {\bibinfo  {journal}
  {Research}\ }\textbf {\bibinfo {volume} {2019}},\ \bibinfo {pages} {8609875}
  (\bibinfo {year} {2019})}\BibitemShut {NoStop}%
\bibitem [{\citenamefont {Serra-Garcia}, \citenamefont {S\"usstrunk},\ and\
  \citenamefont {Huber}(2019)}]{SerraGarcia2019}%
  \BibitemOpen
  \bibfield  {author} {\bibinfo {author} {\bibfnamefont {M.}~\bibnamefont
  {Serra-Garcia}}, \bibinfo {author} {\bibfnamefont {R.}~\bibnamefont
  {S\"usstrunk}}, \ and\ \bibinfo {author} {\bibfnamefont {S.~D.}\ \bibnamefont
  {Huber}},\ }\bibfield  {title} {\enquote {\bibinfo {title} {Observation of
  quadrupole transitions and edge mode topology in an lc circuit network},}\
  }\href {\doibase 10.1103/PhysRevB.99.020304} {\bibfield  {journal} {\bibinfo
  {journal} {Phys. Rev. B}\ }\textbf {\bibinfo {volume} {99}},\ \bibinfo
  {pages} {020304} (\bibinfo {year} {2019})}\BibitemShut {NoStop}%
\bibitem [{\citenamefont {Zangeneh-Nejad}\ and\ \citenamefont
  {Fleury}(2019)}]{Zangeneh-Nejad2019}%
  \BibitemOpen
  \bibfield  {author} {\bibinfo {author} {\bibfnamefont {F.}~\bibnamefont
  {Zangeneh-Nejad}}\ and\ \bibinfo {author} {\bibfnamefont {R.}~\bibnamefont
  {Fleury}},\ }\bibfield  {title} {\enquote {\bibinfo {title} {Nonlinear
  second-order topological insulators},}\ }\href {\doibase
  10.1103/PhysRevLett.123.053902} {\bibfield  {journal} {\bibinfo  {journal}
  {Phys. Rev. Lett.}\ }\textbf {\bibinfo {volume} {123}},\ \bibinfo {pages}
  {053902} (\bibinfo {year} {2019})}\BibitemShut {NoStop}%
\bibitem [{\citenamefont {Hofstadter}(1976)}]{Hofstadter1976}%
  \BibitemOpen
  \bibfield  {author} {\bibinfo {author} {\bibfnamefont {D.~R.}\ \bibnamefont
  {Hofstadter}},\ }\bibfield  {title} {\enquote {\bibinfo {title} {Energy
  levels and wave functions of bloch electrons in rational and irrational
  magnetic fields},}\ }\href {\doibase 10.1103/PhysRevB.14.2239} {\bibfield
  {journal} {\bibinfo  {journal} {Phys. Rev. B}\ }\textbf {\bibinfo {volume}
  {14}},\ \bibinfo {pages} {2239--2249} (\bibinfo {year} {1976})}\BibitemShut
  {NoStop}%
\bibitem [{\citenamefont {Vool}\ and\ \citenamefont
  {Devoret}(2017)}]{Vool2017}%
  \BibitemOpen
  \bibfield  {author} {\bibinfo {author} {\bibfnamefont {U.}~\bibnamefont
  {Vool}}\ and\ \bibinfo {author} {\bibfnamefont {M.}~\bibnamefont {Devoret}},\
  }\bibfield  {title} {\enquote {\bibinfo {title} {Introduction to quantum
  electromagnetic circuits},}\ }\href {\doibase 10.1002/cta.2359} {\bibfield
  {journal} {\bibinfo  {journal} {International Journal of Circuit Theory and
  Applications}\ }\textbf {\bibinfo {volume} {45}},\ \bibinfo {pages}
  {897--934} (\bibinfo {year} {2017})}\BibitemShut {NoStop}%
\bibitem [{\citenamefont {Landauer}(1960)}]{NLTL}%
  \BibitemOpen
  \bibfield  {author} {\bibinfo {author} {\bibfnamefont {R.}~\bibnamefont
  {Landauer}},\ }\bibfield  {title} {\enquote {\bibinfo {title} {Parametric
  amplification along nonlinear transmission lines},}\ }\href {\doibase
  10.1063/1.1735612} {\bibfield  {journal} {\bibinfo  {journal} {Journal of
  Applied Physics}\ }\textbf {\bibinfo {volume} {31}},\ \bibinfo {pages}
  {479--484} (\bibinfo {year} {1960})}\BibitemShut {NoStop}%
\bibitem [{\citenamefont {Ezawa}(2018)}]{Ezawa2018}%
  \BibitemOpen
  \bibfield  {author} {\bibinfo {author} {\bibfnamefont {M.}~\bibnamefont
  {Ezawa}},\ }\bibfield  {title} {\enquote {\bibinfo {title} {Higher-order
  topological insulators and semimetals on the breathing kagome and pyrochlore
  lattices},}\ }\href {\doibase 10.1103/PhysRevLett.120.026801} {\bibfield
  {journal} {\bibinfo  {journal} {Phys. Rev. Lett.}\ }\textbf {\bibinfo
  {volume} {120}},\ \bibinfo {pages} {026801} (\bibinfo {year}
  {2018})}\BibitemShut {NoStop}%
\bibitem [{\citenamefont {Kotwal}\ \emph {et~al.}(2019)\citenamefont {Kotwal},
  \citenamefont {Ronellenfitsch}, \citenamefont {Moseley}, \citenamefont
  {Stegmaier}, \citenamefont {Thomale},\ and\ \citenamefont
  {Dunkel}}]{Kotwal2019}%
  \BibitemOpen
  \bibfield  {author} {\bibinfo {author} {\bibfnamefont {T.}~\bibnamefont
  {Kotwal}}, \bibinfo {author} {\bibfnamefont {H.}~\bibnamefont
  {Ronellenfitsch}}, \bibinfo {author} {\bibfnamefont {F.}~\bibnamefont
  {Moseley}}, \bibinfo {author} {\bibfnamefont {A.}~\bibnamefont {Stegmaier}},
  \bibinfo {author} {\bibfnamefont {R.}~\bibnamefont {Thomale}}, \ and\
  \bibinfo {author} {\bibfnamefont {J.}~\bibnamefont {Dunkel}},\ }\bibfield
  {title} {\enquote {\bibinfo {title} {Active topolectrical circuits},}\ }\href
  {https://arxiv.org/abs/1903.10130} {\bibfield  {journal} {\bibinfo  {journal}
  {arXiv:1903.10130}\ } (\bibinfo {year} {2019})}\BibitemShut {NoStop}%
\bibitem [{\citenamefont {Peano}\ \emph
  {et~al.}(2016{\natexlab{a}})\citenamefont {Peano}, \citenamefont {Houde},
  \citenamefont {Marquardt},\ and\ \citenamefont {Clerk}}]{Peano2016}%
  \BibitemOpen
  \bibfield  {author} {\bibinfo {author} {\bibfnamefont {V.}~\bibnamefont
  {Peano}}, \bibinfo {author} {\bibfnamefont {M.}~\bibnamefont {Houde}},
  \bibinfo {author} {\bibfnamefont {F.}~\bibnamefont {Marquardt}}, \ and\
  \bibinfo {author} {\bibfnamefont {A.~A.}\ \bibnamefont {Clerk}},\ }\bibfield
  {title} {\enquote {\bibinfo {title} {Topological quantum fluctuations and
  traveling wave amplifiers},}\ }\href {\doibase 10.1103/PhysRevX.6.041026}
  {\bibfield  {journal} {\bibinfo  {journal} {Phys. Rev. X}\ }\textbf {\bibinfo
  {volume} {6}},\ \bibinfo {pages} {041026} (\bibinfo {year}
  {2016}{\natexlab{a}})}\BibitemShut {NoStop}%
\bibitem [{\citenamefont {St-Jean}\ \emph {et~al.}(2017)\citenamefont
  {St-Jean}, \citenamefont {Goblot}, \citenamefont {Galopin}, \citenamefont
  {Lema{\^{\i}}tre}, \citenamefont {Ozawa}, \citenamefont {Gratiet},
  \citenamefont {Sagnes}, \citenamefont {Bloch},\ and\ \citenamefont
  {Amo}}]{StJean2017}%
  \BibitemOpen
  \bibfield  {author} {\bibinfo {author} {\bibfnamefont {P.}~\bibnamefont
  {St-Jean}}, \bibinfo {author} {\bibfnamefont {V.}~\bibnamefont {Goblot}},
  \bibinfo {author} {\bibfnamefont {E.}~\bibnamefont {Galopin}}, \bibinfo
  {author} {\bibfnamefont {A.}~\bibnamefont {Lema{\^{\i}}tre}}, \bibinfo
  {author} {\bibfnamefont {T.}~\bibnamefont {Ozawa}}, \bibinfo {author}
  {\bibfnamefont {L.~L.}\ \bibnamefont {Gratiet}}, \bibinfo {author}
  {\bibfnamefont {I.}~\bibnamefont {Sagnes}}, \bibinfo {author} {\bibfnamefont
  {J.}~\bibnamefont {Bloch}}, \ and\ \bibinfo {author} {\bibfnamefont
  {A.}~\bibnamefont {Amo}},\ }\bibfield  {title} {\enquote {\bibinfo {title}
  {Lasing in topological edge states of a one-dimensional lattice},}\ }\href
  {\doibase 10.1038/s41566-017-0006-2} {\bibfield  {journal} {\bibinfo
  {journal} {Nature Photonics}\ }\textbf {\bibinfo {volume} {11}},\ \bibinfo
  {pages} {651} (\bibinfo {year} {2017})}\BibitemShut {NoStop}%
\bibitem [{\citenamefont {Parto}\ \emph {et~al.}(2018)\citenamefont {Parto},
  \citenamefont {Wittek}, \citenamefont {Hodaei}, \citenamefont {Harari},
  \citenamefont {Bandres}, \citenamefont {Ren}, \citenamefont {Rechtsman},
  \citenamefont {Segev}, \citenamefont {Christodoulides},\ and\ \citenamefont
  {Khajavikhan}}]{Parto2018}%
  \BibitemOpen
  \bibfield  {author} {\bibinfo {author} {\bibfnamefont {M.}~\bibnamefont
  {Parto}}, \bibinfo {author} {\bibfnamefont {S.}~\bibnamefont {Wittek}},
  \bibinfo {author} {\bibfnamefont {H.}~\bibnamefont {Hodaei}}, \bibinfo
  {author} {\bibfnamefont {G.}~\bibnamefont {Harari}}, \bibinfo {author}
  {\bibfnamefont {M.~A.}\ \bibnamefont {Bandres}}, \bibinfo {author}
  {\bibfnamefont {J.}~\bibnamefont {Ren}}, \bibinfo {author} {\bibfnamefont
  {M.~C.}\ \bibnamefont {Rechtsman}}, \bibinfo {author} {\bibfnamefont
  {M.}~\bibnamefont {Segev}}, \bibinfo {author} {\bibfnamefont {D.~N.}\
  \bibnamefont {Christodoulides}}, \ and\ \bibinfo {author} {\bibfnamefont
  {M.}~\bibnamefont {Khajavikhan}},\ }\bibfield  {title} {\enquote {\bibinfo
  {title} {Edge-mode lasing in 1{D} topological active arrays},}\ }\href
  {\doibase 10.1103/physrevlett.120.113901} {\bibfield  {journal} {\bibinfo
  {journal} {Phys. Rev. Lett.}\ }\textbf {\bibinfo {volume} {120}},\ \bibinfo
  {pages} {113901} (\bibinfo {year} {2018})}\BibitemShut {NoStop}%
\bibitem [{\citenamefont {Zhao}\ \emph {et~al.}(2018)\citenamefont {Zhao},
  \citenamefont {Miao}, \citenamefont {Teimourpour}, \citenamefont {Malzard},
  \citenamefont {El-Ganainy}, \citenamefont {Schomerus},\ and\ \citenamefont
  {Feng}}]{Zhao2018}%
  \BibitemOpen
  \bibfield  {author} {\bibinfo {author} {\bibfnamefont {H.}~\bibnamefont
  {Zhao}}, \bibinfo {author} {\bibfnamefont {P.}~\bibnamefont {Miao}}, \bibinfo
  {author} {\bibfnamefont {M.~H.}\ \bibnamefont {Teimourpour}}, \bibinfo
  {author} {\bibfnamefont {S.}~\bibnamefont {Malzard}}, \bibinfo {author}
  {\bibfnamefont {R.}~\bibnamefont {El-Ganainy}}, \bibinfo {author}
  {\bibfnamefont {H.}~\bibnamefont {Schomerus}}, \ and\ \bibinfo {author}
  {\bibfnamefont {L.}~\bibnamefont {Feng}},\ }\bibfield  {title} {\enquote
  {\bibinfo {title} {Topological hybrid silicon microlasers},}\ }\href
  {\doibase 10.1038/s41467-018-03434-2} {\bibfield  {journal} {\bibinfo
  {journal} {Nature Communications}\ }\textbf {\bibinfo {volume} {9}},\
  \bibinfo {pages} {981} (\bibinfo {year} {2018})}\BibitemShut {NoStop}%
\bibitem [{\citenamefont {Han}\ \emph {et~al.}(2019)\citenamefont {Han},
  \citenamefont {Lee}, \citenamefont {Callard}, \citenamefont {Seassal},\ and\
  \citenamefont {Jeon}}]{Han2019}%
  \BibitemOpen
  \bibfield  {author} {\bibinfo {author} {\bibfnamefont {C.}~\bibnamefont
  {Han}}, \bibinfo {author} {\bibfnamefont {M.}~\bibnamefont {Lee}}, \bibinfo
  {author} {\bibfnamefont {S.}~\bibnamefont {Callard}}, \bibinfo {author}
  {\bibfnamefont {C.}~\bibnamefont {Seassal}}, \ and\ \bibinfo {author}
  {\bibfnamefont {H.}~\bibnamefont {Jeon}},\ }\bibfield  {title} {\enquote
  {\bibinfo {title} {Lasing at topological edge states in a photonic crystal l3
  nanocavity dimer array},}\ }\href {\doibase 10.1038/s41377-019-0149-7}
  {\bibfield  {journal} {\bibinfo  {journal} {Light: Science \& Applications}\
  }\textbf {\bibinfo {volume} {8}},\ \bibinfo {pages} {40} (\bibinfo {year}
  {2019})}\BibitemShut {NoStop}%
\bibitem [{\citenamefont {Klembt}\ \emph {et~al.}(2018)\citenamefont {Klembt},
  \citenamefont {Harder}, \citenamefont {Egorov}, \citenamefont {Winkler},
  \citenamefont {Ge}, \citenamefont {Bandres}, \citenamefont {Emmerling},
  \citenamefont {Worschech}, \citenamefont {Liew}, \citenamefont {Segev},
  \citenamefont {Schneider},\ and\ \citenamefont {H\"ofling}}]{Klembt2018}%
  \BibitemOpen
  \bibfield  {author} {\bibinfo {author} {\bibfnamefont {S.}~\bibnamefont
  {Klembt}}, \bibinfo {author} {\bibfnamefont {T.~H.}\ \bibnamefont {Harder}},
  \bibinfo {author} {\bibfnamefont {O.~A.}\ \bibnamefont {Egorov}}, \bibinfo
  {author} {\bibfnamefont {K.}~\bibnamefont {Winkler}}, \bibinfo {author}
  {\bibfnamefont {R.}~\bibnamefont {Ge}}, \bibinfo {author} {\bibfnamefont
  {M.~A.}\ \bibnamefont {Bandres}}, \bibinfo {author} {\bibfnamefont
  {M.}~\bibnamefont {Emmerling}}, \bibinfo {author} {\bibfnamefont
  {L.}~\bibnamefont {Worschech}}, \bibinfo {author} {\bibfnamefont {T.~C.~H.}\
  \bibnamefont {Liew}}, \bibinfo {author} {\bibfnamefont {M.}~\bibnamefont
  {Segev}}, \bibinfo {author} {\bibfnamefont {C.}~\bibnamefont {Schneider}}, \
  and\ \bibinfo {author} {\bibfnamefont {S.}~\bibnamefont {H\"ofling}},\
  }\bibfield  {title} {\enquote {\bibinfo {title} {Exciton-polariton
  topological insulator},}\ }\href {\doibase 10.1038/s41586-018-0601-5}
  {\bibfield  {journal} {\bibinfo  {journal} {Nature}\ }\textbf {\bibinfo
  {volume} {562}},\ \bibinfo {pages} {552} (\bibinfo {year}
  {2018})}\BibitemShut {NoStop}%
\bibitem [{\citenamefont {Ohtsubo}(2007)}]{laser_book}%
  \BibitemOpen
  \bibfield  {author} {\bibinfo {author} {\bibfnamefont {J.}~\bibnamefont
  {Ohtsubo}},\ }\href@noop {} {\emph {\bibinfo {title} {Semiconductor Lasers:
  Stability, Instability, and Chaos}}}\ (\bibinfo  {publisher} {Springer},\
  \bibinfo {year} {2007})\BibitemShut {NoStop}%
\bibitem [{\citenamefont {Longhi}, \citenamefont {Kominis},\ and\ \citenamefont
  {Kovanis}(2018)}]{longhi2018b}%
  \BibitemOpen
  \bibfield  {author} {\bibinfo {author} {\bibfnamefont {S.}~\bibnamefont
  {Longhi}}, \bibinfo {author} {\bibfnamefont {Y.}~\bibnamefont {Kominis}}, \
  and\ \bibinfo {author} {\bibfnamefont {V.}~\bibnamefont {Kovanis}},\
  }\bibfield  {title} {\enquote {\bibinfo {title} {Presence of temporal
  dynamical instabilities in topological insulator lasers},}\ }\href {\doibase
  10.1209/0295-5075/122/14004} {\bibfield  {journal} {\bibinfo  {journal}
  {{EPL} (Europhysics Letters)}\ }\textbf {\bibinfo {volume} {122}},\ \bibinfo
  {pages} {14004} (\bibinfo {year} {2018})}\BibitemShut {NoStop}%
\bibitem [{\citenamefont {Longhi}\ and\ \citenamefont
  {Feng}(2018)}]{longhi2018}%
  \BibitemOpen
  \bibfield  {author} {\bibinfo {author} {\bibfnamefont {S.}~\bibnamefont
  {Longhi}}\ and\ \bibinfo {author} {\bibfnamefont {L.}~\bibnamefont {Feng}},\
  }\bibfield  {title} {\enquote {\bibinfo {title} {Invited article: Mitigation
  of dynamical instabilities in laser arrays via non-hermitian coupling},}\
  }\href {\doibase 10.1063/1.5028453} {\bibfield  {journal} {\bibinfo
  {journal} {{APL} Photonics}\ }\textbf {\bibinfo {volume} {3}},\ \bibinfo
  {pages} {060802} (\bibinfo {year} {2018})}\BibitemShut {NoStop}%
\bibitem [{\citenamefont {Schomerus}(2013)}]{Schomerus2013OL}%
  \BibitemOpen
  \bibfield  {author} {\bibinfo {author} {\bibfnamefont {H.}~\bibnamefont
  {Schomerus}},\ }\bibfield  {title} {\enquote {\bibinfo {title} {Topologically
  protected midgap states in complex photonic lattices},}\ }\href {\doibase
  10.1364/ol.38.001912} {\bibfield  {journal} {\bibinfo  {journal} {Optics
  Letters}\ }\textbf {\bibinfo {volume} {38}},\ \bibinfo {pages} {1912}
  (\bibinfo {year} {2013})}\BibitemShut {NoStop}%
\bibitem [{\citenamefont {Weimann}\ \emph {et~al.}(2016)\citenamefont
  {Weimann}, \citenamefont {Kremer}, \citenamefont {Plotnik}, \citenamefont
  {Lumer}, \citenamefont {Nolte}, \citenamefont {Makris}, \citenamefont
  {Segev}, \citenamefont {Rechtsman},\ and\ \citenamefont
  {Szameit}}]{Weimann2016}%
  \BibitemOpen
  \bibfield  {author} {\bibinfo {author} {\bibfnamefont {S.}~\bibnamefont
  {Weimann}}, \bibinfo {author} {\bibfnamefont {M.}~\bibnamefont {Kremer}},
  \bibinfo {author} {\bibfnamefont {Y.}~\bibnamefont {Plotnik}}, \bibinfo
  {author} {\bibfnamefont {Y.}~\bibnamefont {Lumer}}, \bibinfo {author}
  {\bibfnamefont {S.}~\bibnamefont {Nolte}}, \bibinfo {author} {\bibfnamefont
  {K.~G.}\ \bibnamefont {Makris}}, \bibinfo {author} {\bibfnamefont
  {M.}~\bibnamefont {Segev}}, \bibinfo {author} {\bibfnamefont {M.~.~C.}\
  \bibnamefont {Rechtsman}}, \ and\ \bibinfo {author} {\bibfnamefont
  {A.}~\bibnamefont {Szameit}},\ }\bibfield  {title} {\enquote {\bibinfo
  {title} {Topologically protected bound states in photonic
  parity-time-symmetric crystals},}\ }\href {https://doi.org/10.1038/nmat4811}
  {\bibfield  {journal} {\bibinfo  {journal} {Nature Materials}\ }\textbf
  {\bibinfo {volume} {16}},\ \bibinfo {pages} {433} (\bibinfo {year}
  {2016})}\BibitemShut {NoStop}%
\bibitem [{\citenamefont {Malzard}\ and\ \citenamefont
  {Schomerus}(2018)}]{Malzard2018b}%
  \BibitemOpen
  \bibfield  {author} {\bibinfo {author} {\bibfnamefont {S.}~\bibnamefont
  {Malzard}}\ and\ \bibinfo {author} {\bibfnamefont {H.}~\bibnamefont
  {Schomerus}},\ }\bibfield  {title} {\enquote {\bibinfo {title} {Nonlinear
  mode competition and symmetry-protected power oscillations in topological
  lasers},}\ }\href {\doibase 10.1088/1367-2630/aac9e0} {\bibfield  {journal}
  {\bibinfo  {journal} {New Journal of Physics}\ }\textbf {\bibinfo {volume}
  {20}},\ \bibinfo {pages} {063044} (\bibinfo {year} {2018})}\BibitemShut
  {NoStop}%
\bibitem [{\citenamefont {Malzard}, \citenamefont {Cancellieri},\ and\
  \citenamefont {Schomerus}(2018)}]{Malzard2018}%
  \BibitemOpen
  \bibfield  {author} {\bibinfo {author} {\bibfnamefont {S.}~\bibnamefont
  {Malzard}}, \bibinfo {author} {\bibfnamefont {E.}~\bibnamefont
  {Cancellieri}}, \ and\ \bibinfo {author} {\bibfnamefont {H.}~\bibnamefont
  {Schomerus}},\ }\bibfield  {title} {\enquote {\bibinfo {title} {Topological
  dynamics and excitations in lasers and condensates with saturable gain or
  loss},}\ }\href {\doibase 10.1364/OE.26.022506} {\bibfield  {journal}
  {\bibinfo  {journal} {Opt. Express}\ }\textbf {\bibinfo {volume} {26}},\
  \bibinfo {pages} {22506--22518} (\bibinfo {year} {2018})}\BibitemShut
  {NoStop}%
\bibitem [{\citenamefont {Cancellieri}\ and\ \citenamefont
  {Schomerus}(2019)}]{Cancellieri2019}%
  \BibitemOpen
  \bibfield  {author} {\bibinfo {author} {\bibfnamefont {E.}~\bibnamefont
  {Cancellieri}}\ and\ \bibinfo {author} {\bibfnamefont {H.}~\bibnamefont
  {Schomerus}},\ }\bibfield  {title} {\enquote {\bibinfo {title}
  {$\mathcal{PC}$-symmetry-protected edge states in interacting
  driven-dissipative bosonic systems},}\ }\href {\doibase
  10.1103/PhysRevA.99.033801} {\bibfield  {journal} {\bibinfo  {journal} {Phys.
  Rev. A}\ }\textbf {\bibinfo {volume} {99}},\ \bibinfo {pages} {033801}
  (\bibinfo {year} {2019})}\BibitemShut {NoStop}%
\bibitem [{\citenamefont {Longhi}(2018)}]{longhi_supermode}%
  \BibitemOpen
  \bibfield  {author} {\bibinfo {author} {\bibfnamefont {S.}~\bibnamefont
  {Longhi}},\ }\bibfield  {title} {\enquote {\bibinfo {title} {Non-hermitian
  gauged topological laser arrays},}\ }\href {\doibase 10.1002/andp.201800023}
  {\bibfield  {journal} {\bibinfo  {journal} {Annalen der Physik}\ }\textbf
  {\bibinfo {volume} {530}},\ \bibinfo {pages} {1800023} (\bibinfo {year}
  {2018})}\BibitemShut {NoStop}%
\bibitem [{\citenamefont {Gong}\ \emph {et~al.}(2018)\citenamefont {Gong},
  \citenamefont {Ashida}, \citenamefont {Kawabata}, \citenamefont {Takasan},
  \citenamefont {Higashikawa},\ and\ \citenamefont {Ueda}}]{Gong2018}%
  \BibitemOpen
  \bibfield  {author} {\bibinfo {author} {\bibfnamefont {Z.}~\bibnamefont
  {Gong}}, \bibinfo {author} {\bibfnamefont {Y.}~\bibnamefont {Ashida}},
  \bibinfo {author} {\bibfnamefont {K.}~\bibnamefont {Kawabata}}, \bibinfo
  {author} {\bibfnamefont {K.}~\bibnamefont {Takasan}}, \bibinfo {author}
  {\bibfnamefont {S.}~\bibnamefont {Higashikawa}}, \ and\ \bibinfo {author}
  {\bibfnamefont {M.}~\bibnamefont {Ueda}},\ }\bibfield  {title} {\enquote
  {\bibinfo {title} {Topological phases of non-hermitian systems},}\ }\href
  {\doibase 10.1103/PhysRevX.8.031079} {\bibfield  {journal} {\bibinfo
  {journal} {Phys. Rev. X}\ }\textbf {\bibinfo {volume} {8}},\ \bibinfo {pages}
  {031079} (\bibinfo {year} {2018})}\BibitemShut {NoStop}%
\bibitem [{\citenamefont {Longhi}(2019)}]{Longhi2019}%
  \BibitemOpen
  \bibfield  {author} {\bibinfo {author} {\bibfnamefont {S.}~\bibnamefont
  {Longhi}},\ }\bibfield  {title} {\enquote {\bibinfo {title} {Non-hermitian
  topological phase transition in pt-symmetric mode-locked lasers},}\ }\href
  {\doibase 10.1364/OL.44.001190} {\bibfield  {journal} {\bibinfo  {journal}
  {Opt. Lett.}\ }\textbf {\bibinfo {volume} {44}},\ \bibinfo {pages}
  {1190--1193} (\bibinfo {year} {2019})}\BibitemShut {NoStop}%
\bibitem [{\citenamefont {Bahari}\ \emph {et~al.}(2019)\citenamefont {Bahari},
  \citenamefont {Hsu}, \citenamefont {Pan}, \citenamefont {Preece},
  \citenamefont {Ndao}, \citenamefont {Amili}, \citenamefont {Fainman},\ and\
  \citenamefont {Kanté}}]{Bahari_arxiv}%
  \BibitemOpen
  \bibfield  {author} {\bibinfo {author} {\bibfnamefont {B.}~\bibnamefont
  {Bahari}}, \bibinfo {author} {\bibfnamefont {L.-Y.}\ \bibnamefont {Hsu}},
  \bibinfo {author} {\bibfnamefont {S.~H.}\ \bibnamefont {Pan}}, \bibinfo
  {author} {\bibfnamefont {D.}~\bibnamefont {Preece}}, \bibinfo {author}
  {\bibfnamefont {A.}~\bibnamefont {Ndao}}, \bibinfo {author} {\bibfnamefont
  {A.~E.}\ \bibnamefont {Amili}}, \bibinfo {author} {\bibfnamefont
  {Y.}~\bibnamefont {Fainman}}, \ and\ \bibinfo {author} {\bibfnamefont
  {B.}~\bibnamefont {Kanté}},\ }\bibfield  {title} {\enquote {\bibinfo {title}
  {Topological lasers generating and multiplexing topological light},}\ }\href
  {https://arxiv.org/abs/1904.11873} {\bibfield  {journal} {\bibinfo  {journal}
  {arXiv}\ ,\ \bibinfo {pages} {1904.11873}} (\bibinfo {year}
  {2019})}\BibitemShut {NoStop}%
\bibitem [{\citenamefont {Harari}\ \emph {et~al.}(2018)\citenamefont {Harari},
  \citenamefont {Bandres}, \citenamefont {Lumer}, \citenamefont {Rechtsman},
  \citenamefont {Chong}, \citenamefont {Khajavikhan}, \citenamefont
  {Christodoulides},\ and\ \citenamefont {Segev}}]{Harari2018}%
  \BibitemOpen
  \bibfield  {author} {\bibinfo {author} {\bibfnamefont {G.}~\bibnamefont
  {Harari}}, \bibinfo {author} {\bibfnamefont {M.~A.}\ \bibnamefont {Bandres}},
  \bibinfo {author} {\bibfnamefont {Y.}~\bibnamefont {Lumer}}, \bibinfo
  {author} {\bibfnamefont {M.~C.}\ \bibnamefont {Rechtsman}}, \bibinfo {author}
  {\bibfnamefont {Y.~D.}\ \bibnamefont {Chong}}, \bibinfo {author}
  {\bibfnamefont {M.}~\bibnamefont {Khajavikhan}}, \bibinfo {author}
  {\bibfnamefont {D.~N.}\ \bibnamefont {Christodoulides}}, \ and\ \bibinfo
  {author} {\bibfnamefont {M.}~\bibnamefont {Segev}},\ }\bibfield  {title}
  {\enquote {\bibinfo {title} {Topological insulator laser: Theory},}\ }\href
  {\doibase 10.1126/science.aar4003} {\bibfield  {journal} {\bibinfo  {journal}
  {Science}\ }\textbf {\bibinfo {volume} {359}},\ \bibinfo {pages} {eaar4003}
  (\bibinfo {year} {2018})}\BibitemShut {NoStop}%
\bibitem [{\citenamefont {Kartashov}\ and\ \citenamefont
  {Skryabin}(2019)}]{KartashovLaser}%
  \BibitemOpen
  \bibfield  {author} {\bibinfo {author} {\bibfnamefont {Y.~V.}\ \bibnamefont
  {Kartashov}}\ and\ \bibinfo {author} {\bibfnamefont {D.~V.}\ \bibnamefont
  {Skryabin}},\ }\bibfield  {title} {\enquote {\bibinfo {title}
  {Two-dimensional topological polariton laser},}\ }\href {\doibase
  10.1103/PhysRevLett.122.083902} {\bibfield  {journal} {\bibinfo  {journal}
  {Phys. Rev. Lett.}\ }\textbf {\bibinfo {volume} {122}},\ \bibinfo {pages}
  {083902} (\bibinfo {year} {2019})}\BibitemShut {NoStop}%
\bibitem [{\citenamefont {Secli}, \citenamefont {Capone},\ and\ \citenamefont
  {Carusotto}(2019)}]{Secli2019}%
  \BibitemOpen
  \bibfield  {author} {\bibinfo {author} {\bibfnamefont {M.}~\bibnamefont
  {Secli}}, \bibinfo {author} {\bibfnamefont {M.}~\bibnamefont {Capone}}, \
  and\ \bibinfo {author} {\bibfnamefont {I.}~\bibnamefont {Carusotto}},\
  }\bibfield  {title} {\enquote {\bibinfo {title} {Theory of chiral edge state
  lasing in a two-dimensional topological system},}\ }\href
  {https://arxiv.org/abs/1901.01290} {\bibfield  {journal} {\bibinfo  {journal}
  {arXiv}\ ,\ \bibinfo {pages} {1901.01290}} (\bibinfo {year}
  {2019})}\BibitemShut {NoStop}%
\bibitem [{\citenamefont {Guo}\ \emph {et~al.}(2019)\citenamefont {Guo},
  \citenamefont {Ne\ifmmode~\check{c}\else \v{c}\fi{}ada}, \citenamefont
  {Hakala}, \citenamefont {V\"akev\"ainen},\ and\ \citenamefont
  {T\"orm\"a}}]{Guo2019}%
  \BibitemOpen
  \bibfield  {author} {\bibinfo {author} {\bibfnamefont {R.}~\bibnamefont
  {Guo}}, \bibinfo {author} {\bibfnamefont {M.}~\bibnamefont
  {Ne\ifmmode~\check{c}\else \v{c}\fi{}ada}}, \bibinfo {author} {\bibfnamefont
  {T.~K.}\ \bibnamefont {Hakala}}, \bibinfo {author} {\bibfnamefont {A.~I.}\
  \bibnamefont {V\"akev\"ainen}}, \ and\ \bibinfo {author} {\bibfnamefont
  {P.}~\bibnamefont {T\"orm\"a}},\ }\bibfield  {title} {\enquote {\bibinfo
  {title} {Lasing at $k$ points of a honeycomb plasmonic lattice},}\ }\href
  {\doibase 10.1103/PhysRevLett.122.013901} {\bibfield  {journal} {\bibinfo
  {journal} {Phys. Rev. Lett.}\ }\textbf {\bibinfo {volume} {122}},\ \bibinfo
  {pages} {013901} (\bibinfo {year} {2019})}\BibitemShut {NoStop}%
\bibitem [{\citenamefont {Pourjamal}\ \emph {et~al.}(2019)\citenamefont
  {Pourjamal}, \citenamefont {Hakala}, \citenamefont {Nečada}, \citenamefont
  {Freire-Fernández}, \citenamefont {Kataja}, \citenamefont {Rekola},
  \citenamefont {Martikainen}, \citenamefont {Törmä},\ and\ \citenamefont
  {van Dijken}}]{Pourjamal2019}%
  \BibitemOpen
  \bibfield  {author} {\bibinfo {author} {\bibfnamefont {S.}~\bibnamefont
  {Pourjamal}}, \bibinfo {author} {\bibfnamefont {T.~K.}\ \bibnamefont
  {Hakala}}, \bibinfo {author} {\bibfnamefont {M.}~\bibnamefont {Nečada}},
  \bibinfo {author} {\bibfnamefont {F.}~\bibnamefont {Freire-Fernández}},
  \bibinfo {author} {\bibfnamefont {M.}~\bibnamefont {Kataja}}, \bibinfo
  {author} {\bibfnamefont {H.}~\bibnamefont {Rekola}}, \bibinfo {author}
  {\bibfnamefont {J.-P.}\ \bibnamefont {Martikainen}}, \bibinfo {author}
  {\bibfnamefont {P.}~\bibnamefont {Törmä}}, \ and\ \bibinfo {author}
  {\bibfnamefont {S.}~\bibnamefont {van Dijken}},\ }\bibfield  {title}
  {\enquote {\bibinfo {title} {Lasing in ni nanodisk arrays},}\ }\href
  {\doibase 10.1021/acsnano.9b01006} {\bibfield  {journal} {\bibinfo  {journal}
  {ACS Nano}\ }\textbf {\bibinfo {volume} {13}},\ \bibinfo {pages} {5686--5692}
  (\bibinfo {year} {2019})}\BibitemShut {NoStop}%
\bibitem [{\citenamefont {Pilozzi}\ and\ \citenamefont
  {Conti}(2017)}]{Pilozzi2017}%
  \BibitemOpen
  \bibfield  {author} {\bibinfo {author} {\bibfnamefont {L.}~\bibnamefont
  {Pilozzi}}\ and\ \bibinfo {author} {\bibfnamefont {C.}~\bibnamefont
  {Conti}},\ }\bibfield  {title} {\enquote {\bibinfo {title} {Topological
  cascade laser for frequency comb generation in pt-symmetric structures},}\
  }\href {\doibase 10.1364/OL.42.005174} {\bibfield  {journal} {\bibinfo
  {journal} {Opt. Lett.}\ }\textbf {\bibinfo {volume} {42}},\ \bibinfo {pages}
  {5174--5177} (\bibinfo {year} {2017})}\BibitemShut {NoStop}%
\bibitem [{\citenamefont {Yang}\ \emph {et~al.}(2019)\citenamefont {Yang},
  \citenamefont {Lustig}, \citenamefont {Harari}, \citenamefont {Plotnik},
  \citenamefont {Bandres},\ and\ \citenamefont {Segev}}]{YangCLEO}%
  \BibitemOpen
  \bibfield  {author} {\bibinfo {author} {\bibfnamefont {Z.}~\bibnamefont
  {Yang}}, \bibinfo {author} {\bibfnamefont {E.}~\bibnamefont {Lustig}},
  \bibinfo {author} {\bibfnamefont {G.}~\bibnamefont {Harari}}, \bibinfo
  {author} {\bibfnamefont {Y.}~\bibnamefont {Plotnik}}, \bibinfo {author}
  {\bibfnamefont {M.}~\bibnamefont {Bandres}}, \ and\ \bibinfo {author}
  {\bibfnamefont {M.}~\bibnamefont {Segev}},\ }\bibfield  {title} {\enquote
  {\bibinfo {title} {Mode-locked topological laser in synthetic dimensions},}\
  }in\ \href {\doibase 10.1364/CLEO_QELS.2019.FW3D.2} {\emph {\bibinfo
  {booktitle} {Conference on Lasers and Electro-Optics}}}\ (\bibinfo
  {publisher} {Optical Society of America},\ \bibinfo {year} {2019})\ p.\
  \bibinfo {pages} {FW3D.2}\BibitemShut {NoStop}%
\bibitem [{\citenamefont {Suchomel}\ \emph {et~al.}(2018)\citenamefont
  {Suchomel}, \citenamefont {Klembt}, \citenamefont {Harder}, \citenamefont
  {Klaas}, \citenamefont {Egorov}, \citenamefont {Winkler}, \citenamefont
  {Emmerling}, \citenamefont {Thomale}, \citenamefont {H\"ofling},\ and\
  \citenamefont {Schneider}}]{Sucholmel2018}%
  \BibitemOpen
  \bibfield  {author} {\bibinfo {author} {\bibfnamefont {H.}~\bibnamefont
  {Suchomel}}, \bibinfo {author} {\bibfnamefont {S.}~\bibnamefont {Klembt}},
  \bibinfo {author} {\bibfnamefont {T.~H.}\ \bibnamefont {Harder}}, \bibinfo
  {author} {\bibfnamefont {M.}~\bibnamefont {Klaas}}, \bibinfo {author}
  {\bibfnamefont {O.~A.}\ \bibnamefont {Egorov}}, \bibinfo {author}
  {\bibfnamefont {K.}~\bibnamefont {Winkler}}, \bibinfo {author} {\bibfnamefont
  {M.}~\bibnamefont {Emmerling}}, \bibinfo {author} {\bibfnamefont
  {R.}~\bibnamefont {Thomale}}, \bibinfo {author} {\bibfnamefont
  {S.}~\bibnamefont {H\"ofling}}, \ and\ \bibinfo {author} {\bibfnamefont
  {C.}~\bibnamefont {Schneider}},\ }\bibfield  {title} {\enquote {\bibinfo
  {title} {Platform for electrically pumped polariton simulators and
  topological lasers},}\ }\href {\doibase 10.1103/PhysRevLett.121.257402}
  {\bibfield  {journal} {\bibinfo  {journal} {Phys. Rev. Lett.}\ }\textbf
  {\bibinfo {volume} {121}},\ \bibinfo {pages} {257402} (\bibinfo {year}
  {2018})}\BibitemShut {NoStop}%
\bibitem [{\citenamefont {Smirnova}\ \emph
  {et~al.}(2016{\natexlab{a}})\citenamefont {Smirnova}, \citenamefont
  {Khanikaev}, \citenamefont {Smirnov},\ and\ \citenamefont
  {Kivshar}}]{Smirnova2016}%
  \BibitemOpen
  \bibfield  {author} {\bibinfo {author} {\bibfnamefont {D.~A.}\ \bibnamefont
  {Smirnova}}, \bibinfo {author} {\bibfnamefont {A.~B.}\ \bibnamefont
  {Khanikaev}}, \bibinfo {author} {\bibfnamefont {L.~A.}\ \bibnamefont
  {Smirnov}}, \ and\ \bibinfo {author} {\bibfnamefont {Y.~S.}\ \bibnamefont
  {Kivshar}},\ }\bibfield  {title} {\enquote {\bibinfo {title} {Multipolar
  third-harmonic generation driven by optically induced magnetic resonances},}\
  }\href {\doibase 10.1021/acsphotonics.6b00036} {\bibfield  {journal}
  {\bibinfo  {journal} {{ACS} Photonics}\ }\textbf {\bibinfo {volume} {3}},\
  \bibinfo {pages} {1468--1476} (\bibinfo {year}
  {2016}{\natexlab{a}})}\BibitemShut {NoStop}%
\bibitem [{\citenamefont {Shcherbakov}\ \emph {et~al.}(2014)\citenamefont
  {Shcherbakov}, \citenamefont {Neshev}, \citenamefont {Hopkins}, \citenamefont
  {Shorokhov}, \citenamefont {Staude}, \citenamefont {Melik-Gaykazyan},
  \citenamefont {Decker}, \citenamefont {Ezhov}, \citenamefont
  {Miroshnichenko}, \citenamefont {Brener}, \citenamefont {Fedyanin},\ and\
  \citenamefont {Kivshar}}]{Shcherbakov2014}%
  \BibitemOpen
  \bibfield  {author} {\bibinfo {author} {\bibfnamefont {M.~R.}\ \bibnamefont
  {Shcherbakov}}, \bibinfo {author} {\bibfnamefont {D.~N.}\ \bibnamefont
  {Neshev}}, \bibinfo {author} {\bibfnamefont {B.}~\bibnamefont {Hopkins}},
  \bibinfo {author} {\bibfnamefont {A.~S.}\ \bibnamefont {Shorokhov}}, \bibinfo
  {author} {\bibfnamefont {I.}~\bibnamefont {Staude}}, \bibinfo {author}
  {\bibfnamefont {E.~V.}\ \bibnamefont {Melik-Gaykazyan}}, \bibinfo {author}
  {\bibfnamefont {M.}~\bibnamefont {Decker}}, \bibinfo {author} {\bibfnamefont
  {A.~A.}\ \bibnamefont {Ezhov}}, \bibinfo {author} {\bibfnamefont {A.~E.}\
  \bibnamefont {Miroshnichenko}}, \bibinfo {author} {\bibfnamefont
  {I.}~\bibnamefont {Brener}}, \bibinfo {author} {\bibfnamefont {A.~A.}\
  \bibnamefont {Fedyanin}}, \ and\ \bibinfo {author} {\bibfnamefont {Y.~S.}\
  \bibnamefont {Kivshar}},\ }\bibfield  {title} {\enquote {\bibinfo {title}
  {Enhanced third-harmonic generation in silicon nanoparticles driven by
  magnetic response},}\ }\href {\doibase 10.1021/nl503029j} {\bibfield
  {journal} {\bibinfo  {journal} {Nano Lett.}\ }\textbf {\bibinfo {volume}
  {14}},\ \bibinfo {pages} {6488--6492} (\bibinfo {year} {2014})}\BibitemShut
  {NoStop}%
\bibitem [{\citenamefont {Smirnova}\ \emph
  {et~al.}(2016{\natexlab{b}})\citenamefont {Smirnova}, \citenamefont
  {Khanikaev}, \citenamefont {Smirnov},\ and\ \citenamefont
  {Kivshar}}]{Smirnova2016ACS}%
  \BibitemOpen
  \bibfield  {author} {\bibinfo {author} {\bibfnamefont {D.~A.}\ \bibnamefont
  {Smirnova}}, \bibinfo {author} {\bibfnamefont {A.~B.}\ \bibnamefont
  {Khanikaev}}, \bibinfo {author} {\bibfnamefont {L.~A.}\ \bibnamefont
  {Smirnov}}, \ and\ \bibinfo {author} {\bibfnamefont {Y.~S.}\ \bibnamefont
  {Kivshar}},\ }\bibfield  {title} {\enquote {\bibinfo {title} {Multipolar
  third-harmonic generation driven by optically induced magnetic resonances},}\
  }\href {\doibase 10.1021/acsphotonics.6b00036} {\bibfield  {journal}
  {\bibinfo  {journal} {ACS Photonics}\ }\textbf {\bibinfo {volume} {3}},\
  \bibinfo {pages} {1468} (\bibinfo {year} {2016}{\natexlab{b}})}\BibitemShut
  {NoStop}%
\bibitem [{\citenamefont {Camacho-Morales}\ \emph {et~al.}(2016)\citenamefont
  {Camacho-Morales}, \citenamefont {Rahmani}, \citenamefont {Kruk},
  \citenamefont {Wang}, \citenamefont {Xu}, \citenamefont {Smirnova},
  \citenamefont {Solntsev}, \citenamefont {Miroshnichenko}, \citenamefont
  {Tan}, \citenamefont {Karouta}, \citenamefont {Naureen}, \citenamefont
  {Vora}, \citenamefont {Carletti}, \citenamefont {Angelis}, \citenamefont
  {Jagadish}, \citenamefont {Kivshar},\ and\ \citenamefont
  {Neshev}}]{CamachoMorales2016}%
  \BibitemOpen
  \bibfield  {author} {\bibinfo {author} {\bibfnamefont {R.}~\bibnamefont
  {Camacho-Morales}}, \bibinfo {author} {\bibfnamefont {M.}~\bibnamefont
  {Rahmani}}, \bibinfo {author} {\bibfnamefont {S.}~\bibnamefont {Kruk}},
  \bibinfo {author} {\bibfnamefont {L.}~\bibnamefont {Wang}}, \bibinfo {author}
  {\bibfnamefont {L.}~\bibnamefont {Xu}}, \bibinfo {author} {\bibfnamefont
  {D.~A.}\ \bibnamefont {Smirnova}}, \bibinfo {author} {\bibfnamefont {A.~S.}\
  \bibnamefont {Solntsev}}, \bibinfo {author} {\bibfnamefont {A.}~\bibnamefont
  {Miroshnichenko}}, \bibinfo {author} {\bibfnamefont {H.~H.}\ \bibnamefont
  {Tan}}, \bibinfo {author} {\bibfnamefont {F.}~\bibnamefont {Karouta}},
  \bibinfo {author} {\bibfnamefont {S.}~\bibnamefont {Naureen}}, \bibinfo
  {author} {\bibfnamefont {K.}~\bibnamefont {Vora}}, \bibinfo {author}
  {\bibfnamefont {L.}~\bibnamefont {Carletti}}, \bibinfo {author}
  {\bibfnamefont {C.~D.}\ \bibnamefont {Angelis}}, \bibinfo {author}
  {\bibfnamefont {C.}~\bibnamefont {Jagadish}}, \bibinfo {author}
  {\bibfnamefont {Y.~S.}\ \bibnamefont {Kivshar}}, \ and\ \bibinfo {author}
  {\bibfnamefont {D.~N.}\ \bibnamefont {Neshev}},\ }\bibfield  {title}
  {\enquote {\bibinfo {title} {Nonlinear generation of vector beams from
  {AlGaAs} nanoantennas},}\ }\href {\doibase 10.1021/acs.nanolett.6b03525}
  {\bibfield  {journal} {\bibinfo  {journal} {Nano Letters}\ }\textbf {\bibinfo
  {volume} {16}},\ \bibinfo {pages} {7191--7197} (\bibinfo {year}
  {2016})}\BibitemShut {NoStop}%
\bibitem [{\citenamefont {Frizyuk}\ \emph {et~al.}(2019)\citenamefont
  {Frizyuk}, \citenamefont {Volkovskaya}, \citenamefont {Smirnova},
  \citenamefont {Poddubny},\ and\ \citenamefont {Petrov}}]{Frizyuk2019}%
  \BibitemOpen
  \bibfield  {author} {\bibinfo {author} {\bibfnamefont {K.}~\bibnamefont
  {Frizyuk}}, \bibinfo {author} {\bibfnamefont {I.}~\bibnamefont
  {Volkovskaya}}, \bibinfo {author} {\bibfnamefont {D.}~\bibnamefont
  {Smirnova}}, \bibinfo {author} {\bibfnamefont {A.}~\bibnamefont {Poddubny}},
  \ and\ \bibinfo {author} {\bibfnamefont {M.}~\bibnamefont {Petrov}},\
  }\bibfield  {title} {\enquote {\bibinfo {title} {Second-harmonic generation
  in mie-resonant dielectric nanoparticles made of noncentrosymmetric
  materials},}\ }\href {\doibase 10.1103/PhysRevB.99.075425} {\bibfield
  {journal} {\bibinfo  {journal} {Phys. Rev. B}\ }\textbf {\bibinfo {volume}
  {99}},\ \bibinfo {pages} {075425} (\bibinfo {year} {2019})}\BibitemShut
  {NoStop}%
\bibitem [{\citenamefont {Poddubny}\ \emph {et~al.}(2014)\citenamefont
  {Poddubny}, \citenamefont {Miroshnichenko}, \citenamefont {Slobozhanyuk},\
  and\ \citenamefont {Kivshar}}]{Poddubny2014majorana}%
  \BibitemOpen
  \bibfield  {author} {\bibinfo {author} {\bibfnamefont {A.}~\bibnamefont
  {Poddubny}}, \bibinfo {author} {\bibfnamefont {A.}~\bibnamefont
  {Miroshnichenko}}, \bibinfo {author} {\bibfnamefont {A.}~\bibnamefont
  {Slobozhanyuk}}, \ and\ \bibinfo {author} {\bibfnamefont {Y.}~\bibnamefont
  {Kivshar}},\ }\bibfield  {title} {\enquote {\bibinfo {title} {Topological
  {M}ajorana states in zigzag chains of plasmonic nanoparticles},}\ }\href
  {\doibase 10.1021/ph4000949} {\bibfield  {journal} {\bibinfo  {journal} {ACS
  Photonics}\ }\textbf {\bibinfo {volume} {1}},\ \bibinfo {pages} {101}
  (\bibinfo {year} {2014})}\BibitemShut {NoStop}%
\bibitem [{\citenamefont {Slobozhanyuk}\ \emph {et~al.}(2015)\citenamefont
  {Slobozhanyuk}, \citenamefont {Poddubny}, \citenamefont {Miroshnichenko},
  \citenamefont {Belov},\ and\ \citenamefont {Kivshar}}]{Slobozhanyuk2015}%
  \BibitemOpen
  \bibfield  {author} {\bibinfo {author} {\bibfnamefont {A.~P.}\ \bibnamefont
  {Slobozhanyuk}}, \bibinfo {author} {\bibfnamefont {A.~N.}\ \bibnamefont
  {Poddubny}}, \bibinfo {author} {\bibfnamefont {A.~E.}\ \bibnamefont
  {Miroshnichenko}}, \bibinfo {author} {\bibfnamefont {P.~A.}\ \bibnamefont
  {Belov}}, \ and\ \bibinfo {author} {\bibfnamefont {Y.~S.}\ \bibnamefont
  {Kivshar}},\ }\bibfield  {title} {\enquote {\bibinfo {title} {Subwavelength
  topological edge states in optically resonant dielectric structures},}\
  }\href {\doibase 10.1103/PhysRevLett.114.123901} {\bibfield  {journal}
  {\bibinfo  {journal} {Phys. Rev. Lett.}\ }\textbf {\bibinfo {volume} {114}},\
  \bibinfo {pages} {123901} (\bibinfo {year} {2015})}\BibitemShut {NoStop}%
\bibitem [{\citenamefont {Slobozhanyuk}\ \emph
  {et~al.}(2016{\natexlab{c}})\citenamefont {Slobozhanyuk}, \citenamefont
  {Poddubny}, \citenamefont {Sinev}, \citenamefont {Samusev}, \citenamefont
  {Yu}, \citenamefont {Kuznetsov}, \citenamefont {Miroshnichenko},\ and\
  \citenamefont {Kivshar}}]{Slobozhanyuk2016LPR}%
  \BibitemOpen
  \bibfield  {author} {\bibinfo {author} {\bibfnamefont {A.~P.}\ \bibnamefont
  {Slobozhanyuk}}, \bibinfo {author} {\bibfnamefont {A.~N.}\ \bibnamefont
  {Poddubny}}, \bibinfo {author} {\bibfnamefont {I.~S.}\ \bibnamefont {Sinev}},
  \bibinfo {author} {\bibfnamefont {A.~K.}\ \bibnamefont {Samusev}}, \bibinfo
  {author} {\bibfnamefont {Y.~F.}\ \bibnamefont {Yu}}, \bibinfo {author}
  {\bibfnamefont {A.~I.}\ \bibnamefont {Kuznetsov}}, \bibinfo {author}
  {\bibfnamefont {A.~E.}\ \bibnamefont {Miroshnichenko}}, \ and\ \bibinfo
  {author} {\bibfnamefont {Y.~S.}\ \bibnamefont {Kivshar}},\ }\bibfield
  {title} {\enquote {\bibinfo {title} {Enhanced photonic spin {H}all effect
  with subwavelength topological edge states},}\ }\href {\doibase
  10.1002/lpor.201600042} {\bibfield  {journal} {\bibinfo  {journal} {Laser
  {\&} Photonics Reviews}\ }\textbf {\bibinfo {volume} {10}},\ \bibinfo {pages}
  {656} (\bibinfo {year} {2016}{\natexlab{c}})}\BibitemShut {NoStop}%
\bibitem [{\citenamefont {Kruk}\ \emph {et~al.}(2017)\citenamefont {Kruk},
  \citenamefont {Slobozhanyuk}, \citenamefont {Denkova}, \citenamefont
  {Poddubny}, \citenamefont {Kravchenko}, \citenamefont {Miroshnichenko},
  \citenamefont {Neshev},\ and\ \citenamefont {Kivshar}}]{Kruk2017zigzag}%
  \BibitemOpen
  \bibfield  {author} {\bibinfo {author} {\bibfnamefont {S.}~\bibnamefont
  {Kruk}}, \bibinfo {author} {\bibfnamefont {A.}~\bibnamefont {Slobozhanyuk}},
  \bibinfo {author} {\bibfnamefont {D.}~\bibnamefont {Denkova}}, \bibinfo
  {author} {\bibfnamefont {A.}~\bibnamefont {Poddubny}}, \bibinfo {author}
  {\bibfnamefont {I.}~\bibnamefont {Kravchenko}}, \bibinfo {author}
  {\bibfnamefont {A.}~\bibnamefont {Miroshnichenko}}, \bibinfo {author}
  {\bibfnamefont {D.}~\bibnamefont {Neshev}}, \ and\ \bibinfo {author}
  {\bibfnamefont {Y.}~\bibnamefont {Kivshar}},\ }\bibfield  {title} {\enquote
  {\bibinfo {title} {Edge states and topological phase transitions in chains of
  dielectric nanoparticles},}\ }\href {\doibase 10.1002/smll.201603190}
  {\bibfield  {journal} {\bibinfo  {journal} {Small}\ }\textbf {\bibinfo
  {volume} {13}},\ \bibinfo {pages} {1603190} (\bibinfo {year}
  {2017})}\BibitemShut {NoStop}%
\bibitem [{\citenamefont {Shalaev}\ \emph
  {et~al.}(2018{\natexlab{b}})\citenamefont {Shalaev}, \citenamefont {Desnavi},
  \citenamefont {Walasik},\ and\ \citenamefont {Litchinitser}}]{Shalaev2018}%
  \BibitemOpen
  \bibfield  {author} {\bibinfo {author} {\bibfnamefont {M.~I.}\ \bibnamefont
  {Shalaev}}, \bibinfo {author} {\bibfnamefont {S.}~\bibnamefont {Desnavi}},
  \bibinfo {author} {\bibfnamefont {W.}~\bibnamefont {Walasik}}, \ and\
  \bibinfo {author} {\bibfnamefont {N.~M.}\ \bibnamefont {Litchinitser}},\
  }\bibfield  {title} {\enquote {\bibinfo {title} {Reconfigurable topological
  photonic crystal},}\ }\href {\doibase 10.1088/1367-2630/aaac04} {\bibfield
  {journal} {\bibinfo  {journal} {New J. Phys.}\ }\textbf {\bibinfo {volume}
  {20}},\ \bibinfo {pages} {023040} (\bibinfo {year}
  {2018}{\natexlab{b}})}\BibitemShut {NoStop}%
\bibitem [{\citenamefont {Shalaev}, \citenamefont {Walasik},\ and\
  \citenamefont {Litchinitser}(2019)}]{Shalaev2019}%
  \BibitemOpen
  \bibfield  {author} {\bibinfo {author} {\bibfnamefont {M.~I.}\ \bibnamefont
  {Shalaev}}, \bibinfo {author} {\bibfnamefont {W.}~\bibnamefont {Walasik}}, \
  and\ \bibinfo {author} {\bibfnamefont {N.~M.}\ \bibnamefont {Litchinitser}},\
  }\bibfield  {title} {\enquote {\bibinfo {title} {Optically tunable
  topological photonic crystal},}\ }\href {\doibase 10.1364/optica.6.000839}
  {\bibfield  {journal} {\bibinfo  {journal} {Optica}\ }\textbf {\bibinfo
  {volume} {6}},\ \bibinfo {pages} {839} (\bibinfo {year} {2019})}\BibitemShut
  {NoStop}%
\bibitem [{\citenamefont {Kudyshev}\ \emph
  {et~al.}(2019{\natexlab{a}})\citenamefont {Kudyshev}, \citenamefont
  {Kildishev}, \citenamefont {Boltasseva},\ and\ \citenamefont
  {Shalaev}}]{Kudyshev2019termal}%
  \BibitemOpen
  \bibfield  {author} {\bibinfo {author} {\bibfnamefont {Z.~A.}\ \bibnamefont
  {Kudyshev}}, \bibinfo {author} {\bibfnamefont {A.~V.}\ \bibnamefont
  {Kildishev}}, \bibinfo {author} {\bibfnamefont {A.}~\bibnamefont
  {Boltasseva}}, \ and\ \bibinfo {author} {\bibfnamefont {V.~M.}\ \bibnamefont
  {Shalaev}},\ }\bibfield  {title} {\enquote {\bibinfo {title} {Photonic
  topological phase transition on demand},}\ }\href {\doibase
  10.1515/nanoph-2019-0043} {\bibfield  {journal} {\bibinfo  {journal}
  {Nanophotonics}\ }\textbf {\bibinfo {volume} {8}},\ \bibinfo {pages}
  {1349--1356} (\bibinfo {year} {2019}{\natexlab{a}})}\BibitemShut {NoStop}%
\bibitem [{\citenamefont {Kudyshev}\ \emph
  {et~al.}(2019{\natexlab{b}})\citenamefont {Kudyshev}, \citenamefont
  {Kildishev}, \citenamefont {Boltasseva},\ and\ \citenamefont
  {Shalaev}}]{Kudyshev2019oxides}%
  \BibitemOpen
  \bibfield  {author} {\bibinfo {author} {\bibfnamefont {Z.~A.}\ \bibnamefont
  {Kudyshev}}, \bibinfo {author} {\bibfnamefont {A.~V.}\ \bibnamefont
  {Kildishev}}, \bibinfo {author} {\bibfnamefont {A.}~\bibnamefont
  {Boltasseva}}, \ and\ \bibinfo {author} {\bibfnamefont {V.~M.}\ \bibnamefont
  {Shalaev}},\ }\bibfield  {title} {\enquote {\bibinfo {title} {Tuning topology
  of photonic systems with transparent conducting oxides},}\ }\href {\doibase
  10.1021/acsphotonics.8b01355} {\bibfield  {journal} {\bibinfo  {journal}
  {{ACS} Photonics}\ }\textbf {\bibinfo {volume} {6}},\ \bibinfo {pages}
  {1922--1930} (\bibinfo {year} {2019}{\natexlab{b}})}\BibitemShut {NoStop}%
\bibitem [{\citenamefont {Blanco-Redondo}\ \emph {et~al.}(2018)\citenamefont
  {Blanco-Redondo}, \citenamefont {Bell}, \citenamefont {Oren}, \citenamefont
  {Eggleton},\ and\ \citenamefont {Segev}}]{BlancoRedondo2018}%
  \BibitemOpen
  \bibfield  {author} {\bibinfo {author} {\bibfnamefont {A.}~\bibnamefont
  {Blanco-Redondo}}, \bibinfo {author} {\bibfnamefont {B.}~\bibnamefont
  {Bell}}, \bibinfo {author} {\bibfnamefont {D.}~\bibnamefont {Oren}}, \bibinfo
  {author} {\bibfnamefont {B.~J.}\ \bibnamefont {Eggleton}}, \ and\ \bibinfo
  {author} {\bibfnamefont {M.}~\bibnamefont {Segev}},\ }\bibfield  {title}
  {\enquote {\bibinfo {title} {Topological protection of biphoton states},}\
  }\href {\doibase 10.1126/science.aau4296} {\bibfield  {journal} {\bibinfo
  {journal} {Science}\ }\textbf {\bibinfo {volume} {362}},\ \bibinfo {pages}
  {568--571} (\bibinfo {year} {2018})}\BibitemShut {NoStop}%
\bibitem [{\citenamefont {Wang}\ \emph
  {et~al.}(2019{\natexlab{b}})\citenamefont {Wang}, \citenamefont {Doyle},
  \citenamefont {Bell}, \citenamefont {Collins}, \citenamefont {Magi},
  \citenamefont {Eggleton}, \citenamefont {Segev},\ and\ \citenamefont
  {Blanco-Redondo}}]{Wang2019entangled}%
  \BibitemOpen
  \bibfield  {author} {\bibinfo {author} {\bibfnamefont {M.}~\bibnamefont
  {Wang}}, \bibinfo {author} {\bibfnamefont {C.}~\bibnamefont {Doyle}},
  \bibinfo {author} {\bibfnamefont {B.}~\bibnamefont {Bell}}, \bibinfo {author}
  {\bibfnamefont {M.~J.}\ \bibnamefont {Collins}}, \bibinfo {author}
  {\bibfnamefont {E.}~\bibnamefont {Magi}}, \bibinfo {author} {\bibfnamefont
  {B.~J.}\ \bibnamefont {Eggleton}}, \bibinfo {author} {\bibfnamefont
  {M.}~\bibnamefont {Segev}}, \ and\ \bibinfo {author} {\bibfnamefont
  {A.}~\bibnamefont {Blanco-Redondo}},\ }\bibfield  {title} {\enquote {\bibinfo
  {title} {Topologically protected entangled photonic states},}\ }\href
  {\doibase 10.1515/nanoph-2019-0058} {\bibfield  {journal} {\bibinfo
  {journal} {Nanophotonics}\ }\textbf {\bibinfo {volume} {8}},\ \bibinfo
  {pages} {1327--1335} (\bibinfo {year} {2019}{\natexlab{b}})}\BibitemShut
  {NoStop}%
\bibitem [{\citenamefont {Blanco-Redondo}\ \emph {et~al.}(2016)\citenamefont
  {Blanco-Redondo}, \citenamefont {Andonegui}, \citenamefont {Collins},
  \citenamefont {Harari}, \citenamefont {Lumer}, \citenamefont {Rechtsman},
  \citenamefont {Eggleton},\ and\ \citenamefont
  {Segev}}]{blanco2016topological}%
  \BibitemOpen
  \bibfield  {author} {\bibinfo {author} {\bibfnamefont {A.}~\bibnamefont
  {Blanco-Redondo}}, \bibinfo {author} {\bibfnamefont {I.}~\bibnamefont
  {Andonegui}}, \bibinfo {author} {\bibfnamefont {M.~J.}\ \bibnamefont
  {Collins}}, \bibinfo {author} {\bibfnamefont {G.}~\bibnamefont {Harari}},
  \bibinfo {author} {\bibfnamefont {Y.}~\bibnamefont {Lumer}}, \bibinfo
  {author} {\bibfnamefont {M.~C.}\ \bibnamefont {Rechtsman}}, \bibinfo {author}
  {\bibfnamefont {B.~J.}\ \bibnamefont {Eggleton}}, \ and\ \bibinfo {author}
  {\bibfnamefont {M.}~\bibnamefont {Segev}},\ }\bibfield  {title} {\enquote
  {\bibinfo {title} {Topological optical waveguiding in silicon and the
  transition between topological and trivial defect states},}\ }\href {\doibase
  10.1103/PhysRevLett.116.163901} {\bibfield  {journal} {\bibinfo  {journal}
  {Phys. Rev. Lett.}\ }\textbf {\bibinfo {volume} {116}},\ \bibinfo {pages}
  {163901} (\bibinfo {year} {2016})}\BibitemShut {NoStop}%
\bibitem [{\citenamefont {Marino}\ \emph {et~al.}(2019)\citenamefont {Marino},
  \citenamefont {Solntsev}, \citenamefont {Xu}, \citenamefont {Gili},
  \citenamefont {Carletti}, \citenamefont {Poddubny}, \citenamefont {Rahmani},
  \citenamefont {Smirnova}, \citenamefont {Chen}, \citenamefont {Lema\^{i}tre},
  \citenamefont {Zhang}, \citenamefont {Zayats}, \citenamefont {Angelis},
  \citenamefont {Leo}, \citenamefont {Sukhorukov},\ and\ \citenamefont
  {Neshev}}]{Marino2019}%
  \BibitemOpen
  \bibfield  {author} {\bibinfo {author} {\bibfnamefont {G.}~\bibnamefont
  {Marino}}, \bibinfo {author} {\bibfnamefont {A.~S.}\ \bibnamefont
  {Solntsev}}, \bibinfo {author} {\bibfnamefont {L.}~\bibnamefont {Xu}},
  \bibinfo {author} {\bibfnamefont {V.~F.}\ \bibnamefont {Gili}}, \bibinfo
  {author} {\bibfnamefont {L.}~\bibnamefont {Carletti}}, \bibinfo {author}
  {\bibfnamefont {A.~N.}\ \bibnamefont {Poddubny}}, \bibinfo {author}
  {\bibfnamefont {M.}~\bibnamefont {Rahmani}}, \bibinfo {author} {\bibfnamefont
  {D.~A.}\ \bibnamefont {Smirnova}}, \bibinfo {author} {\bibfnamefont
  {H.}~\bibnamefont {Chen}}, \bibinfo {author} {\bibfnamefont {A.}~\bibnamefont
  {Lema\^{i}tre}}, \bibinfo {author} {\bibfnamefont {G.}~\bibnamefont {Zhang}},
  \bibinfo {author} {\bibfnamefont {A.~V.}\ \bibnamefont {Zayats}}, \bibinfo
  {author} {\bibfnamefont {C.~D.}\ \bibnamefont {Angelis}}, \bibinfo {author}
  {\bibfnamefont {G.}~\bibnamefont {Leo}}, \bibinfo {author} {\bibfnamefont
  {A.~A.}\ \bibnamefont {Sukhorukov}}, \ and\ \bibinfo {author} {\bibfnamefont
  {D.~N.}\ \bibnamefont {Neshev}},\ }\bibfield  {title} {\enquote {\bibinfo
  {title} {Spontaneous photon-pair generation from a dielectric nanoantenna},}\
  }\href {\doibase 10.1364/OPTICA.6.001416} {\bibfield  {journal} {\bibinfo
  {journal} {Optica}\ }\textbf {\bibinfo {volume} {6}},\ \bibinfo {pages}
  {1416--1422} (\bibinfo {year} {2019})}\BibitemShut {NoStop}%
\bibitem [{\citenamefont {Peano}\ \emph
  {et~al.}(2016{\natexlab{b}})\citenamefont {Peano}, \citenamefont {Houde},
  \citenamefont {Brendel}, \citenamefont {Marquardt},\ and\ \citenamefont
  {Clerk}}]{Peano2016b}%
  \BibitemOpen
  \bibfield  {author} {\bibinfo {author} {\bibfnamefont {V.}~\bibnamefont
  {Peano}}, \bibinfo {author} {\bibfnamefont {M.}~\bibnamefont {Houde}},
  \bibinfo {author} {\bibfnamefont {C.}~\bibnamefont {Brendel}}, \bibinfo
  {author} {\bibfnamefont {F.}~\bibnamefont {Marquardt}}, \ and\ \bibinfo
  {author} {\bibfnamefont {A.~A.}\ \bibnamefont {Clerk}},\ }\bibfield  {title}
  {\enquote {\bibinfo {title} {Topological phase transitions and chiral
  inelastic transport induced by the squeezing of light},}\ }\href {\doibase
  10.1038/ncomms10779} {\bibfield  {journal} {\bibinfo  {journal} {Nature
  Communications}\ }\textbf {\bibinfo {volume} {7}},\ \bibinfo {pages} {10779}
  (\bibinfo {year} {2016}{\natexlab{b}})}\BibitemShut {NoStop}%
\end{thebibliography}

%

\end{document}